%% file: main.tex
\documentclass[letterpaper]{article}

\usepackage[T1]{fontenc}

\usepackage{geometry}
\usepackage{setspace}

\usepackage[usetitle=true]{achemso}
\usepackage{url}

\usepackage{graphicx}
\usepackage{float}
\newfloat{scheme}{htbp}{los}
\floatname{scheme}{Scheme}
\floatname{chart}{Chart}
\newfloat{graph}{htbp}{loh}

\usepackage{chemformula} 
\usepackage[version = 4]{mhchem} 

\usepackage{authblk}
\author[1]{Thang Pham}
\author[2,3,4]{Vsevolod Ivanov}
\author[5]{Dominic P. Goronzy}
\author[5]{Abhiram Devata}
\author[5]{Joshua Feldon}
\author[5]{David Barton}
\author[6]{You Zhou}
\affil[1]{Department of Materials Science and Engineering, Virginia Tech, Blacksburg, Virginia 24061, USA}
\affil[2]{Virginia Tech National Security Institute, Blacksburg, Virginia 24060, USA}
\affil[3]{Department of Physics, Virginia Tech, Blacksburg, Virginia 24061, USA}
\affil[4]{Virginia Tech Center for Quantum Information Science and Engineering, Blacksburg, Virginia 24061, USA}
\affil[5]{Department of Materials Science and Engineering, Northwestern University, Evanston, Illinois 60208, USA}
\affil[6]{Department of Materials Science and Engineering, University of Maryland, College Park, Maryland 20742, USA}

\title{Materials for Quantum Information Science: Roles in the Quantum Evolution 2.0}
\date{*Email: thangpham@vt.edu}

\begin{document}

\maketitle

\begin{abstract}
  Quantum information science is entering a second phase, the Quantum Evolution 2.0, in which the challenge has shifted from demonstrating coherent control of individual quantum states to building scalable multi-qubit processors and networks. This transition places materials science at the center of the field. Across superconducting circuits, quantum defects, quantum photonic devices, and emerging materials platforms, including two-dimensional materials and heterostructures, performance is now limited less by device design than by poorly controlled surfaces, buried interfaces, and defects whose atomic identities remain incompletely known. This review surveys the materials challenges of these quantum platforms together with the characterization methods needed to resolve them. For each platform we identify the dominant decoherence mechanisms, the current state of materials understanding, and the most pressing open materials problems. A cross-platform comparison then reveals a shared structure-coherence problem. The implicated material chemistry recurs across platforms, involving light elements in disordered or buried environments, yet no platform can quantitatively connect a specific atomic-scale structure to a measured change in coherence. We close by identifying three needs, mechanistic understanding of decoherence at the atomistic level, high-throughput proxy metrics predictive of device performance, and characterization tools built for quantum materials, whose resolution would advance coherence, scalability, and integration across all platforms.
\end{abstract}

\pagebreak

\setcounter{secnumdepth}{3}
\setcounter{section}{0}

\input{1.introduction.tex}
\input{2.superconducting_qubits.tex}
\input{3.quantum_defects.tex}

\input{4.quantum_photonics.tex}
\input{5.two-dimensional_materials.tex}

\input{6.materials_characterization.tex}

\input{7.outlook.tex}

\section*{Author Contributions} T.P. conceived the review, defined its scope, and coordinated the collaboration. D.P.G. drafted Section 2, V.I. drafted Section 3, A.D., J.F., and D.B. drafted Section 4, Y.Z. drafted Section 5, and T.P. drafted Section 6, the Introduction, and Section 7 with input from all co-authors. All authors contributed to the review, editing, and revision of the complete manuscript and approved its final version.

\section*{Acknowledgments}
The authors thank M. C. Hersam and V. P. Dravid for encouragement for discussions that led to this manuscript. 
Funding: T. P. is supported by Virginia Tech's College of Engineering and the department of Materials Science and Engineering. V. I. acknowledges support from the U.S. National Science Foundation Growing Convergence Research Grant OIA-2428507, and support by the U.S. Department of Energy, Office of Science, Office of High Energy Physics Quantum Technology Outposts in Fundamental Physics under Award Number DE-SC0026492. Y.Z. acknowledges support from the Army Research Office W911NF2510066, the U.S. Department of Energy, Office of Science, Office
of Basic Energy Sciences, Award No. DE-SC-0022885 and the U.S. National Science Foundation OSI-2553574. 
Conflict of interest: The authors declare no competing interests.



\bibliography{merged.bib}

\end{document}

%% file: 1.introduction.tex


\section{Introduction}

The ability to harness quantum mechanical phenomena, namely superposition, entanglement, and quantum coherence, for information gathering, transmission, and  processing represents one of the most consequential technological transitions of our era. Quantum systems promise computational advantages for problems that are intractable on classical hardware, from simulating quantum chemistry and materials to executing Shor's algorithm for cryptographic applications. Beyond computation, quantum coherence underpins quantum communication protocols and quantum sensing with sensitivities that surpass classical limits. Together, these capabilities define what has been termed the second quantum revolution: a shift from passively exploiting quantum effects, as in the transistor or the laser, to actively engineering and manipulating individual quantum states.

\begin{figure}[htbp]
    \centering
    \includegraphics[width=0.7\textwidth]{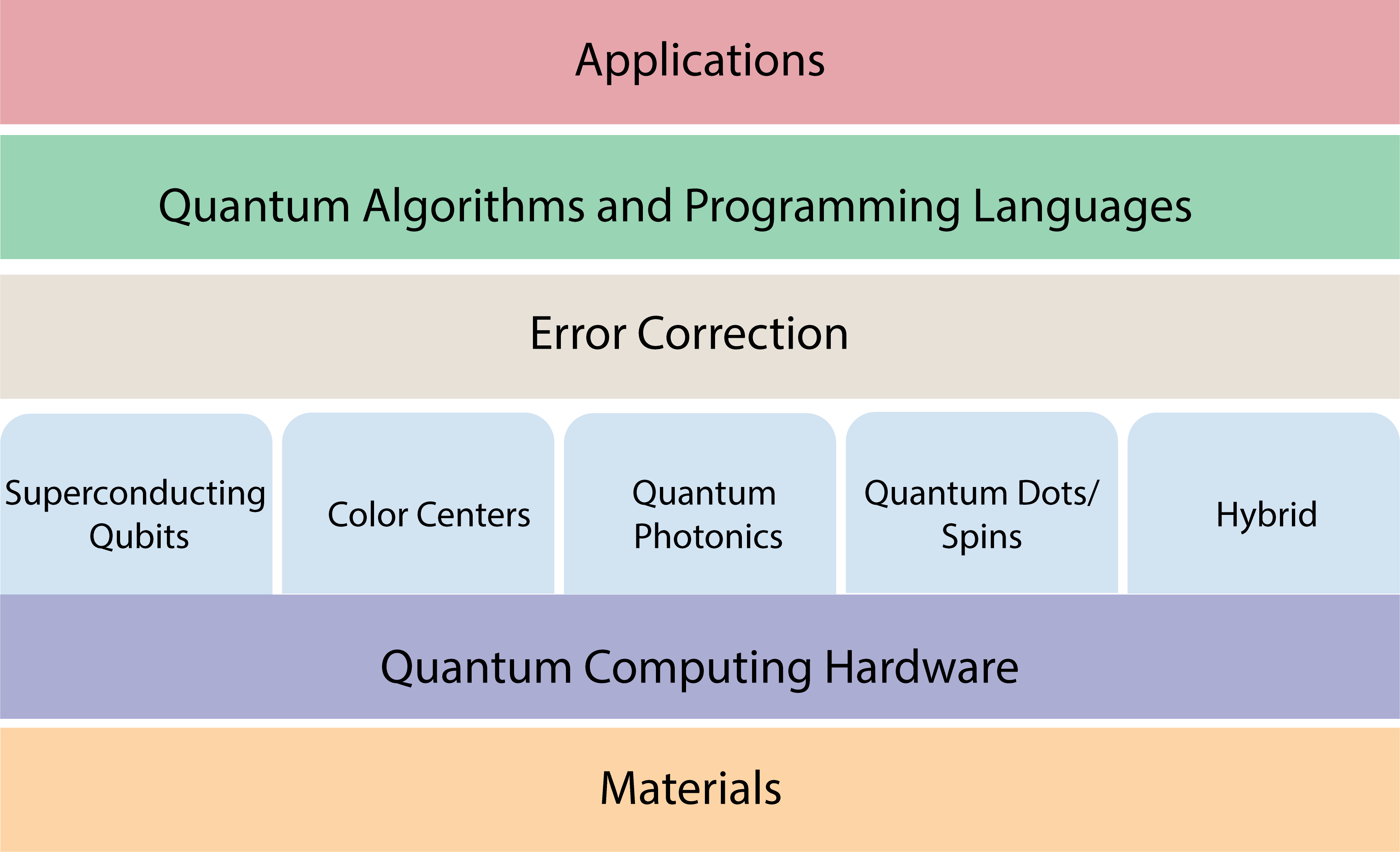}
    \caption{Ecosystem of Quantum Evolution 2.0. The "Materials" block is the focus of this review. Images adapted and modified with permission from ref \cite{Murray2021}
    .}
    \label{fig:Overview}
\end{figure}

Progress over the past two decades has been substantial with superconducting processors, color-center spin qubits, semiconductor quantum dots, trapped ions, and emerging topological platforms all having demonstrated proof-of-principle quantum operations. Coherence times have improved by orders of magnitude, from nanoseconds to milliseconds in superconducting qubits, driven largely by advances in circuit design, qubit geometry, and device engineering. Recent demonstrations include quantum processors operating beyond the reach of classical simulation and multi-qubit gate fidelities exceeding 9x\% in several platforms. These achievements mark the culmination of a first phase of QIS development, one defined by establishing that coherent quantum control is physically realizable (Figure \ref{fig:Overview}).

We are now entering a second phase, the so-called “Quantum Evolution 2.0”, in which the challenge has shifted from demonstrating isolated quantum operations to building functional multi-qubit processors capable of scalable, fault-tolerant computation. This transition demands a new wave of research different from what drove earlier progress. Improvements in pulse sequences, circuit topology, and measurement protocols can optimize the performance of a given physical system, but they cannot overcome the fundamental materials constraints that now limit qubit quality, reproducibility, and integration. As de Leon et al. articulated in a 2021 landmark review \cite{Leon2021}, materials science has so far informed QIS primarily through the down-selection of favorable material platforms, for example the kinetically limited AlO$_x$ tunnel barrier in Josephson junctions (JJs), the isotopically purified $^28$Si host for spin qubits, the high-purity CVD diamond for color centers, but comparatively little work has been directed at using the full toolkit of materials science to systematically improve and scale quantum hardware. The implication is clear: the gains available from device engineering are approaching saturation, and further progress requires a bottom-up materials approach that controls the atomic structure of interfaces, surfaces, defects, and thin films across all relevant platforms (Figure \ref{fig:Overview}).

A critical and underappreciated aspect of this materials challenge is that it is not platform-specific. Despite the profound differences in physical implementation, such as a Josephson junction circuit cooled to millikelvin temperatures, a spin defect in a diamond lattice, or a proximitized semiconductor nanowire, the materials problems that limit performance share a common detractor. Poorly controlled surfaces and interfaces introduce electric and magnetic field noise that degrades coherence across all solid-state platforms. Structural and chemical disorder at buried interfaces, for instance between metal and substrate, between superconductor and semiconductor, between 2D layers, creates loss channels whose microscopic origins remain incompletely identified in every quantum platform. Defects and impurities in bulk and thin-film materials set bounds on coherence that cannot be circumvented by device design alone. And scaling to larger qubit systems introduces new materials problems, namely wafer-scale uniformity, interface reproducibility, 3D integration, that are largely invisible at the single-qubit level. Recognizing these commonalities is not merely an observation: it opens the possibility of cross-platform learning, where insights developed in one community, such as the microwave loss metrology of superconducting qubits, the surface chemistry of diamond, the epitaxial interface engineering of semiconductor heterostructures, can accelerate progress in others.

This framing motivates two parallel goals for this review (Figure 1). The first is platform-specific: for each materials system covered, we aim to identify the dominant decoherence mechanisms, assess the current state of materials understanding, and highlight the most pressing open challenges. The second is cross-platform: by surveying these systems together, we seek to identify the common materials problems that plague QIS broadly, and to propose how addressing them holistically through shared characterization methodologies, transferable synthesis strategies, and cross-disciplinary collaboration leading to acceleration of the field in ways that platform-siloed efforts cannot.

Concretely, the review addresses three interconnected areas. 
We first discuss superconducting qubits from a materials perspective, tracing the connection from thin-film deposition and interface chemistry through microwave loss mechanisms to the structure–coherence correlation problem (Section 2). We then examine quantum defects in bulk solid-state materials, focusing on spin and single-photon emitter properties and the emerging role of high-throughput computational screening in accelerating defect discovery (Section 3). Section 4 covers quantum photonic platform, spanning from a review of materials selection, to state-of-the-art device fabrication as well as the challenges of the field in scalable light-based quantum devices. We examine two-dimensional (2D) van der Waals (vdW) materials and heterostructures as an emerging quantum platform, encompassing single-photon emitters, spin qubits, and proximitized topological systems (Section 5). Finally, a dedicated section on materials characterization surveys the techniques, from scanning transmission electron microscopy, scanning probe methods to x-ray spectroscopy, that translate structural observables into coherence metrics across platforms (Section 6). The review closes with an outlook that draws cross-platform lessons and identifies the materials challenges whose resolution will most directly enable the Quantum Evolution 2.0.

%% file: 2.superconducting_qubits.tex

\section{Superconducting Qubits: From Materials to Coherence}

Superconducting circuits, like many of its peer technologies in the field of Quantum Information Sciences (QIS), have seen tremendous growth over the last three decades, which has only accelerated in the last few years. Furthermore, similar to other QIS technology implementations, advancement in superconducting hardware is increasingly limited by materials challenges, including incomplete atomistic understanding of decoherence mechanisms, limited control of surfaces and buried interfaces, fabrication-induced defects, and the lack of materials proxy measurements predictive of qubit performance \cite{Leon2021}. A superconducting qubit operates on the basis of using superconducting circuits elements to create an anharmonic oscillator to which quantum information can be encoded. This setup offers the advantages of leveraging well established deposition, lithography, and other fabrication techniques from the semiconductor chip industry and provides high tunability in qubit parameters through circuit layout and design \cite{Kjaergaard2020, Krantz2019}. From the first demonstration of a Cooper-pair box by Nakamura \textit{et al.} in 1999 to the cutting edge of device performance today, we have seen coherence times improve by six orders of magnitude \cite{Nakamura1999, Bland2025}. A significant portion of these improvements has come with improved circuit design, particularly the development of the transmon, a capacitor shunted by a Josephson junction, which is now the preeminent circuit design in the field \cite{Koch2007}. However, even in the case of the transmon and other contemporary circuit designs, like the fluxonium and the mergemon, the field has reached the point where performance is now dictated by materials limitations \cite{Siddiqi2021, Murray2021, Zhao2020, Somoroff2023}. Consequently, a central challenge facing the field is establishing quantitative relationships between materials structure and properties, fabrication process history, and measured qubit coherence.

When it comes to assessing the performance of a superconducting qubit and its underlying materials, we are interested in the decoherence time (T$_{2}$), where decoherence is the loss of information in our qubit. Decoherence is a function of both energy relaxation (T$_{1}$) and dephasing (T$_\Phi$). In the context of resonators, we are also interested in the quality factor (Q), which can be related to T$_1$. Resonators are a powerful tool in evaluating materials without investing in the full qubit fabrication process \cite{McRae2020_resonators}. Broadly speaking there are two categories of noise or loss that drive decoherence: low-power loss and power-independent loss. Low-power loss is typically dielectric loss or two-level system (TLS) loss, where aberrant TLS in the materials couple to the qubit and extract energy resulting in decay. TLS loss is dominant at low power where the TLS are unsaturated, but since this is also the operating range of a qubit, this tends to be the dominant loss source driving decoherence. Another factor to consider in dielectric loss is not only the intrinsic loss of a material, but where that material is located within the qubit circuit layout and the level of participation that region has in the overall electric field of the qubit. Power-independent loss accounts for a multitude of sources that are both based in the materials and related to extrinsic factors. One last noise source to consider is $1/f$ noise, however the source of this noise is poorly understood and is possibly driven by material properties but is an area that requires further research. There are several high-quality reviews that have written in detail about these loss sources and the role materials play in qubit decoherence \cite{Oliver2013, Siddiqi2021, Murray2021, McRae2020_resonators}. Rather than providing a comprehensive review of superconducting qubit decoherence mechanisms, here we focus on recent advances in understanding how materials selection and synthesis, interface chemistry, thin film microstructure, and Josephson junction fabrication influence quantum coherence. Particular emphasis is placed on the emerging challenge of establishing predictive relationships between measurable materials properties and device performance and new avenues of materials characterization that need to be pursued further.
\begin{figure}[htbp]
    \centering
    \includegraphics[width=1\textwidth]{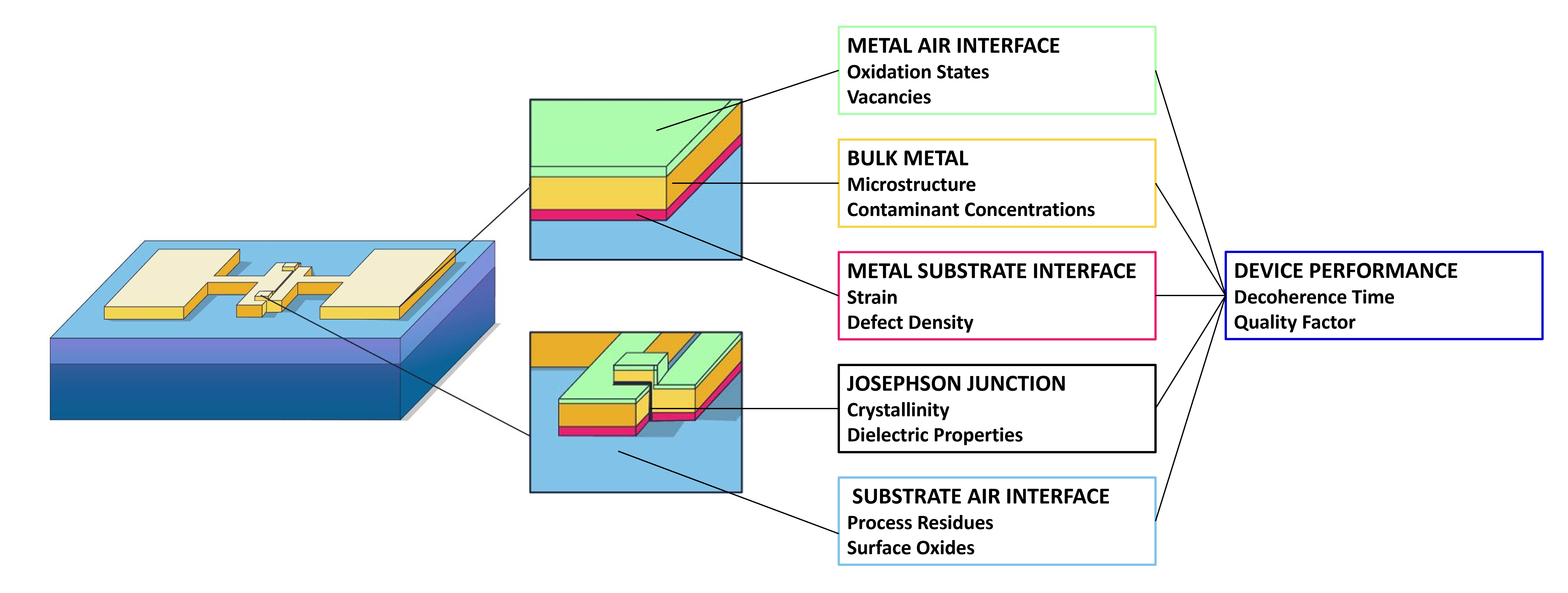}
    \caption{Superconducting qubit performance is a function of the properties of the constituent materials and the surfaces and interfaces within the qubit device. The future of improved performance lies in understand these structure-property-performance relationships and developing novel materials and fabrication strategies to mitigate decoherence. }
    \label{super_q_overview}
\end{figure}

\subsection{Surface Chemistry and Interface Engineering}

The substantial improvements in superconducting qubit coherence achieved over the last several years have increasingly indicated that coherence is often limited by defects residing within nanometer-scale surface oxides and buried interfaces, rather than the bulk superconducting films themselves \cite{wang_surface_2015, Dial2016, Gambetta2017, Woods2019, McRae2020}. Superconducting thin films such as aluminum (Al), niobium (Nb), tantalum (Ta), and titanium nitride (TiN) all exhibit sufficiently low intrinsic microwave loss for quantum circuit applications and in particular Nb has established itself as a favorable material, given its compatibility with scalable manufacturing, high transition temperature, and low kinetic inductance \cite{Murray2021, Verjauw2022, Nersisyan2019, Tolpygo2015, Oliver2013, Charpentier2025}. However, recent studies have identified Ta as a promising alternative to Nb, with otherwise identical qubit designs exhibiting substantially improved coherence when the Nb wiring is replaced with Ta \cite{Wang2022_Ta, Place2021}. Furthermore, Bland \textit{et al.} recently demonstrated T$_1$ lifetimes in excess of 1 ms in a Ta on silicon (Si) based transmon qubit \cite{Bland2025}. 

The origin of these improvements remains an active area of investigation, but increasing evidence implicates differences in oxide chemistry between Nb and Ta. Native oxides are unavoidable in ambient environments and occupy regions of high electric-field participation at metal-air and metal-substrate interfaces. While both Nb and Ta form amorphous pentoxide-rich surface layers, detailed materials characterization has shown important differences in oxide composition and structure. In particular, native Nb oxides commonly exhibit an oxygen concentration gradient containing multiple suboxide phases, whereas Ta oxides are generally more chemically uniform and contain fewer suboxide species \cite{Premkumar2021, Crowley2023, Mun2023, McLellan2023}. Recent measurements of deposited amorphous pentoxides further suggest that Nb$_2$O$_5$ itself exhibits a higher dielectric loss tangent than Ta$_2$O$_5$, indicating that differences in the pentoxide phase may contribute to the superior performance of Ta circuits in addition to the reduced prevalence of suboxide species \cite{Goronzy2026}. Later in this review, in the section on materials characterization, we discuss more in-depth the structural differences of Nb and Ta oxide and how that deeper understanding can guide the field in identifying atomistic sources of TLS and decoherence. Understanding the microscopic origins of loss in Nb and Ta oxides has also become an active area of computational research, with studies examining the origin of TLSs in the amorphous matrix, the role of tunneling hydrogen in the oxides, and differences in paramagnetic states in Nb and Ta pentoxides \cite{Gorgichuk2023, Mendez2026, Pritchard2025}.

The benefits of the Ta surface oxide have also been demonstrated as an encapsulation scheme, showing that similar enhancement in performance can be achieved with a Nb base layer that is \textit{in situ} capped with Ta to prevent formation of Nb oxide \cite{Bal2024}. It is also noteworthy that a similar encapsulation of Ta with noble metals, such as Au and AuPd, showed no significant difference between capped and uncapped devices, indicating that these devices are more loss dominated by oxides formed on the sidewalls of etched circuit elements rather than the top surface \cite{Chang2025}. This finding is also supported by participation-ratio analyses showing that the substrate-air and metal sidewall-air interfaces often contribute substantially more electric-field participation than the top metal surface \cite{Goronzy2026}. To this end there have been a limited number of attempts at conformal passivation strategies that can simultaneously address the top, sidewall, and substrate surfaces utilizing strategies such as self-assembled monolayer surface functionalization or atomic layer etching \cite{Alghadeer2022, Gupta2026, Mahuli2025}.

While considerable attention has focused on the metal-air interface, losses associated with the substrate-air interface are often equally, if not more important due to both the large electric-field participation and the tendency of substrates such as Si to form native dielectric layers \cite{Altoe2022}. Fluorinated wet etchants, such as hydrofluoric acid (HF) and buffered oxide etch (BOE), are widely used throughout qubit fabrication to remove native oxides from both superconducting films and substrate surfaces as well as residual process contamination. Their implementation has repeatedly been associated with reductions in decoherence-related loss channels \cite{Altoe2022, Gingras2025}. Furthermore, employing sequential etch protocols of HF and ammonium fluoride to modify the surface structure and termination of the Si surface at various points in the qubit fabrication process flow has demonstrated tangible improvements in materials properties and device performance \cite{Berti2023, Kopas2024}. However, it has also been shown that these wet chemical treatments can drive the incorporation of hydrogen and the formation of hydrides in the metal base layer in both the case of Nb and Ta and these hydrides are a source of power-independent loss \cite{TorresCastanedo2024, Lozano2025}. These studies highlight a recurring challenge in superconducting qubit materials engineering: process steps introduced to mitigate one source of loss frequently create new defects that introduce different decoherence channels. Such tradeoffs are only revealed by systematic studies examining individual sources of decoherence and understanding the mechanisms of loss from a materials standpoint. Building on that fundamental information, qubit fabrication processes can be optimized, which in this case is finding the appropriate amount of wet chemical etching to mitigate lossy surface oxides without introducing new loss from excess hydrides. Improvements in device performance increasingly require understanding not only the chemistry of individual interfaces, but also how entire fabrication workflows modify competing defect populations.

\subsection{Buried Interfaces and the Josephson Junction}

While the role of exposed surfaces in superconducting qubit decoherence has become increasingly apparent, buried interfaces present a distinct and often more challenging materials characterization and fabrication problem. Unlike the metal-air or substrate-air interfaces, which can be modified through cleaning, passivation, or encapsulation strategies, buried interfaces are largely inaccessible in the downstream fabrication processing. Furthermore, these interfaces have a significant portion of electric field participation, making the defects that reside in them potentially strong drivers of decoherence. Consequently, as losses associated with exposed surfaces are progressively mitigated, buried interfaces are emerging as an increasingly important frontier in the pursuit of higher coherence superconducting devices.

The most prominent example of such a buried interface is the Josephson junction, which serves as both a defining circuit element of most superconducting qubit architectures and one of the more complex materials systems present within the device \cite{Krantz2019, Kjaergaard2020, Murray2021}. In its most frequent implementation, the Josephson junction consists of an ultrathin amorphous AlO$_x$ tunnel barrier confined between polycrystalline Al electrodes. From a spatial perspective, the dimensions of the Josephson junction make it incredibly difficult to perform materials characterization on. This challenge and the current state of understanding of the materials structure of the junction are discussed further in the materials characterization section of this review. Despite decades of study, the extent to which the tunnel barrier itself limits the performance of state-of-the-art superconducting qubits remains an active topic of discussion. One approach to understanding the relative importance of junction defects has been to directly probe strongly coupled TLS and infer their spatial distribution within the device. Lisenfeld \textit{et al.} used electric field bias and mechanical strain to tune TLS defects and found that qubit decoherence is still largely dominated by defects on surfaces and other interfaces while $\sim 40$\% of defects occur in parasitic junctions where the Josephson junction electrodes contact the capacitive pads and only $\sim 3$\% of defects are associated with the tunnel barrier \cite{Lisenfeld2019}. Looking more closely at the strongly coupled TLS in the tunnel barrier, a later study by the same group found that these defects seem to be evenly distributed within the bulk of the dielectric and are not edge or surface defects, while also appearing to present similarly across varying junction fabrication processes \cite{Bilmes2022}. Moreover, other studies have shown that reducing the junction size is a critical parameter to control that can achieve a reduction in losses \cite{ColaoZanuz2025}. However, a recent study by Pfaff and coworkers, looking at a large dataset of junction defects and fabrication parameters, identified key correlations between TLS density and Al electrode thickness and lateral Al grain size suggesting that further optimization of junction fabrication parameters is required to achieve higher coherence devices \cite{Wolff2026}. Given the disordered, amorphous nature of the AlO$_x$ tunnel barrier the specific nature of defects is challenging to characterize and here we heavily rely on computational modeling. These theoretical studies have shown that incomplete oxidation and variation in the oxide stoichiometry can lead to issues of increased quasiparticle current flow and leakage current and cause large scale deviations in junction resistance and critical current \cite{Bayros2024, Lapham2022}. In addition to introducing potential decoherence channels, these defects contribute to junction-to-junction variability and present a significant challenge for frequency targeting and scalable manufacturing.

Beyond the junction itself, characterization of other buried interfaces within the qubit architecture is an evolving area of research. The metal-substrate interface is an area of significant potential defect induced loss given the high electric field participation ratio arising from the high dielectric constant of the substrate \cite{Wisbey2010, Calusine2018, Woods2019}. Nb on Si is one of the most prolific materials stacks in the field and there have been several studies to characterize the intermixing region that forms at the Nb/Si interface and associated potential sources of loss, but this remains a poorly understood area \cite{Lu2022, Oh2022, Kopas2020}. Sapphire is also a popular substrate of choice, especially with the growing usage of Ta as a superconducting base layer, given the need for elevated temperature deposition and lattice orientation matching to achieve the preferred phase of Ta \cite{Place2021, Wang2022_Ta}. Work by McFadden \textit{et al.} examined the interface of both Nb and Ta with the sapphire substrate as a function of deposition temperature and substrate surface preparation and found increased loss channels specifically in the case of epitaxial Ta grown directly on sapphire and surprisingly no similar losses were found with Nb \cite{McFadden2025}. Collectively these studies illustrate that there is still a significant information gap in understanding the materials defects driving loss at these buried interfaces. In contrast to exposed surfaces, the inaccessibility of these interfaces for characterization and experimentation results in that the correlation between buried interface structure and quantum performance remains substantially less developed.

\subsection{Fabrication Effects on Quantum Coherence}

The surface and buried interfaces discussed in the previous sections do not arise solely from the intrinsic properties of the constituent materials, but also from the sequence of processing steps used to fabricate a superconducting quantum circuit. As a result, the final defect landscape of a qubit is determined not only by material selection, but also by substrate preparation, thin-film deposition, lithographic patterning, etching, cleaning, and Josephson junction fabrication. Over the last several years, it has become increasingly clear that fabrication-induced defects can contribute as significantly to decoherence as the intrinsic defects inherent to the materials themselves. Furthermore, processing steps introduced to mitigate one source of loss often simultaneously modify or create alternative loss channels, complicating optimization efforts. Consequently, understanding the evolution of materials defects throughout the fabrication workflow has emerged as a critical challenge for both achieving higher coherence and reducing device-to-device variability.

The fabrication process naturally starts with the substrate which critically has by far the highest electric field participation ratio of any part of the device and it has already been well established the substrate selection is key to device performance \cite{Murray2021}. Notably, in a recent study Sipahigil and coworkers showed that acceptor dopants in the Si substrate, like boron, create a TLS bath that can drive decoherence, demonstrating the need for ultrapure substrates as the field pushes the frontier of device performance \cite{Zhang2024}. Similarly, issues have been observed with sapphire substrates where studies have demonstrated difference in dielectric loss based on the growth method of the substrate crystal, with HEMEX sapphire showing the lowest loss \cite{Read2023}. These results highlight that as the field further improves the loss coming from other elements of the qubit device, the substrate will play a greater role in limiting coherence time.

As discussed in the previous sections there are several forms of substrate preparation including annealing, chemical passivation, and oxide removal that can affect the subsequent device performance. As such, the optimization of the base metal layer deposition is an active area of investigation that goes beyond just materials selection. Sputter deposition is one of the most common physical vapor deposition techniques used in the field and a recent study compared high-power impulse magnetron sputtering (HiPIMS) and direct current (DC) magnetron sputtering at varying powers and demonstrated that high power DC sputtering, which also achieves the highest rate, provides that largest Nb grain size and the best device performance \cite{Oh2024}. The role of the film microstructure on performance is still poorly understood and there are several avenues being explored, including the role of oxygen segregation at grain boundaries, vortex states, or magnetic flux penetration \cite{Lee2026, Park2025, Datta2026}. A noteworthy recent finding demonstrated that by using krypton, as opposed to the standard argon process gas, the desired phase of Ta could be achieved at a lower deposition temperature and device performance improved \cite{Olszewski2026}. Alternatively, there have been multiple studies that have demonstrated that e-beam deposition of Nb, despite the significantly slower deposition rate can produce devices with even higher quality factors \cite{Kowsari2021, Zheng2022}. Accordingly, along with microstructure the question of impurities introduced during the fabrication process is also a persistent one. Comparative studies across multiple fabrication facilities have further shown that deposition and lithography processes affect the chemistry of both exposed and buried interfaces, introducing impurities including hydrogen, carbon, oxygen, and halogens into the final device structure \cite{Murthy2022}. These studies highlight an important challenge in superconducting qubit materials engineering: deposition conditions simultaneously influence microstructure, impurity incorporation, oxide formation, and interfacial chemistry, making it difficult to isolate which materials attributes and fabrication processes are most strongly correlated with decoherence.

Several more studies have tried to understand the role that lithography and etching procedures have on the materials and the subsequent device performance. Multiple works have compared wet etching and dry etching for patterning in the fabrication of superconducting qubits and consistently found that devices made utilizing dry etching have higher performance \cite{Place2021, Wang2022_Ta, Chudakova2024}. However this can add additional difficulty when working with Ta-based devices, which required chlorine-based reactive ion etching, as this capability is not as universally available as other forms of dry etching. In addition to patterning, etching also defines the exposed sidewall and trench geometry of superconducting circuits. Differences in sidewall angle and trench depth across a wafer have been correlated with device-to-device variation \cite{Murthy2026}. Furthermore, researchers have also looked at the effect of physical ion milling on devices and found that varying decoherence mechanisms depending on the underlying superconductor, which are sometimes reversible with subsequent wet chemical oxide etching \cite{VanDamme2023}. A further critical factor during patterning processes is the use of polymer based resists, and indeed several studies have linked organic residues remaining on device surfaces after fabrication with additional loss \cite{Murthy2022, Alghadeer2025}. This resist issue is relevant both to the patterning of the base wiring layer as well as the fabrication of the Josephson junction. In particular for the latter case, several groups have explored pathways to eliminate the use of organic resists by either using deep trenches for shadow evaporation or using inorganic hard masks \cite{Banerjee2026, Gingras2025, Hanna2026}. In addition to minimizing fabrication-induced defects, recent efforts have explored post-fabrication processing strategies to improve junction parameter reliability. Targeted annealing of a junction post fabrication can modify the resistance and thereby tune the frequency of the qubit more precisely aligned with design parameters than what is normally achieved in standard fabrication. In some cases, annealing has also been shown to reduce TLS defects in the junction and improve coherence. This targeted annealing can be achieved through the use of a laser or through the application of an alternating bias \cite{Hertzberg2021, Pappas2024}.

The landscape of fabrication induced decoherence defects is incredibly complex and our understanding of these relationships is constantly evolving. One of the significant challenges that the field faces is that there are no standardized processes. Researchers are fabricating qubits using different procedures on dissimilar equipment and as a result identifying the most critical fabrication details relevant for performance is increasingly difficult.

\subsection{Predictive Metrics, Scaling, and Emerging Materials}

Despite remarkable improvements in coherence over the last two decades, superconducting qubit materials research remains largely empirical. Significant advances have been achieved through improved materials selection, interface engineering, and fabrication optimization, yet the resulting gains are often difficult to generalize across different device architectures, fabrication facilities, and materials platforms. In most cases, device performance is only ascertained after complete fabrication and cryogenic characterization, limiting the ability to rapidly evaluate new materials and processing strategies. Consequently, the next major challenge facing the field is not simply identifying additional sources of decoherence but establishing quantitative structure-property-performance relationships that can guide fabrication, predict device behavior, and accelerate materials discovery. Addressing this challenge will be essential for improving reproducibility, enabling scalable manufacturing, and accelerating the development of next-generation superconducting quantum technologies.

Despite significant progress in understanding qubit decoherence, broadly applicable materials descriptors that reliably predict device performance remain elusive. However, there are several advances in this area that have arisen in the last few years, some of which have already been highlighted in the previous sections. One such approach is to create large datasets and track a wide range of materials, fabrication, and performance metrics from a reasonably large pool of devices, in order to identify the strongest correlations \cite{Murthy2026, Wolff2026}. Another avenue is to track specific materials metrics or bring new materials characterization methods to bear, including residual resistivity ratio (RRR) measurements, phase boundary measurements, magneto-optical imaging, and terahertz spectroscopy \cite{Premkumar2021, Ryan2022, Datta2026, Park2025}. Lastly, proxy devices, such as coplanar waveguide resonators or tripole resonators, that are easier to fabricate and measure, can be used to more rapidly screen materials in an iterative manner and can provide more detailed information about loss sources and specific interfaces \cite{McRae2020_resonators, Ganjam2024, Vallieres2025}. Certainly all three of these approaches must be further pursued to develop new materials strategies to mitigate decoherence. Additionally, the field must adopt unified materials characterization and fabrication procedures to develop a more standardized high-throughput structure-property-TLS pipeline, but only few such frameworks have been proposed \cite{Wolff2026, Dravid2026}.

Beyond understanding and combating decoherence in current state-of-the-art structures, there are two paths that are pushing the frontier of the field in terms of materials strategies: the development of processes to reproducibly fabricate high-performance qubits at scale and new materials that can be implemented to address decoherence. Some of the issues limiting scalability revolve around better understanding qubit-to-qubit variation which has already been addressed above. Post-fabrication tuning approaches, including the targeted junction annealing discussed previously, provide a potential route to reducing device-to-device variability and improving frequency yield in large-scale systems. A significant remaining hurdle is translating the qubit fabrication process to industrial scale equipment and moving away from tools and methods optimized for academic exploration. A few considerations that have been explored here are the viability of qubit fabrication on the 300 mm wafer scale and transitioning away from the conventional double angle shadow evaporation Josephson junction fabrication method \cite{Verjauw2022, VanDamme2024, Ke2026, Sethi2025}. The junction is also a focal point with regards to emerging materials because, as has been discussed, one key challenge with the conventional Josephson junctions is the amorphous, lossy tunnel barrier. As a result, significant effort has been devoted to replacing the conventional amorphous AlO$_x$ barrier and implementing epitaxial or crystalline materials. Several process examples have been produced in the form of semiconductor interfaces, nitrides, crystalline phases of alumina, and also Van der Waals materials \cite{OConnellYuan2021, Bhatia2025, wang2026all, Kim2021, GarciaWetten2026, Balgley2025, Wang2022_hBN}. This is a particularly active area of research and over the next few years we should expect to see the implementation of new intrinsically low-loss materials and device architectures exploiting those materials.

%% file: 3.quantum_defects.tex

\section{Quantum Defects in Solid-State Materials}
\label{sec:quantum_point_defects}

Quantum point defects (QPDs) are atomic-scale imperfections in a host lattice consisting of vacancies, interstitials, substitutional atoms, or combinations thereof. Such defects can create localized electronic states that host optical and/or spin degrees of freedom and have been used as single-photon sources, quantum sensors, quantum memories and network nodes, and spin qubits \cite{weber2010, bassett2019, wolfowicz2021, zhang2020}. In this section, we discuss the state of the art for quantum point defects in solid-state materials and their relevant properties for quantum information science applications. We will not attempt to provide a complete historical account of QPD research, and instead, we refer the reader to several comprehensive reviews covering defect design, first-principles theory, and major material platforms \cite{doherty2013, gali2019, gali2023, wolfowicz2021, fang2026}.

\subsection{Applications and Design Principles}
\label{subsec:qpd_design}

Each combination of defect structure and material host offers a distinct set of physical properties, enabling a paradigm of bespoke defects tailored to specific quantum functions \cite{bassett2019,wolfowicz2021}. For instance, quantum communication protocols, including quantum-key-distribution and photonic entanglement-distribution schemes, require robust sources of single photons, while more demanding network architectures additionally require photons that are mutually indistinguishable and coherently interfaced with a stationary quantum memory \cite{ruf2021, humphreys2018, raissi2024}. For these applications, useful defect single-photon emitters should therefore have good photostability and quantum efficiency, a large fraction of emission into a narrow zero-phonon line (ZPL), a short optical lifetime, low spectral diffusion, and an emission wavelength compatible with the intended photonic platform. Particularly for long-distance fiber networks, emission in or near the telecommunications bands is particularly advantageous because it reduces propagation loss. In contrast, applications that rely primarily on local sensing or chip-scale photon collection may place greater weight on brightness, operating temperature, or compatibility with nanophotonic cavities than on the absolute emission wavelength.

For defects used as spin qubits, the ground-state manifold must contain a well-isolated paramagnetic state that can be initialized, coherently controlled, and read out with high fidelity. Long longitudinal relaxation and coherence times, conventionally characterized by $T_1$, $T_2$, and $T_2^\ast$, are critical, but must be considered together with the ability to reproducibly address the relevant spin transitions and with the sensitivity of the defect to electrical, magnetic, strain, and thermal fluctuations \cite{wolfowicz2021}. For spin--photon interfaces, it is also desirable to have a spin-conserving optical transition with a strong optical contrast between the spin states. Furthermore, the surrounding nuclear-spin environment can be extremely important. A low concentration of host nuclear spins is generally desirable to reduce dephasing, but on the other hand, a small number of nearby addressable nuclei can provide exceptionally long-lived ancillary qubits for quantum memories and error-correction protocols.

These examples highlight the importance of simultaneous co-design of the defect structure and its host environment. Hosts with wide band gaps can accommodate deep localized levels while reducing unwanted hybridization with extended bulk states and, in some cases, can suppress ionization of the optically active charge state \cite{bassett2019, wolfowicz2021}.. A low natural abundance of magnetic isotopes can reduce hyperfine-induced dephasing, while high thermal conductivity, chemical stability, and mature crystal growth are advantageous for device fabrication. At the level of the defect itself, strong orbital localization, favorable selection rules, and high local symmetry are often desirable. In particular, inversion symmetry can suppress the first-order Stark response of optical transitions and thereby reduce sensitivity to fluctuating electric fields \cite{bradac2019}. This symmetry protection is not sufficient by itself, however, as the orbitally degenerate electronic states of the group-IV vacancy centers can couple strongly to lattice phonons. In SiV$^-$, for example, single-phonon transitions between the two ground-state orbital branches produce rapid orbital relaxation and strongly limit the spin coherence at temperatures of a few kelvin \cite{jahnke2015, pingault2017, sukachev2017}. The properties of a useful QPD therefore cannot generally be understood from the isolated defect alone, and must instead be considered together with the host phonons, nuclear-spin bath, charge environment, surfaces, and the device in which the defect is embedded.

\subsection{Established and Emerging Quantum-Defect Systems}
\label{qpd_platforms}

\begin{figure}[htbp]
    \centering
    \includegraphics[width=1.0\textwidth]{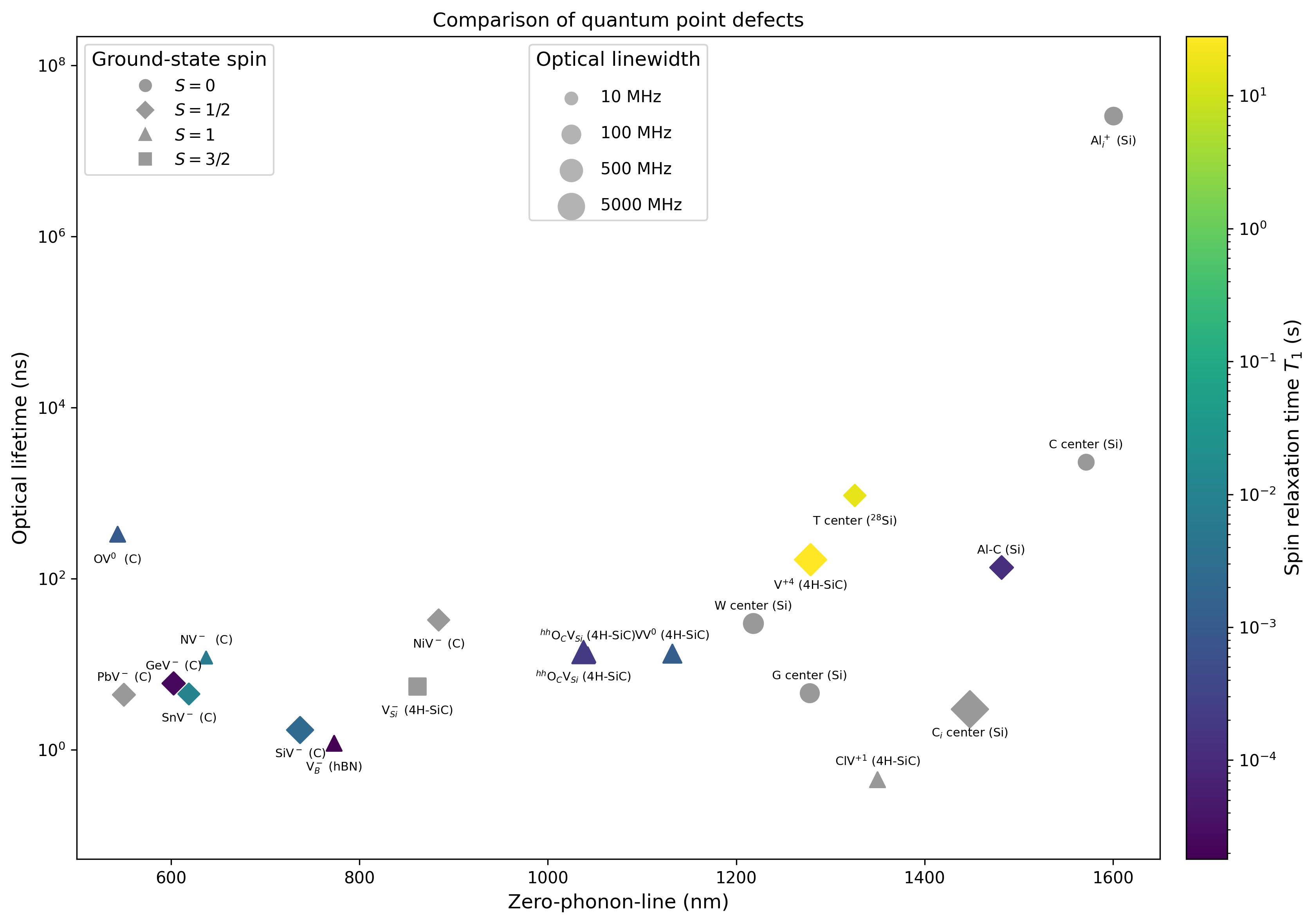}
    \caption{Comparison of some experimentally validated color center defect properties in diamond (C), 4H silicon carbide (4H-SiC), silicon (Si), and hexagonal boron nitride (hBN). The plot compares the optical radiative lifetime against the zero-phonon-line emission for each defect. For each defect, the marker shape indicates the spin state, the marker size represents the smallest experimentally measured linewidth on a logarithmic scale, and the marker color corresponds to the $T_1$ spin-relaxation time (if known).}
    \label{qpd_overview}
\end{figure}

\subsubsection{Diamond}

Historically, the negatively charged nitrogen-vacancy (NV$^-$) center in diamond has been the most extensively studied and widely integrated QPD. Its favorable properties include optical spin polarization into the $m_s=0$ sublevel of an $S=1$ ground state, room-temperature optically detected magnetic resonance, and long spin coherence enabled by the rigid diamond lattice and the possibility of isotopic purification \cite{doherty2013, gali2019}. The breadth of experimental and theoretical characterization of the NV center has established it as a benchmark against which many newer QPDs are compared. The NV center is not without limitations, however. Only a small fraction of its optical emission occurs in the ZPL, and its optical transition is susceptible to local electric-field fluctuations produced by bulk impurities, nearby surfaces, and charge traps. These effects complicate the generation of transform-limited and mutually indistinguishable photons, particularly for shallow centers and nanostructured devices.

The group-IV vacancy centers, silicon-vacancy (SiV), germanium-vacancy,  (GeV), tin-vacancy, and lead-vacancy (PbV), despite their similarity in name to the NV-center, possess a different local structure that leads to a distinct set of functional properties \cite{bradac2019}. In their negatively charged states, these defects adopt a split-vacancy geometry with approximately $D_{3d}$ symmetry, in which the impurity lies between two carbon vacancies. The inversion symmetry of this structure strongly suppresses the first-order first-order Stark response of the optical transition. Additionally, this family of defects generally have a significantly larger fraction of their optical emission within the ZPL; 
the silicon-vacancy for instance, emits about 70\% of its photons in the ZPL compared with only  a few percent for NV$^-$ \cite{bradac2019}. In fact, the SiV$^-$ was the first of this family to be developed into a coherent spin--photon interface, and subsequent demonstrations of GeV$^-$, SnV$^-$, and PbV$^-$ have progressively expanded the accessible range of spin-orbit splittings and operating temperatures \cite{rogers2014, iwasaki2015, iwasaki2017, trusheim2020, wang2024pbv}. A central limitation of the lighter group-IV centers is phonon-mediated orbital relaxation, which can strongly limit spin coherence unless the device is cooled to sufficiently low temperature. Moving to heavier impurities increases the ground-state orbital splitting and suppresses thermally activated orbital relaxation; SnV$^-$, for example, has shown coherent optical transitions together with useful spin properties at temperatures of a few kelvin \cite{trusheim2020}. PbV centers extend this trend further and have exhibited near-transform-limited optical emission above 10 K \cite{wang2024pbv}. The group-IV series therefore illustrates a general design principle: symmetry can protect optical coherence, while the detailed spin-orbit and phonon energy scales determine whether that protection can coexist with a long-lived spin.

Beyond these established systems, computational searches continue to identify new diamond defects with properties that are difficult to obtain simultaneously in the NV or group-IV families. Group-III vacancy complexes have been predicted to support stable spin-triplet states with inversion-symmetric optical transitions \cite{harris2020}, while automated high-throughput calculations have uncovered high-spin sodium-related defects \cite{davidsson2024na}. More recent screening has targeted spectral stability directly and identified, among other candidates, zinc-vacancy complexes with favorable combinations of symmetry and optical properties \cite{xiong2025}. First-principles calculations are also helping to recover the microscopic identities of historically observed color centers. A notable example is the split nickel-vacancy defect, whose calculated magneto-optical signatures connect the NiV$^-$ structure to the long-studied 1.4-eV optical and NIRIM-2 EPR centers \cite{collins1983, thiering2021}. This type of ``rediscovery'' is valuable because decades of spectroscopy can become immediately relevant to modern quantum-device design once the underlying atomic structure is understood.

A related recent example is the WAR5 spin center in diamond \cite{mukherjee2026}, for which millisecond-scale $T_1$ relaxation has been measured at room temperature. The center has been associated with a neutral oxygen-vacancy configuration and shows optical spin polarization, making it a potentially viable alternative to NV$^-$. In this case, however, the microscopic and optical assignments are considerably newer than for NV$^-$ or the group-IV centers and should be treated accordingly. These examples emphasize that the discovery of useful quantum behavior and the microscopic identification of the defect responsible for that behavior are distinct problems.

\subsubsection{Silicon Carbide}

Silicon carbide (SiC) has emerged as one of the most important alternatives to diamond because it combines optically addressable defect spins with a technologically mature wide-band-gap semiconductor platform. SiC is available in high-quality wafers, supports established doping and microfabrication processes, and can be integrated with photonic and electronic devices. Its multiple structural polymorphs have multiple distinct lattice and interstitial sites lead a larger number of distinct defects, which provides more flexibility for defect engineering but at the same time makes spectroscopic assignment more complex \cite{castelletto2020}.

Perhaps the most well-characterized spin defect in SiC is the neutral divacancy in 4H-SiC. It possesses an $S=1$ ground state, can be optically polarized and read out, and has demonstrated millisecond-scale coherence in appropriate material and temperature regimes \cite{christle2015}. Another common family of defects is the set of negatively charged silicon vacancies, which have an $S=3/2$ ground state and have been coherently controlled at room temperature down to the single-defect level \cite{koehl2011, widmann2015}. 
Recently, the commonly observed photoluminescence lines PL5-PL8 have been associated with $S=1$ oxygen-vacancy defect structures through a combination of experimental characterization \cite{4hsic-ov1, chen2026oxygenvacancyquantumspindefects, Hu2026_high_yield} and first principles kinetic formation calculations \cite{4hsic-ov2_formation}. This family of defects exhibit narrow optical transitions, high ODMR contrast at room temperature, and improved robustness against resonant photoionization \cite{he2024}. 
Transition-metal impurities provide yet another route. In particular, the vanadium V$^{4+}$ defect in 4H-SiC combines has a $S=1/2$ ground state with near-telecom optical transitions and very long spin relaxation at cryogenic temperature \cite{spindlberger2019}. Together, this set of defects make SiC and attractive platform for wafer-scalable quantum technologies based on optically addressable spin defects.

\subsubsection{Silicon}

From a defect engineering perspective, silicon presents significantly different opportunities and challenges. Its relatively small band gap makes the stabilization of deep, optically addressable defect states more challenging than in diamond or SiC, but this limitation is offset by mature crystal growth, isotopically purified $^{28}$Si, CMOS-compatible processing, and an unrivaled silicon-photonics infrastructure. Several radiation- and impurity-induced centers emit directly in the near-infrared telecommunications bands, making silicon particularly attractive for scalable quantum networks \cite{zhiyenbayev2025}.

The G-center is one of the most extensively studied silicon color centers and is composed of two carbon atoms and one silicon atom. It is generally formed when a mobile interstitial carbon, $\mathrm{C_i}$, associates with substitutional carbon, $\mathrm{C_s}$. The resulting center has a ZPL near 969 meV ($\sim1278$~nm) and monoclinic $C_{1h}$ symmetry \cite{thonke1981gcenter,song1990gcenter,ivanov2022localization}. An important complication is that the same elemental composition can form more than one closely related atomic configuration. In the historically proposed type-A structure, the defect forms a bent $\mathrm{C_s}$-$\mathrm{Si_s}$-$\mathrm{C_i}$ arrangement. In the type-B structure, the central silicon atom is displaced from its lattice site and becomes interstitial, producing a $\mathrm{C_s}$-$\mathrm{Si_i}$-$\mathrm{C_s}$ complex in which the two carbon atoms occupy substitutional sites. Charge-state-dependent switching between related configurations has been observed experimentally, while more recent calculations and spectroscopy generally favor the type-B structure for the optically active neutral ground state \cite{song1990gcenter, ivanov2022localization}. The low coordination and strong relaxation of the central $\mathrm{Si_i}$ atom also make the G center particularly sensitive to local structural perturbations. This sensitivity is reflected in calculations of the excited-state localization and zero-field splitting, and provides a microscopic explanation for why its optical properties depend strongly on the local strain and damage environment \cite{ivanov2022localization, redjem2023ionpulses}. At the same time, this sensitivity can be used constructively: individual G centers integrated into silicon photonic structures have been spectrally tuned and addressed, demonstrating that the local environment can provide a practical control parameter once the emitter has been formed \cite{prabhu2023gcenter}.

The W-center is structurally very different despite emitting only $\sim 60$ nm away from the G center. It is an intrinsic radiation-damage center containing only silicon atoms and is associated with a compact cluster of three self-interstitials. Early uniaxial-stress measurements established that the 1018-meV ($\sim 1218$ nm) optical transition occurs at a trigonal center between nondegenerate spin-singlet states \cite{davies1987wcenter}. Several tri-interstitial configurations were subsequently proposed, including structures commonly labeled I$_3$-I, I$_3$-II, and I$_3$-V. Calculations of their formation energies, local vibrational modes, and symmetry showed that no single criterion was sufficient to determine the structure; for instance, the lowest-energy configuration does not necessarily reproduce the local vibrational sideband observed experimentally \cite{carvalho2005wcenter, ivanov2022localization}. More recent single-defect spectroscopy and first-principles calculations support a compact trigonal tri-interstitial structure and show an unusual electronic mechanism for its luminescence: the W-center can bind an exciton through the local Coulomb potential even when conventional single-particle calculations do not produce a localized defect level in the silicon band gap \cite{baron2022wcenter}. In this sense, the W-center is an important counterexample to the simple design rule that a useful optical defect must necessarily possess two isolated in-gap single-particle states.

The different structures of the G and W centers also influence how they are formed and how they respond to processing. Both can be created at the single-emitter level by implantation, and focused-ion-beam methods have demonstrated controllable formation of individual centers at specified locations \cite{hollenbach2022wafer}. Under intense pulsed ion irradiation, G and W centers can form directly without a separate post-implantation anneal \cite{redjem2023ionpulses}. The G-center linewidth broadens more strongly with increasing local damage than the W-center linewidth, consistent with the comparatively flexible, low-coordination $\mathrm{C_s}$-$\mathrm{Si_i}$-$\mathrm{C_s}$ structure of the G center and the more rigid all-Si interstitial cluster of the W center \cite{redjem2023ionpulses}. These results provide a direct example of how microscopic structural information can be used to understand, and eventually optimize, the synthesis of a quantum defect.

For applications that require a stationary spin in addition to a single-photon transition, the T center is particularly important. Isotope substitution, stress spectroscopy, and magnetic-field measurements identify the T center as a carbon--carbon--hydrogen complex with monoclinic symmetry, in which an interstitial carbon-hydrogen unit binds to a substitutional carbon atom \cite{safonov1996tcenter}. The ground state contains a localized spin-$1/2$ electron, while optical excitation produces a defect-bound exciton. First-principles calculations find strong carbon-$p$ character in the localized state and show that the stability of the T center is particularly sensitive to hydrogenation and dehydrogenation, explaining the narrow annealing window required for its efficient formation \cite{dhaliah2022tcenter}. Single T-center spins have now been optically resolved in silicon, establishing a direct telecommunications-band interface between a localized electron spin and the silicon photonic environment \cite{bergeron2020, higginbottom2022}. Because the dominant isotope $^{28}$Si carries no nuclear spin, isotopic purification can additionally provide an exceptionally quiet magnetic environment.

Other, less mature silicon centers broaden the range of available structures and optical mechanisms. Carbon-interstitial-related centers can be created and modified through implantation and local laser processing, providing a useful system for studying programmable defect formation \cite{jhuria2024}. The C-center, generally associated with a $\mathrm{C_i}$-$\mathrm{O_i}$ complex, has an optical transition in the L-band and has been proposed as a defect that combines telecom emission with a long-lived internal state \cite{udvarhelyi2022}. Very recent experiments have also identified Al-related telecom centers, including an interstitial-Al center with an optically resolved metastable triplet state and an Al-C-related center that behaves as a bright spin-photon interface \cite{woolverton2026, crosta2026}. These systems are substantially less mature than the G-, W-, or T- centers, but they illustrate the large chemical and structural space that remains accessible even in a material whose conventional defect physics has been studied for many decades.

\subsubsection{Other and Emerging Host Materials}

Hexagonal boron nitride (hBN) has become a major platform for quantum emission because its layered structure permits defects to reside within a few atomic planes of a surface. This geometry can be advantageous for coupling to nanophotonic structures and for sensing fields or chemical species outside the host. Bright single-photon emission has been observed from exfoliated and grown hBN over a broad spectral range \cite{tran2016}. At the same time, the apparent simplicity of a two-dimensional host does not imply a simple defect landscape. Several different vacancies, antisites, substitutional impurities, and defect complexes can give rise to optical emission in similar spectral ranges, and the microscopic identities of many hBN emitters remain unsettled \cite{hayee2020}. Nevertheless, coherent single-spin control at ambient conditions has now been demonstrated for a carbon-related $S=1$ defect, showing that hBN is developing from primarily a single-photon-emitter platform into a spin-qubit and quantum-sensing host \cite{stern2024}.

GaN provides another promising wide-band-gap semiconductor platform. Single-defect optically detected magnetic resonance has been demonstrated at room temperature \cite{luo2024}, and ODMR has more recently been observed from telecom-wavelength single-photon emitters \cite{eng2025}. As in hBN, however, the precise microscopic structures responsible for several optically active GaN families are not yet uniquely established. This uncertainty is important because emitters with similar ZPLs can have different charge states, spin multiplicities, and formation mechanisms.

Low-dimensional transition-metal dichalcogenides provide still greater access to individual atomic sites. A particularly instructive example is a substitutional defect in WS$_2$ that was selected through high-throughput computation and subsequently created by site-selective scanning-tunneling-microscope manipulation \cite{thomas2024}. Rare-earth ions in solids form a related but electronically distinct class of substitutional point defects. Their shielded $4f$ electronic states can produce extremely narrow optical transitions and long-lived internal states, making them important candidates for quantum memories and network nodes \cite{tittel2025}. Finally, first-principles searches have proposed promising defects in simple oxides and other wide-band-gap compounds \cite{davidsson2024oxides}. Many of these systems remain primarily theoretical candidates, and it is useful to distinguish between a calculated defect with favorable intrinsic properties and an experimentally identified center whose structure, charge state, and reproducible synthesis have been established.

The relative performance of several experimentally characterized QPDs is compared in Fig.~\ref{qpd_overview}. The spread of optical lifetimes, linewidths, spin states, and $T_1$ times illustrates the wide range of available defect property combinations that are becoming available, enabling bespoke defects with properties tailored to a particular application to be selected.

\subsection{High-Throughput Discovery of Quantum Defects}
\label{qpd_screening}

Many of the canonical QPDs were initially discovered through spectroscopy of impurities or radiation damage rather than through deliberate quantum-materials design. This approach has produced several of the most successful defect platforms, but it samples only a small fraction of the available materials and chemical space. The search problem is intrinsically combinatorial: each host may contain vacancies, interstitials, antisites, substitutional impurities, and multi-atom complexes, and every defect can occur in multiple charge, spin, and metastable structural configurations. High-throughput discovery is therefore naturally suited to a hierarchical approach in which inexpensive criteria are first used to reduce the search space, followed by increasingly accurate calculations of the properties that directly determine quantum performance.

At the host level, materials databases can be screened for wide band gaps, favorable nuclear-isotope composition, chemical and structural stability, and other properties that are conducive to localized and coherent defect states. Ferrenti \textit{et al.} demonstrated this type of host-first strategy by identifying candidate materials before specifying the defect itself \cite{ferrenti2020}. Once a host is selected, automated first-principles workflows can enumerate defect structures and charge states and calculate formation energies, thermodynamic charge-transition levels, spin states, hyperfine interactions, zero-field splittings, and optical transition energies. The ADAQ framework provides one implementation for automated magneto-optical defect calculations \cite{davidsson2021adaq}, while the QPOD database extends large-scale screening to point defects in two-dimensional materials \cite{bertoldo2022}.

A complementary approach is to explicitly enumerate large numbers of defects across several technologically relevant semiconductors. In one such study, high-throughput calculations were performed for more than 50,000 defects in diamond, SiC, and silicon, including formation energies, spin states, transition dipole moments, and ZPLs \cite{ivanov2023database}. In silicon alone, thousands of composite defects that are stable under intrinsic conditions were then filtered for optically bright telecom emitters and spin-qubit candidates. An important feature of this approach is that the relaxed structures and calculated properties are stored in a searchable database, allowing the screening criteria to be changed as the application changes. For example, the set of defects preferred for a telecom single-photon source need not be the same as the set preferred for a long-lived spin memory.

The most technologically important properties are often much more expensive to calculate than ground-state formation energies. High-throughput defect discovery is consequently a multi-fidelity problem. Semilocal density-functional theory can be used to rapidly eliminate clearly unfavorable structures, but shortlisted candidates may require hybrid functionals, constrained excited-state calculations, configuration interaction, quantum embedding, or explicit calculations of electron--phonon and spin--phonon coupling. Recent searches in simple oxides illustrate this progression from broad defect enumeration to more accurate calculations of spin and coherence properties \cite{davidsson2024oxides}. Likewise, recent screening in diamond has incorporated application-specific targets such as optical spectral stability, rather than relying only on generic descriptors such as the host band gap or the existence of a localized state \cite{xiong2025}. Machine-learning and active-learning methods can further accelerate this process by identifying relationships between expensive target observables and lower-cost descriptors such as orbital localization, symmetry, defect-level separation, local chemical environment, or hybridization \cite{fang2026}.

High-throughput discovery need not be exclusively computational. Automated confocal localization, wide-field imaging, resonant spectroscopy, and feedback-controlled measurements can characterize large populations of emitters under reproducible conditions. Sutula \textit{et al.} demonstrated large-scale automated optical characterization of solid-state emitters, including resonant measurements performed at rates far beyond conventional point-by-point characterization \cite{sutula2023}. This type of statistical information is critical because the performance of a defect platform is determined not only by its best measured emitter, but also by the distributions of charge-state stability, optical linewidth, spectral diffusion, orientation, spatial placement, and conversion yield. The most useful future screening workflows will therefore combine computational prediction with automated synthesis and characterization, allowing experimental results to be fed back into subsequent calculations and candidate selection.

\paragraph{Defect Identification.}
An important part of this discovery process is determining the microscopic structure responsible for an observed optical or spin signature. A reproducible ZPL, photon antibunching, or even an ODMR spectrum does not by itself uniquely determine the atomic structure of a defect. Different complexes can produce optical transitions in similar energy ranges, while strain, local electric fields, isotope composition, and charge state can shift the spectrum of a nominally identical center. This problem is particularly apparent in hBN, where correlated optical and electron microscopy has shown that several structurally distinct classes of defects contribute to quantum emission \cite{hayee2020}.

The long histories of the G-, W-, and Ni-related centers provide similar examples in bulk three-dimensional materials. The optical lines associated with the G and W centers were characterized long before their microscopic structures could be constrained to the present level, and even today details of their electronic excitation mechanisms remain under active study \cite{ivanov2022localization, baron2022wcenter}. Conversely, modern first-principles calculations can connect older spectroscopic signatures to specific microscopic structures, as in the assignment of the diamond 1.4-eV/NIRIM-2 center to NiV$^-$ \cite{thiering2021}. Reliable defect identification therefore benefits from combining several independent signatures: optical polarization and selection rules, EPR or ODMR, zero-field splitting, hyperfine and isotope fingerprints, local vibrational modes, charge-state dependence, response to strain and electric field, and, where possible, direct structural microscopy. First-principles calculations can then provide a common microscopic model against which these observables are compared. Establishing this correspondence is not merely a matter of labeling a spectral line; without a reliable atomic structure, it is difficult to predict formation pathways, stabilize the desired charge state, or intentionally modify the defect.

\subsection{Theory-Guided Defect Engineering and Controlled Synthesis}
\label{qpd_engineering}

Theoretical calculations can contribute to defect engineering at two different levels. The first is to determine whether a given structure has useful intrinsic optical or spin properties. The second, and generally more difficult, problem is to determine how that structure can be formed and stabilized in a real material. Under equilibrium conditions, defect formation energies and thermodynamic charge-transition levels provide a map of stability as a function of elemental chemical potentials and Fermi level \cite{freysoldt2014}. These quantities can identify chemical conditions that favor a desired impurity complex, determine which charge states are thermodynamically accessible, and reveal competition with compensating defects. However, many QPDs are created by implantation, irradiation, laser excitation, or post-growth annealing and are therefore formed under strongly nonequilibrium conditions. Formation energy alone is not sufficient to predict the final defect concentration.

A predictive description of synthesis must consequently include defect kinetics. Vacancy and interstitial migration barriers, association and dissociation energies, competing sinks, charge-state-dependent diffusion, and annealing temperatures determine whether the species introduced during growth or irradiation can actually reach the desired microscopic configuration. Minimum-energy-path calculations and kinetic models can be used to identify bottlenecks and competing reactions, and can therefore suggest implantation sequences, co-dopants, or annealing windows that increase the formation probability of the target defect. The silicon G and T centers provide useful examples. The G center is formed by the association of mobile interstitial carbon with substitutional carbon, while the stability of the T center depends sensitively on hydrogen incorporation and can be limited by dehydrogenation outside a relatively narrow processing window \cite{dhaliah2022tcenter}. These considerations are invisible if only the final relaxed defect structure is calculated.

Experimental synthesis methods already cover a wide range of nonequilibrium conditions, including in-situ incorporation during growth, ion implantation, electron and neutron irradiation, laser writing, thermal annealing, and, in low-dimensional materials, direct atomic manipulation. Focused implantation has enabled controlled formation of individual G and W centers in silicon \cite{hollenbach2022wafer}, while intense laser-driven ion pulses can produce these defects directly through a combined sequence of displacement damage, implantation, and local heating \cite{redjem2023ionpulses}. The programmable formation of carbon-related emitters by local laser processing provides another example in which the relative populations of defect structures can be modified after growth \cite{jhuria2024}. At the atomic limit, the site-selective creation of a computationally screened defect in WS$_2$ demonstrates a direct connection between theoretical defect selection and controlled fabrication \cite{thomas2024}. Together, these approaches suggest that synthesis conditions themselves can become targets of computational optimization, rather than being treated as a separate empirical step after a promising defect has been identified.

Charge-state control is a central part of this problem. A defect can possess an ideal spin and optical level structure in one charge state but spend only a small fraction of time in that state under realistic illumination or near a surface. Thermodynamic charge-transition levels determine the equilibrium tendency of a defect to exchange carriers with the host, but optical excitation introduces additional photoionization and carrier-capture pathways. The NV center provides a canonical example: illumination can convert between NV$^-$ and NV$^0$, and the resulting charge dynamics directly modify fluorescence and spin-readout contrast \cite{aslam2013,yuan2020}. Similar considerations apply to essentially every optically driven defect for which one or more charge states lie within an experimentally accessible energy range.

Photoionization physics is therefore an important design criterion rather than a secondary correction to the optical spectrum. The relevant thresholds depend on the position of the ground and excited defect states relative to the valence- and conduction-band edges, and optical excitation can ionize a defect through one- or multi-photon pathways even when the equilibrium charge state is stable. Once a carrier is emitted, nearby impurities, surfaces, or other defects can capture it and prevent immediate recovery of the desired state. Surface-induced band bending can further shift the local Fermi level and make charge instability particularly severe for shallow defects and nanostructures. Modified divacancies in SiC provide an instructive example in which changing the local chemical structure can improve stability under resonant excitation \cite{he2024}. Theory can contribute by calculating thermodynamic and optical ionization energies, carrier-capture coefficients, and the response of defect levels to local electrostatic boundary conditions. Experimentally, co-doping, electrostatic gates, surface termination, and deliberately introduced charge reservoirs provide corresponding means of stabilizing the desired charge state.

The host lattice and device geometry provide additional engineering degrees of freedom. Isotopic purification can reduce magnetic noise, while deliberate placement of selected nuclear spins can create ancillary registers. Strain and electric fields can tune optical transition energies and lift degeneracies, and calculations can be used to determine both the useful tuning coefficients and the sensitivity to uncontrolled fluctuations. Nanophotonic cavities can enhance emission into a selected transition through the Purcell effect, compensating in part for a modest bare radiative rate or Debye--Waller factor. At the same time, etched surfaces and interfaces can introduce charge traps, strain gradients, and paramagnetic noise that degrade the properties of the same defect. For this reason, the appropriate design target is not simply an isolated point defect with a favorable set of calculated properties, but a defect whose structure, formation pathway, charge environment, and device integration can all be controlled simultaneously.

\subsection{Open Challenges}
\label{qpd_outlook}

Despite major progress, a substantial gap remains between identifying a promising isolated defect and producing scalable defect-based quantum technology. On the theory side, many of the properties that are most important experimentally are also among the most difficult to calculate. Ground-state formation energies, relaxed structures, and spin densities are now comparatively routine, whereas excited-state multiplet ordering, spin--orbit coupling, intersystem crossing, nonradiative recombination, photoionization, electron--phonon coupling, optical line shapes, and spin--lattice relaxation require increasingly sophisticated methods \cite{gali2023}. Strong electronic correlation and near-degenerate localized orbitals can invalidate a simple single-determinant description, while finite-temperature phonons and local environmental disorder can qualitatively modify behavior that appears favorable in an ideal bulk calculation. The central theoretical challenge is therefore to retain sufficient accuracy for these many-body and dynamical properties while extending the calculations to the very large configurational spaces required for defect discovery.

The structure of the defect itself can also remain uncertain even after extensive spectroscopy. As illustrated by the G and W centers in silicon, several structural models may have similar formation energies while reproducing different subsets of the measured symmetry, vibrational, optical, or spin properties \cite{ivanov2022localization}. Similar identification problems persist for many hBN and GaN emitters. A stronger connection between optical spectroscopy, spin resonance, isotope engineering, microscopy, and first-principles calculations will be required to unambiguously determine these structures. This is particularly important for high-throughput discovery, where a calculated candidate is only useful if an experimentally observed center can be mapped back onto the same microscopic structure.

On the experimental and materials side, deterministic fabrication remains a major bottleneck. It is necessary to control not only whether a defect is created, but also its position, orientation, microscopic configuration, local strain, and charge state while minimizing collateral lattice damage. High conversion yield must be achieved without sacrificing optical linewidth or spin coherence. For shallow defects and nanostructures, surface disorder can dominate both spectral diffusion and magnetic noise. In addition, the processing conditions that maximize defect yield are not necessarily those that minimize residual damage. The contrasting response of G and W centers to intense ion irradiation is a simple example of this structure-dependent tradeoff \cite{redjem2023ionpulses}. Developing predictive kinetic models that connect implantation or growth conditions to defect yield and device-quality coherence therefore remains an important open problem.

A related issue is the lack of standardized benchmarking across defect platforms. Reported values of $T_1$, $T_2$, optical linewidth, and radiative efficiency can differ by orders of magnitude with temperature, magnetic field, isotope composition, excitation protocol, and whether a measurement probes a single center or an ensemble. Likewise, a measured photoluminescence decay time is not necessarily equal to the purely radiative lifetime, and a narrow ensemble distribution is not equivalent to a transform-limited single-emitter linewidth. Meaningful comparisons should therefore report the measurement conditions together with the quantum efficiency, Debye--Waller factor, spectral diffusion, charge-state stability, photon indistinguishability, and fabrication yield. Standardized data of this type would also provide substantially better training and validation sets for high-throughput and machine-learning models.

Finally, no known defect simultaneously optimizes all of the properties required for every quantum application. The NV center offers exceptional spin control at ambient conditions but has a relatively weak ZPL. Inversion-symmetric group-IV centers provide improved optical stability but can require cryogenic operation to suppress phonon-mediated orbital relaxation. SiC supports a broad set of spin-active defects together with mature wafer-scale processing. Silicon provides direct access to telecom photonics and an exceptionally quiet isotopic host, but its small band gap places stronger constraints on defect localization and charge stability. hBN and GaN offer new opportunities for near-surface sensing and semiconductor integration, while their defect identification and materials uniformity continue to mature. The likely outcome is therefore not a single universally optimal QPD, but a collection of defect--host combinations selected for different applications.

The field is consequently moving from predominantly serendipitous discovery toward a more systematic materials-design framework. Application requirements can first be translated into quantitative target properties. Databases and automated calculations can then be used to identify candidate hosts and defects, using higher-level electronic-structure and many-body methods can validate the most promising candidates. Further, thermodynamic and kinetic calculations can suggest routes for formation and charge stabilization. Automated experiments can then measure statistically meaningful distributions of the resulting properties. As these components become more closely connected, theory should increasingly be able to address not only which defect has favorable intrinsic properties, but also how that defect can be formed reproducibly, stabilized in the correct charge state, and incorporated into a device without losing the properties that motivated its selection. Once predictive and reliable theoretical methods exist, it will become possible to design tailored defects for individual scalable quantum components.

%% file: 4.quantum_photonics.tex

\section{Materials for Quantum Photonics \label{mats_q_photonics}}
Integrated photonic elements can form a variety of critical components in quantum hardware, including elements that can integrate and route single photons, act as detection units, generate quantum resources, and interface signals across octaves of bandwidth. As such, some critical components may be passive or active, depending on their use. These materials face similar challenges with unwanted defects in the bulk, surfaces, and interfaces, that can degrade their performance. Several reviews\cite{dutt2024nonlinear, moody20222022, bogdanov2016material} have discussed the variety of opportunities different materials can have in quantum photonics, including active\cite{aharonovich2026programmable} and nonlinear photonics and those leveraging single photon emitters. For active devices in particular, optical resonances and narrow bandwidths are often required, meaning that defects, surfaces, and interfaces may play increasingly strong roles on device operation. This section will focus on only a few prominent material systems for active or passive integrated photonic devices, providing a perspective on a materials-centric approach to understanding sources of nonidealities in composite devices. Many of these important device nonidealities can be elucidated with materials spectroscopies and microscopies\cite{taheriniya2026atomic}. A focus on some unknown aspects of specific material properties or defects as related to device performance will be emphasized, attempting to connect device and system-level challenges to fundamental material structure. Finally, a brief discussion of how materials characterization may be relevant in single photon detectors will be provided, but a more extensive discussion of intrinsic and extrinsic factors impacting their performance are left to more specific reviews\cite{holzman2019superconducting,venza2025research}.

\begin{figure}[htbp]
    \centering
    \includegraphics[width=0.75\textwidth]{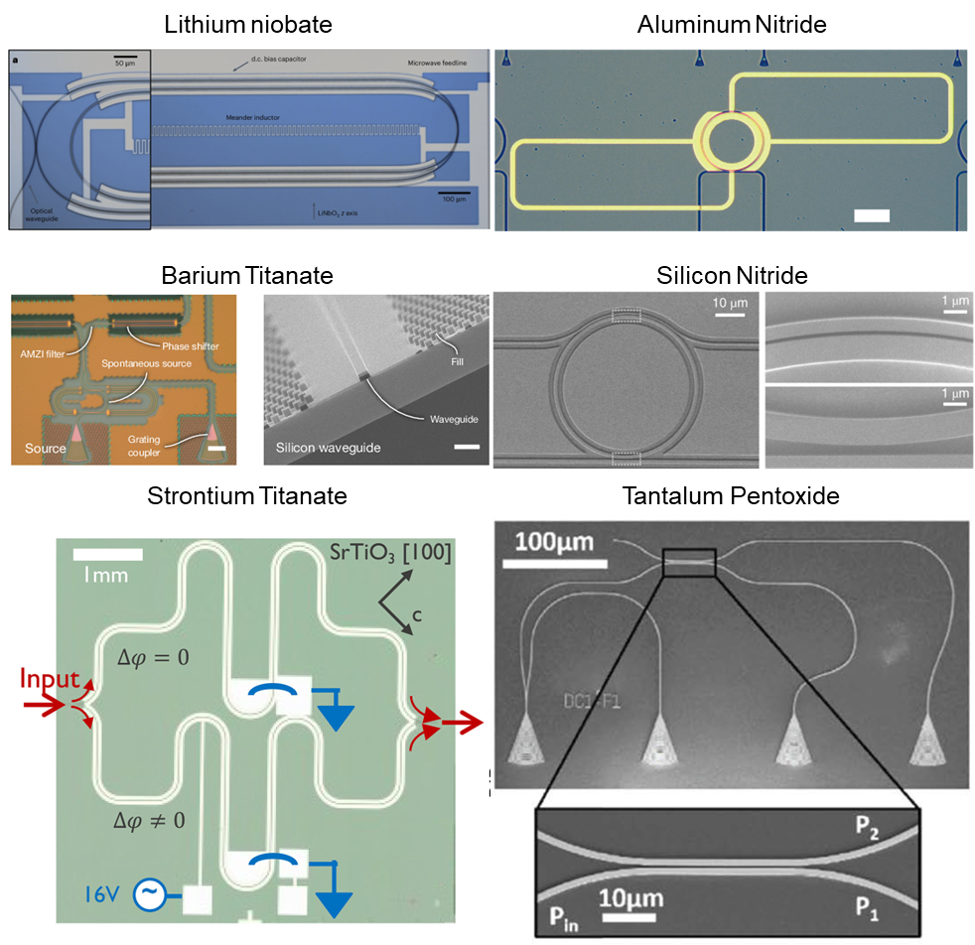}
    \caption{Integrated waveguide materials primarily discussed in this review. Active materials include wide bandgap oxides and nitrides that feature nonlinear susceptibilities that can leverage external electric fields to control optical properties. Passive materials, including tantalum pentoxide, offer potential advantages as low loss and low noise quantum photonic systems.  Images taken adapted with permission from: \cite{warner2025coherent}(lithium niobate), \cite{psiquantum2025manufacturable}(barium titanate), \cite{ulrich2025engineering}(strontium titanate), \cite{fan2018superconducting}(aluminum nitride),  \cite{lu2019chip}(silicon nitride), and \cite{splitthoff2020tantalum}(tantalum pentoxide). }
    \label{fig:S4_photonic_devices}
\end{figure}

\subsection{Active Materials for Nonlinear and Quantum Optics}
Photon-photon interactions can be mediated through $\chi^{(2)}$ or $\chi^{(3)}$ interactions in solids, where photons can mix with each other through appropriate dispersion engineering of the underlying guided modes. This sort of light-matter interaction is intrinsically weak, and the nonlinear susceptibilities are orders of magnitude weaker than the linear susceptibility (refractive index). These are quite important, however, to create entangled photons through spontaneous parametric downconversion (SPDC), four-wave mixing, or to interface microwave and optical photons through an electro-optic effect. A non-zero $\chi^{(2)}$ nonlinearity can only occur in materials that lack inversion symmetry. This nonlinear susceptibility means that electric fields, whether from photons or electrodes, can contribute to the induced polarization in the material, allowing for generation terms at different frequencies, like in SPDC, sum- and difference-frequency generation, and second harmonic generation. It can additionally be used (via the Pockel’s effect) to control the refractive index of a material at DC or microwave frequencies. The lack of centrosymmetry additionally makes these materials pyroelectric and piezoelectric, in which temperature fluctuations or strain can induce charge accumulation, which then can modify the refractive index through the Pockel’s (electro-optic) effect. As a result, materials with this functionality can have many degrees of freedom to control the response of an optical component, change the state of a qubit, or create quantum resources. Conversely, this means that they are very sensitive to small changes in their environment, and defects within the material can lead to significant changes to their overall response. Surfaces, which explicitly break inversion symmetry, can also exhibit a second order effect and is responsible for effects like two photon absorption in otherwise centrosymmetric materials like silicon\cite{bristow2007two}. All materials will exhibit a third order nonlinear susceptibility $\chi^{(3)}$, which comes from the anharmonic potential inside of solid materials. This is useful in creating optical sources in resonators based on four-wave mixing and can be used to create stable ultrafast sources through soliton generation. These effects are even weaker than $\chi^{(2)}$ effects, typically requiring resonant enhancement. Over long distances, other third order effects like Raman scattering can play a role in wavelength conversion, though typically reducing this effect is desirable as it acts as a source of noise and loss that reduces fidelity of weak quantum signals over long distances. We will focus here on $\chi^{(2)}$ materials and will leave higher order nonlinearities to other reviews.

Because the nonlinear susceptibility is small, quantum devices that require high cooperativities and strong responses at the single photon level often demand significant resonant engineering at the frequencies of interest. This can include creating low mode volume optical cavities to enhance light-matter interactions, resonator structures and dispersion engineered waveguides to improve the nonlinear response, and long optical path lengths in periodically poled ferroelectric materials for nonlinear generation. As a result, great care is taken to create resonators at specific optical frequencies, as the enhanced cavity lifetime necessarily reduces the bandwidth over which these devices operate. These devices are therefore very sensitive to small changes in geometry or refractive index, making device engineering a co-design challenge with materials selection.

\subsection{Materials Selection}
Presently available wafer-scale $\chi^{(2)}$ materials used in quantum photonic devices are predominantly wide bandgap materials like lithium niobate\cite{warner2025coherent} (Fig. \ref{fig:S4_photonic_devices}) and lithium tantalate, with barium titanate\cite{psiquantum2025manufacturable}, strontium titanate\cite{ulrich2025engineering}, and aluminum nitride\cite{fan2018superconducting} additionally seeing increasing use. A variety of review papers discuss potential applications of these sorts of materials, with a particularly detailed reference in reference \cite{zhu2021integrated}. All these materials feature the same fundamental advantages and are potentially susceptible to the same underlying challenges. First and foremost, the nonlinear susceptibility depends on crystal orientation and polarization of the applied fields, making crystal structure and domain poling critical. One benefit of lithium niobate and lithium tantalate is their historical development over the last 40 years and high Curie temperature. Grown and poled as large boules before being ion sliced and bonded to carrier substrates\cite{levy1998fabrication}, these materials can have specified crystal and ferroelectric domain orientations that simplify device and system design. Materials like barium titanate feature significantly lower Curie temperatures, and thin films are typically grown at temperatures where the film is in its paraelectric phase. Upon cooling, a structural re-arrangement and the paraelectric-ferroelectric transition means that these thin films have no net nonlinear response over typical waveguide length scales, which is composed of ferroelectric domains with 4-fold symmetry in-plane\cite{abel2019large, little1955dynamic, zhang2006charge,  zuo2014domain, bednyakov2023charged}. Even when poled after device fabrication, domain walls can additionally create conductive pathways \cite{zhang2006charge, zuo2014charge} that can modify local electric field distributions during operation. The electro-optic coefficient of barium titanate can be significantly higher than lithium niobate and tantalate, and is being widely considered for quantum photonic technologies\cite{psiquantum2025manufacturable}. The soft ferroelectric nature of this material additionally means that a positive bias field typically needs to be applied during operating to avoid depoling individual domains, which would lead to a hysteretic response\cite{catala2026high, geler2022ferroelectric} and modify operation over time. Additionally, a tetragonal-rhombohedral phase transition occurs at moderate to low temperatures\cite{eltes2020integrated}, which reduces the overall electro-optic response compared to operation at room temperature. 

Strontium titanate can exhibit a quantum paraelectric phase transition at low temperatures\cite{muller1979srti} that can be modified with oxygen isotope engineering\cite{itoh1999ferroelectricity}, which makes it a potentially ideal material to interface with spin defects or superconducting qubits, as its nonlinearity grows at cryogenic temperatures, rather than reducing. The associated phase transition amplifies the dielectric constant, but electro-optic coefficients exceeding several thousand pm/V have been reported\cite{ulrich2025engineering, anderson2025quantum}. Extensive studies on its stability and the ability to create low loss resonators in this platform are required for it to see significant gains in quantum photonic devices. 

Materials like aluminum nitride are additionally of interest because of their large piezoelectric response. Their electro-optic properties are weaker than other materials\cite{xiong2012low, xiong2012aluminum}, but useful devices like quantum transducers have been implemented in the system\cite{fan2018superconducting}. Extensive theory work and experimental demonstrations have shown that this can be enhanced by doping with scandium\cite{yang2024unveiling}. Near phase transitions, the dielectric constant increases, which leads to an associated increase in the nonlinear susceptibility. However, high quality growth of thin films with high scandium content has faces challenges with optical losses. \cite{liu2026hybrid, wang2024cmos}. Increased scandium content has been associated with small inclusions of the unwanted rocksalt phase\cite{yang2024unveiling} and greater incorporation of oxygen\cite{casamento2020oxygen, gallardo2025surface} that dope and increase absorption. Future developments that include layered ordering of dopants and strain\cite{wang2025towards} can further boost the nonlinearity in these materials. Improved quality of growth at high Sc compositions, as well as a deeper understanding of how to control grain sizes and reduce unwanted phase inclusions, can increase the quality of these films for quantum photonics. Other future electro-optic materials may also be defined by their proximity to phase boundaries, which can be controlled with composition (like Sc-doped AlN) or strain. Straining common ferroelectric materials like barium titanate have demonstrated that electro-optic coefficients can be enhanced dramatically\cite{suceava2026colossal, ross2026quantum} in a manner similar to strontium titanate at low temperatures, with electro-optic coefficients being measured several times higher than unstrained films at room temperature. Integrating these materials as thin films onto foundry-compatible substrates through wafer-scale bonding and polishing could be one route to large-scale device development.

\subsection{Material Challenges: Optical and Electronic Drift}
\begin{figure}[htbp]
    \centering
    \includegraphics[width=\textwidth]{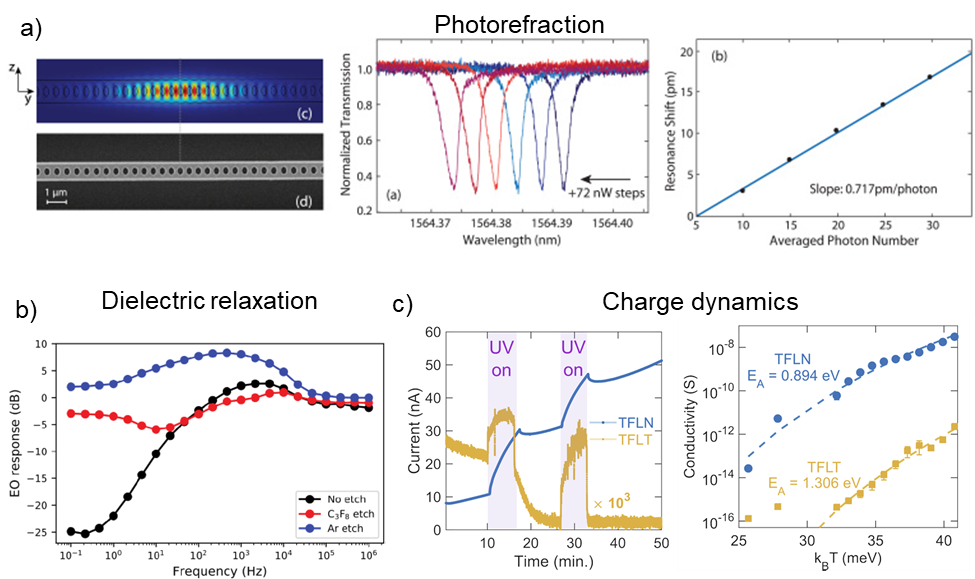}
    \caption{photorefraction and drift in thin film lithium niobate. a) photorefraction is observed in nanocavities with nearly single photon level tuning. b) dielectric relaxation, as measured as a frequency dependence of the electro-optic response of a modulator, depends on both device geometry and chemistries used to etch and define the contact interface. c) Charge dynamics play a significant role in unwanted time-dependent responses in modulators. UV light modifies conductivity and reported lower activation energies in differing materials defines the magnitude of drift. These phenomena motivate a materials-level characterization of device response improvements, correlating device outcomes to fundamental material properties. Images are adapted with permission from (a)\cite{li2019photon}, (b)\cite{holzgrafe2024relaxation}, and (c)\cite{sayem2026unveiling}}
    \label{fig:S4_drift_relaxation}
\end{figure}

Electro-optics is of particular interest for modulators and quantum transducers, which require an electric field to induce changes to the refractive index. This can be used to encode information, upconvert quantum signals, or tune the response of a device for long periods of time. As a result, electric field transients can disturb this response, commonly referred to as DC drift or electro-optic relaxation\cite{holzgrafe2024relaxation}.  This has been extensively reported and discussed in bulk modulators\cite{yamada1981dc, becker1985circuit, gee1985minimizing, nagata1994possibility}, but these effects are enhanced in thin film devices. In these materials, the fundamental underlying mechanism is unknown, although contributions from thermal history, alkali metal diffusion, metal-ferroelectric interfaces, buried oxide layers, and point defects have all been proposed\cite{yeh2025interface}.  In each case, unwanted dead layers, charge buildup, trap states, unintended interfacial conductivity, and charge injection across barriers may all contribute to a time-dependent electro-optic effect\cite{yeh2025interface}. Understanding and mitigating these issues is fundamentally a materials characterization and processing problem. Lithium niobate and lithium tantalate are the most studied materials in this regard Presently, correlative analyses have been performed that suggest processing improvements to DC drift, but the defining defects (potentially niobium antisite defects, lithium or oxygen vacancies, etc.), their location (in the bulk, sub-surface, etched surfaces, bonded interfaces) and definitive methods of eliminating these defects are not known. This includes thermal processing conditions, which may undo irradiation damage from lithography and reduce oxygen vacancies that can increase conductivity, but it can also lead to additional out-diffusion of lithium, providing another avenue and timescale for drift\cite{yeh2025interface}. Even otherwise innocuous choices like etch chemistries to define the metal contact interface with lithium niobate has been shown to modify the low frequency response of these devices, suggesting that surfaces and interfaces are important to consider\cite{holzgrafe2024relaxation} (fig. \ref{fig:S4_drift_relaxation}b). Current lithium tantalate devices have reduced drift and stability challenges than their niobate counterpart\cite{wang2024optical, chen2026thin, powell2024dc, sayem2026unveiling} (Fig. \ref{fig:S4_drift_relaxation}c), but the reason for this remains unclear. One hypothesis is the lower concentration of defects or a change in activation energy of relevant defects\cite{sayem2026unveiling}. This has been somewhat verified in lithium niobate, where stoichiometric thin films exhibit a lower DC bias drift than congruently grown films\cite{zhang2026fabrication}. Impacts at the bonding interface need to be better studied, as both strain and composition can vary. Further mechanistic insights are needed to determine unambiguously the defects at play.

Several of these mechanisms are electronic and defect-mediated in origin, which means that their impact is significantly reduced at low temperatures. However, other nonidealities persist and occur even down to the single photon level\cite{li2019photon} (fig. \ref{fig:S4_drift_relaxation}a). Photorefraction is a common mechanism that leads to optically-induced drift in these materials. Here, charges in trap states can be optically excited to the conduction band, which then propagate along the ferroelectric domain axis before occupying a new trap site. The charge density imbalance leads to an electric field, which modifies the refractive index through the electro-optic effect. The strength of this effect and its persistence is related to the underlying defect concentration, optical power, and conductivity of the material, and has been observed down to the single photon level in low mode volume cavities. Combined with tight confinement in waveguides and resonant structures, this effect can be strong in thin film devices\cite{jiang2017fast, sun2017nonlinear} This, combined with electro-optic drift, can lead to unwanted hysteretic responses that are challenging to correct during device operation. At low temperatures, photorefraction effects are persistent due to reduced material conductivity, which poses challenges in measure devices\cite{lange2026photorefraction}. Counterpropagating modes caused by unwanted reflections and surface roughness can create optical standing waves in the material; photorefraction that occurs because of this creates a bragg grating in the material\cite{xu2021photorefraction} that can act as a selective reflector.  

Varying reports regarding the role of thermal annealing\cite{xu2021mitigating, shams2022reduced}, the existence of substrate and cladding oxides\cite{xu2021mitigating}, and various other processing methods have been reported to reduce photorefraction and improve optical loss. Increasing operating temperatures have shown improved stability for modulators operating in the visible spectrum\cite{celik2024roles}. It's not presently known whether there is a higher concentration of defects at etched surfaces or subsurfaces, and what types of defects are present. More generally, the ion slicing approach used for lithium niobate and lithium tantalate may induce additional radiation damage within the thin films, while wafer bonding may induce local strain that can relax over time during device operation via the piezoelectric effect. Thermal annealing has again been proposed to reduce this phenomena in lithium niobate, but the underlying defect are not clearly delineated. In addition, pyroelectric charges can accumulate in these materials during cooldown to cryogenic temperatures\cite{thiele2024pyroelectric}, leading to unknown changes to optical properties that are difficult to correct for. As these are all nominally related to electronic defects, approaches to locally increase the conductivity to help sweep out unwanted charges have shown some promise in reducing device drift\cite{shi2025alleviation}; however, this is usually at the expense of optical loss due to increased carriers or surface roughness.

Given the breadth of device and material types used in quantum photonic devices, a formalized metric that defines drift rates in a device-agnostic manner (interferometer, resonator, etc.) and compared against processing and material interactions will help advance the field, where direct structure-property-processing relations can be effectively established and compared with other material systems. Converting device metrics like GHz/hr in resonators or dB/hr in interferometers to an index change per unit time can allow for more direct comparisons. This must be done at a variety of temperature ranges to directly compare materials that will be used at cryogenic or room temperature. Notably, measurements of DC or photorefractive drift are sparse at cryogenic temperatures, with little data available for materials like barium titanate, strontium titanate, and aluminum nitride. Additionally, correlative spectroscopies and microscopies should be utilized to elucidate the root causes of these effects in device-relevant contexts. For example, cooling rates in the growth of barium titanate thin films have been shown to be important for the resulting electro-optic coefficient, which can be correlated with transmission electron microscopy to analyze local structure and defects formed during growth\cite{reynaud2025enhancement} (fig. \ref{fig:S4_characterization}a). With increased surface area and potential for damage at etched surfaces, structural and chemical characterization of optical waveguides and key interfaces can provide a wealth of information to pin down the root causes of these effects. Strain analysis may also be important near these etched surfaces and interfaces due to piezoelectric coupling. Finally, as new materials develop, these approaches will be critical to understand their properties. For example, the polar texture of strontium titanate at low temperatures was recently visualized using electron microscopy\cite{zhang2026imaging} and hosts complex phenomena that can potentially be exploited or engineered for further improvements in quantum photonic devices.

\subsection{Challenges in Quantum Transduction}
One particularly challenging quantum photonic device that leverages these material nonlinearities is quantum transduction Interfacing quantum resources across orders of magnitude in frequency is a central challenge in quantum networking and distributed computing. While one prominent candidate for quantum computing relies on superconducting qubits, it is extraordinarily challenging to route single microwave photons across long distances with high fidelity due to thermal noise at these frequencies and attenuation of microwave signals over long distances. As a result, optical transducers that upconvert this signal into single photons that can leverage modern telecommunications networks is ideal\cite{awschalom2021development, o2009photonic}. These structures typically leverage architectures with several resonances\cite{holzgrafe2020cavity, mckenna2020cryogenic, xu2021bidirectional}, coupling either purely through optical and microwave means or through optomechanical structures\cite{jiang2020efficient}. In quantum electro-optic devices, a triply resonant structure could be used to either drive quantum devices or to read out single microwave photons. Here, a microwave resonator must be resonant with the frequency of a qubit (a few GHz) as well as the detuning frequency between two optical modes. When the lower energy optical mode is pumped and a microwave photon interacts with the cavity, a single optical photon can be upconverted to the second optical mode, with the quantum information retained in the higher energy photon. Conversely, microwave photons can be generated by beating two optical tones together in the same cavity; using a microwave resonator at the frequency of a qubit, this generated microwave photon can then drive the state of the qubit\cite{warner2025coherent}. This process depends nonlinearly with the loss rates of each resonator, placing a quadratic scaling on the optical loss in the cavities. As a result, the highest efficiency and cooperativity devices that use this architecture feature very high optical quality factors (intrinsic quality factors of order 10 million at telecommunications wavelengths), meaning that the operating bandwidth is additionally extraordinarily narrow, and the detunings of each cavity mode must be precisely placed. This can be done with an electric field through the pockels effect in these materials, but non-idealities require active stabilization that is challenging in these devices. Indeed, photorefraction is often explicitly listed as a mechanism that limits device efficiency\cite{xu2021bidirectional}. Additional power and device overhead is required to continually monitor and adjust the response of these devices over time, which is clearly undesirable in future quantum hardware, such that exploring materials with lower drift is an area of active exploration \cite{axline2026stable}. Understanding defects and processing methods that eliminate this drift at the material level will therefore provide a massive enhancement for quantum devices and systems. 

\subsection{Challenges in Nonlinear Optics: Phase Matching} 
\begin{figure}[htbp]
    \centering
    \includegraphics[width=\textwidth]{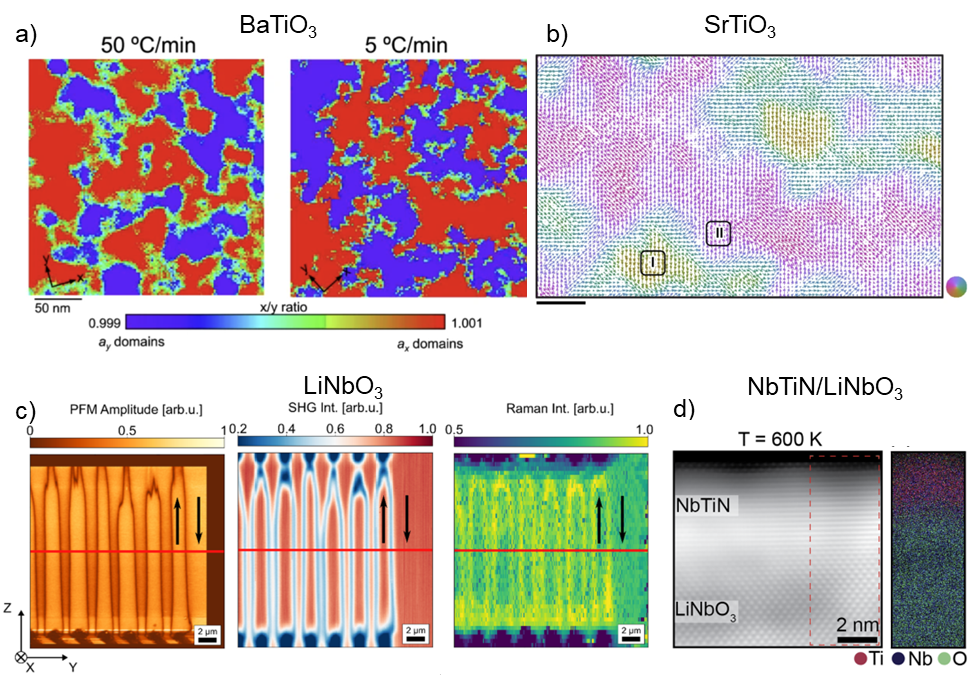}
    \caption{Materials characterization approaches that enhance understanding of material properties. a) Barium titanate domain mapping after growth on oxidized silicon can be correlated to electro-optic coefficients in bulk films. b) Strontium titanate polar domains at low temperature are mapped using electron microscopy, which are vitally important to understand and modify the electro-optic effect for quantum devices. c) a variety of surface probe and nonlinear optical measurements can be used to correlate domain movement and dynamics to periodically pole thin film lithium niobate, relevant for nonlinear optical generation. d) superconductor growth on thin films waveguide materials is important to optimize material properties and integrate detectors into active nanophotonic structures. Images adapted with permission from refs. \cite{reynaud2025enhancement}(a), \cite{zhang2026imaging}(b), \cite{reitzig2021seeing}(c), and \cite{telkamp2026high}(d).}
    \label{fig:S4_characterization}
\end{figure}

Entangled photons can be generated via spontaneous parametric downconversion, a process in which two photons are generated from one photon of higher energy. Energy and momentum conservation requirements make this effect challenging to achieve purely through modal engineering due to the natural chromatic dispersion of materials across large wavelength ranges. Quasi-phase matching has been developed for several decades\cite{fejer1992quasi} to overcome this effect, where a periodic modulation of the sign of the $\chi^{(2)}$ tensor through domain inversion can provide the additional momentum kick required for this process. In materials like lithium niobate, the high electric fields required for domain inversion create additional challenges in this poling process, and the optimal configuration of electrodes, poling waveform, cladding material, and other experimental requirements have been optimized through extensive experimentation\cite{younesi2021periodic, rao2019actively, rao2016second,miller1998periodically}. Small changes to these variables can impact domain inversion dynamics, manifesting as changes to uniformity or shape of the inverted domains\cite{reitzig2021seeing}. This is vitally important to understand and control to create scalable devices, as high conversion efficiency typically requires both long (mm-cm)\cite{maeder2026programmable, bollmers2025segmented, babel2026high} and very uniform periodically poled waveguides to minimize momentum walk-off. The reason for this is still unclear, but is assumed to be related to the metal-ferroelectric interface potentially hosting additional defects that modify the transient circuit-level response. While metrology-focused fabrication\cite{xin2025wavelength} and adaptive poling approaches\cite{chen2024adapted} have yielded large-scale and efficient integrated nonlinear sources, more is required for periodic poling-based quantum devices that may require several identical sources on the same chip. Electronic characterization methods in conjunction with structural characterization via electron microscopy or other optical means (fig. \ref{fig:S4_characterization}c) after poling can help elucidate this effect.

\subsection{Other Materials Opportunities and Challenges}

Materials that lack an intrinsic $\chi^{(2)}$, such as silicon nitride and tantalum pentoxide, can also serve as promising platforms for quantum devices\cite{lu2019chip, splitthoff2020tantalum}.  Silicon nitride in particular provides a complementary platform for quantum photonics that features ultra low optical loss, broad transparency, and established wafer-scale processing   \cite{moss_new_2013, puckett_422_2021}. Its main challenges are related to thermal processing to mitigate strain and optical loss, as well as unwanted fluorescence. Stoichiometric silicon nitride deposited by low pressure chemical vapor deposition (LPCVD) exhibits substantial tensile stress when grown on silicon dioxide, limiting the maximum film thickness to around 200 nm to avoid film cracking \cite{daldosso_fabrication_2004}. This can be insufficient, as several device applications require micron-thick films for appropriate dispersion engineering \cite{okawachi_bandwidth_2014, mcnulty_overcoming_2025}.  Stress-relief trenches \cite{wu_stress-released_2020, wu_integrated_2021}, substrate prepatterning \cite{pfeiffer_photonic_2016, pfeiffer_photonic_2018}, and multilayer \cite{mcnulty_overcoming_2025} or damascene \cite{pfeiffer_photonic_2016, pfeiffer_photonic_2018} approaches can confine or suppress crack formation to overcome this challenge. 

\begin{figure}[htbp]
    \centering
    \includegraphics[width=0.8\textwidth]{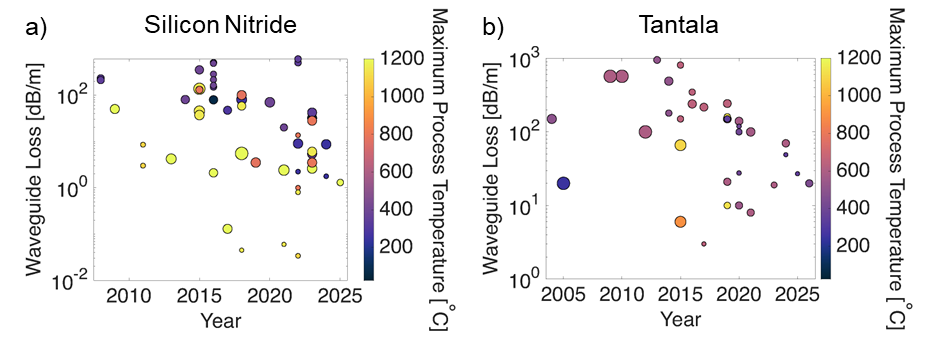}
    \caption{Material loss by year and post-growth annealing temperature in (a) silicon nitride and (b) tantala waveguides. Lower temperatures are required in tantala waveguides to avoid crystallization.}
    \label{fig:S4_loss}
\end{figure}

A separate materials limitation arises from the chemical composition of deposited silicon nitride. Hydrogen-containing precursors can leave Si-H and N-H bonds in the film, whose vibrational overtones produce absorption near telecommunications wavelengths\cite{chia_optical_2022, osinsky_optical_2002}.  This limits the propagation loss achievable in waveguides and thus reduces the quality factor of optical resonators \cite{chia_optical_2022}.  High temperature annealing of these films after deposition can reduce this loss (fig. \ref{fig:S4_loss}a), but require temperatures approaching 1200$^{\circ}$C\cite{shaw_fabrication_2005, bruyere_annealing_1992, chia_optical_2022}.  While this strategy is effective in standalone silicon nitride films, this thermal budget can be problematic when it must be integrated with other materials or back end of line (BEOL) CMOS processing, which places an upper bound on temperature processing to approximately 400$^{\circ}$C. Recent strategies that use low temperature deposition\cite{bose_anneal-free_2024}  and  deuterated\cite{chia_optical_2022}  precursors can overcome these challenges, which are more amenable to heterogeneous integration. This is one approach that should be explored in more detail to interface with other active materials, like the ones discussed above\cite{bose_250c_2022}. 

Even as thermal processing and strain issues are being resolved, the microscopic origin of optical processes in these films should be better understood.  Autofluorescence and parasitic quantum emission have been observed, yet the atomic structures responsible for these defects remain elusive, limiting their deterministic control through materials processing\cite{senichev_room-temperature_2021, meher_origin_2026}.  The sensitivity of these emissions to composition and fabrication conditions suggests that local chemistry and interfaces are more important than the photonic behavior determined by the bulk film\cite{senichev_silicon_2022}.  More recent work linking hydroxyl-containing species to optical loss further emphasizes that low-loss silicon nitride may require control of chemical states at surfaces and interfaces\cite{giesriegl_hf_2026, cavalleri_reduction_2026, paik_reducing_2010}.   Correlating deposition approaches that minimize optical loss with fluorescence characteristics may help elucidate the underlying defects responsible for this, and iterative analysis of both of these concerns should be evaluated when working with these materials and devices. Reduced luminescence has been observed in nitrogen rich silicon nitride, and proposals of both autofluorescence and Raman scattering from different deposition conditions have been reported. Low temperature optical measurements, combined with spectroscopic assignment of local bonding environments may help unambiguously identify the root cause. For example, a recent parameter extraction approach has seen a clear correlation between optical loss and fluorescence in foundry-fabricated silicon nitride waveguides, and background signal intensity varies between samples and annealing procedures. Combining this sort of measurement approach with defect-sensitive spectroscopies and microscopies can help identify these defects and provide appropriate working principles for quantum device s\cite{coleto_extracting_2025}.  

Tantalum pentoxide (tantala) provides another promising platform for QIS photonics, combining broad optical transparency \cite{belt_ultra-low-loss_2017, jung_tantala_2021} and a low thermo-optic coefficient \cite{zhao_low-loss_2020} with the ability to support thick films for dispersion engineering \cite{jung_tantala_2021}. In contrast to stoichiometric silicon nitride, tantala films hundreds of nanometers thick can be deposited without the same severe cracking issues, providing greater flexibility for confining optical modes, and reports of reduced autofluorescence suggest that this amorphous material may be an ideal addition to quantum photonic devices \cite{irvine_comparative_2025}. However, achieving low optical loss generally requires maintaining tantala in an amorphous state. Since film crystallization onsets at 650$^{\circ}$C, this upper bounds the thermal budget available for subsequent processing (fig. \ref{fig:S4_loss}b) or heterogeneous integration\cite{degnan_reducing_2026, appel_design_2015, ren_annealing_2021}.  Zr doping of tantala can shift this phase transition to higher temperature by frustrating the structural ordering\cite{abernathy_exploration_2021}.  Development of amorphous tantala films that simultaneously provide low optical loss, processing stability, and structural uniformity will determine whether tantala can support the same heterogeneous thermal budget constraints already identified for silicon nitride, without sacrificing its native optical performance.

The amorphous structure that enables low-temperature processing also introduces local chemical disorder that results in optical loss. Oxygen deficiency and energetic oxygen-ion bombardment during deposition can increase oxygen vacancy populations and degrade optical performance, demonstrating that oxygen chemical potential and deposition conditions directly influence loss\cite{wang_reduction_2025}.  More locally, individual oxygen defects have been identified and selectively modified to improve cavity performance, linking short-range chemical structure to optical quality factor at longer wavelengths \cite{kong_physical_2021}.  Disorder within the longer-range amorphous network produces Urbach tail absorption, which more strongly absorbs higher energy photons \cite{kong_physical_2021}.  Thus, reducing tantala loss requires controlling both the local defect population and the broader structural disorder of the amorphous network. How tantala deposition chemistry and processing determine the short and long range order, as well as their contributions to optical loss across the relevant QIS spectral ranges, remains an open challenge. 

These optimized optical properties must also be preserved through the fabrication steps required to form photonic devices on tantala. Surface and interface chemistry can introduce additional defect populations, as demonstrated in superconducting tantalum devices where interfacial TLS contribute to microwave loss \cite{kong_physical_2021}.  Its hardness and brittleness complicate dry etching, and produce rough sidewalls that introduce scattering loss\cite{sim_tantalum_2024, jung_tantala_2021}.   

\subsection{Superconducting Nanowire Single Photon Detectors}
Single-photon detectors based on superconducting nanowires (SNSPDs) are an attractive element in quantum systems because of their high conversion efficiency, ability to act as number-resolving detectors, and speed \cite{Divochiy_Marsili_Bitauld_Gaggero_Leoni_Mattioli_Korneev_Seleznev_Kaurova_Minaeva_et_al, You_2020}. They are produced using thin films of superconductors and patterned into meander or hairpin nanowire structures. When biased near their superconducting switching current, a single photon absorption event breaks Cooper pairs and creates a region of higher resistance by suppressing the superconducting state. This leads to a voltage pulse in the device that can be read out to indicate the existence of a photon \cite{Natarajan_Tanner_Hadfield_2012, You_2020}. Here, both crystalline and amorphous materials have been considered with a variety of transition temperatures, critical currents, superconducting gap energies, and kinetic inductance values. Common device metrics like dark count rates, detection efficiency, and jitter are composed of both material properties and device design. The most commonly used materials for these detectors fall into two classes: polycrystalline nitrides and amorphous silicides.  NbN was the first material used in an SNSPD where the device was used to detect single infrared photons and NbN is still the most widely studied SNSPD material to date \cite{Goltsman_Okunev_Chulkova_Lipatov_Semenov_Smirnov_Voronov_Dzardanov_Williams_Sobolewski_2001, Cucciniello_Lee_Feng_Yang_Zeng_Patibandla_Zhu_Jia_2022}.. Since these growths can be somewhat challenging to make reproducibly at the wafer scale, NbTiN is actively being explored as an alternative material, as it  combines polycrystalline-like device performances with high film uniformity \cite{Ma_Shu_Zhang_Yu_Jia_Xiao_Yu_Liu_Li_Eklund_et_al_2022}. 
Among the polycrystalline nitrides discussed, NbN has impressive detection efficiencies (often exceeding 90\%), a higher transition temperature, and a larger volume of research.\cite{Goltsman_Okunev_Chulkova_Lipatov_Semenov_Smirnov_Voronov_Dzardanov_Williams_Sobolewski_2001, Pernice_Schuck_Minaeva_Li_Goltsman_Sergienko_Tang_2012}. As a result, NbN detectors are often fast and efficient. However, achieving wafer-scale uniformity remains challenging due to narrow growth windows and restrictive lattice matching constraints that often limit the choice of substrate to substrates like MgO or sapphire \cite{knehr2021wafer}. These challenges make it difficult to integrate NbN detectors into silicon photonic systems \cite{Cucciniello_Lee_Feng_Yang_Zeng_Patibandla_Zhu_Jia_2022}. TiN, however, is much more tolerant to non-lattice-matched substrates. The superconducting critical temperature in this system, however, is generally lower than that of NbN. NbTiN is a compromise between NbN and TiN, offering the ability to produce a polycrystalline films in which the critical temperature and lattice matching sensitivity can be tuned between the two extremes by altering its stoichiometry \cite{Ma_Shu_Zhang_Yu_Jia_Xiao_Yu_Liu_Li_Eklund_et_al_2022}.

In the amorphous MoSi and WSi alloys, performance is governed chiefly by stoichiometry rather than crystal structure. For MoSi, the transition temperature is maximized with a metal-rich ratio, despite the enhanced likelihood of crystallite defect formation \cite{Bosworth_Sahonta_Hadfield_Barber_2015b}. Common defects often result from stoichiometric variation: metal or Si-rich clusters can nucleate crystalline structures which decrease overall uniformity and can degrade performance by acting as a source of dark counts. Further, these devices are highly susceptible to oxidation-related defects, where oxides form in the film structure and locally destabilize the superconducting state, reducing the critical temperature. Capping layers and careful interface engineering may help to prevent oxidation. 

Integrated quantum devices often require detectors grown on top of waveguides, making the materials selection aspect of both underlying active material and detector important considerations, and can also be a major driver of integration complexity \cite{Ferrari_Schuck_Pernice_2017}. For crystalline superconductor materials, this means that there is often not an ideal epitaxial substrate for growth. Polycrystalline superconductors like the nitrides can therefore have different grain sizes that depend on both growth conditions and underlying crystal structure of the substrate. The relative size of grains to the width of the nanowire strip may be an important parameter that impacts critical device metrics like dark counts and jitter, and can act as regions for enhanced sensitivity to oxidation. As a result, research into epitaxial SNSPD materials seeks to mitigate this \cite{Cheng_Wright_Xing_Jena_Tang_2020}.  Grains that are either significantly larger or smaller than the nanowire size may be ideal in this case. Furthermore, effects at the film-substrate interface may create a different local structure that can act as a superconducting dead layer. For example, integrating a nitride-based superconductor like NbTiN onto an oxide platform could create an interfacial oxynitride if the growth conditions are not ideal, whose superconducting properties are not the same as the desired film. Electron microscopy has been used to confirm the high quality growth of nitride superconductors on lithium niobate\cite{telkamp2026high} (fig. \ref{fig:S4_characterization}d), and this will likely be an important consideration as the palette of integrated materials grows. 

Often, the high temperatures required to obtain crystalline SNSPDs can be sidestepped entirely by amorphous detector platforms like MoSi or WSi \cite{Marsili_Verma_Stern_Harrington_Lita_Gerrits_Vayshenker_Baek_Shaw_Mirin_et_al_2013, Bosworth_Sahonta_Hadfield_Barber_2015b, Verma_Korzh_Bussières_Horansky_Dyer_Lita_Vayshenker_Marsili_Shaw_Zbinden_et_al_2015}. These detector materials can be deposited on a variety of substrates at low temperature. This structure is in principle easier to grow and provides greater film homogeneity that can improve device yield and consistent device metrics. NbN, TaN, or TiN  boast high critical temperatures and low kinetic inductances, allowing for relaxed cryogenic constraints and fast switching \cite{Goltsman_Okunev_Chulkova_Lipatov_Semenov_Smirnov_Voronov_Dzardanov_Williams_Sobolewski_2001}.

Similar to other thin films discussed in this review, it is clear that surfaces, grain boundaries, interfaces, and defects all play a significant role in device performance. Electron microscopy of not only high quality growth of superconducting materials\cite{telkamp2026high, wang2026all}, but also of fabricated devices, will prove to be increasingly important as detectors become more integrated with emerging thin film photonic structures. Because these devices also depend quite a bit on geometric factors like thickness, deposition type and uniformity will likely be a significant consideration in scalable systems.  Approaches like ALD for both amorphous and nitride materials will likely be an important method in future devices. However, it has been observed that wafer-scale analysis can lead to significant cross-wafer variation in material properties and device performance\cite{linzen2017structural, knehr2021wafer}. Correlating growth conditions and uniformity, device performance metrics, and structural analysis of fabricated devices can provide important insights into heterogeneity in performance of nominally identical devices. 

%% file: 5.two-dimensional_materials.tex

\section{Two-Dimensional Materials and Heterostructures}

2D materials and their heterostructures have emerged as a novel materials platform for quantum information science, offering interfacial control that is difficult to achieve in conventional three-dimensional solids. In bulk materials, surfaces and interfaces are often prone to defects such as dangling bonds, charged defects, dislocations, introduced during growth and fabrication. By contrast, layered materials can be isolated down to a single atomic plane and assembled into van der Waals heterostructures with clean interfaces, even without strict lattice matching \cite{Geim2013, Ajayan2016}.

This capability enables the combination of chemically and electronically distinct materials, including semiconductors, superconductors, magnets, insulators, and topological materials, into designer heterostructures with atomically abrupt interfaces. At the same time, the absence of chemical bonding across the interface can, in principle, eliminate many defect-generation pathways. These features are particularly attractive for the ``Quantum Evolution 2.0'', where device performance is increasingly limited by microscopic disorder and interface nonideality. For example, recent work shows that crystalline hBN can serve as a low-loss dielectric for high-quality-factor superconducting resonators and qubits \cite{Wang2022_hBN}. The improved performance of hBN-based devices is attributed to a reduced density of two-level systems, which are often associated with interfacial oxides, although direct microscopic studies are still needed to clarify the mechanisms.

In addition, their electronic and optical properties are highly tunable by electrostatic gating, strain, layer number, twist angle, and proximity effects, enabling precise control over their functional properties. The strong interactions among elementary excitations, such as electrons and excitons, lead to novel electronic phases \cite{Smolenski2021, Zhou2021, Zhang2023, Zhang2026}, and strong nonlinear effects that could be useful even at the quantum level \cite{Gu2024, Gu2025nonlinear}. Last but not least, the assembled 2D heterostructures can be readily integrated with electronic and photonic structures \cite{deAbajo2025, Sarkar2025}, enabling heterogeneous integration of quantum systems and providing new opportunities for quantum sensing at or near interfaces. This chapter reviews recent progress in the use of 2D materials in QE 2.0 devices and the challenges still facing these materials.

\begin{figure}[htbp]
    \centering
    \includegraphics[width=0.75\textwidth]{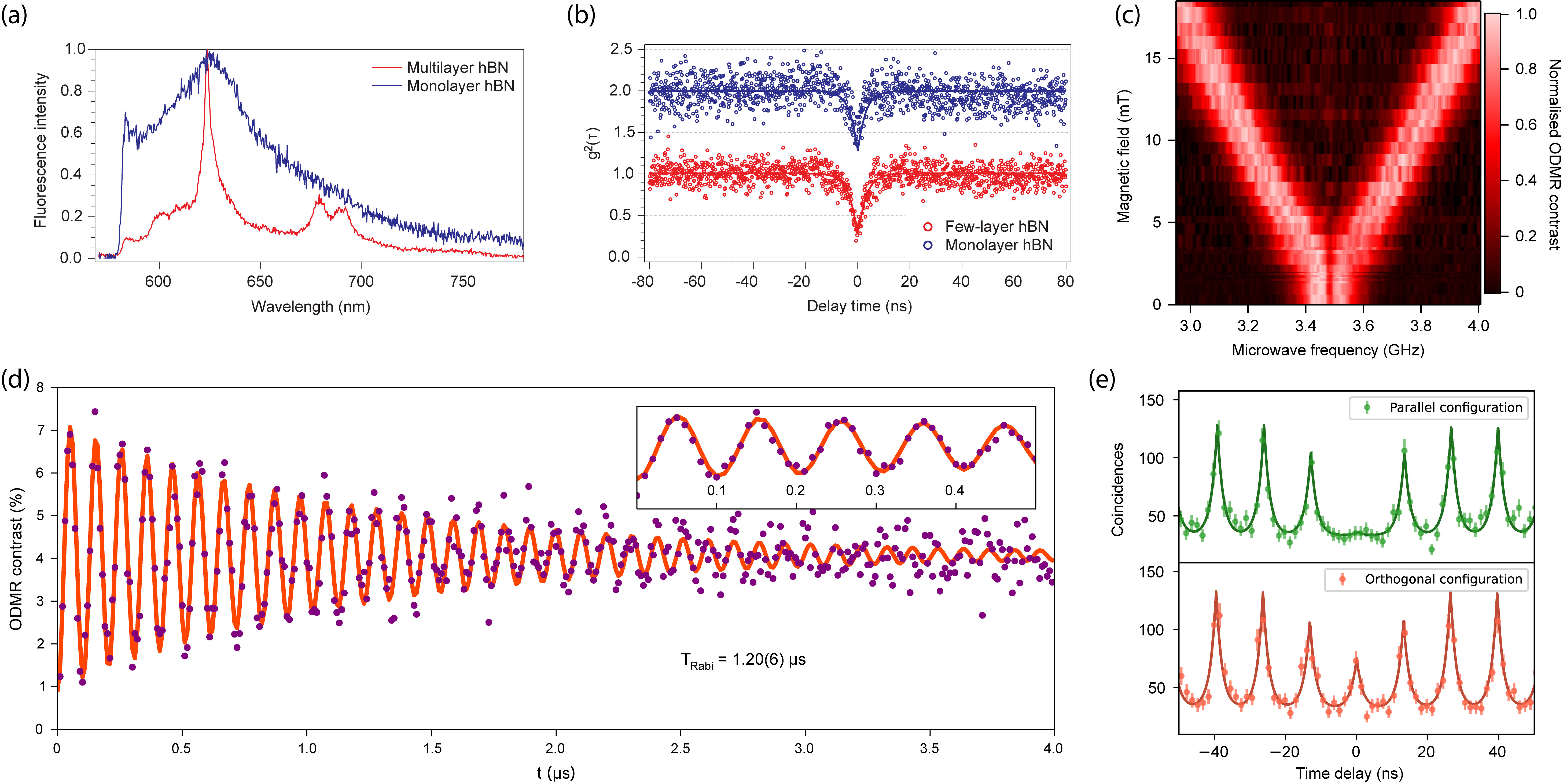}
    \caption{Single-photon emission and spin coherence of quantum emitters in 2D materials.(a) Room-temperature photoluminescence spectra of a defect centre in monolayer (blue) and multilayer (red) h-BN. (b) Second-order correlation function g2($\tau$) of an individual defect centre in monolayer (blue circles) and few-layer (red circles) h-BN; g2(0) < 0.5 confirms single-photon emission. (c) Optically detected magnetic resonance (ODMR) contrast of ensemble VB$^-$ spin defects in h-BN as a function of magnetic field and microwave frequency, showing the Zeeman splitting of the ground-state spin triplet at room temperature. (d) Room-temperature Rabi oscillations of a single carbon-related spin defect in h-BN measured at high microwave power (purple circles; TRabi = 1.20(6) $\mu$s). (e) Two-photon (Hong–Ou–Mandel) interference of consecutive photons emitted by a resonantly driven B centre in h-BN: coincidence histograms measured with the two photons in parallel (green) and orthogonal (orange) polarization configurations. The suppression of the zero-delay peak in the parallel configuration demonstrates photon indistinguishability. Images taken adapted with permission from: \cite{tran2016}(a and b), \cite{Gottscholl2020}(c), \cite{stern2024}(d), and \cite{Gerard2026}(e). }
    \label{fig:2D SPE}
\end{figure}

\subsection{Single-Photon Emitters in Two-Dimensional Materials}

A single-photon source is an optically addressable system that emits one photon at a time with high brightness, spectral stability, and, ideally, photon indistinguishability. Defects, such as color centers in diamond and rare-earth-doped crystals, are among the most widely studied solid-state single-photon emitters. 2D materials offer a complementary route, where quantum emitters can be embedded in an atomically thin host, making them highly accessible for photonic integration, local strain engineering, electrostatic control, and near-field probes \cite{Aharonovich2016}. Recent comprehensive reviews summarize progress across these material platforms and benchmark them against other solid-state single-photon emitter systems \cite{Esmann2024, Chen2025}.

Hexagonal boron nitride has become a central material in this area. Its wide band gap, chemical stability, and ability to host room-temperature quantum emitters make it highly attractive \cite{tran2016, Grosso2017}. Defect-related emitters in hBN have shown bright single-photon emission across the visible and near-infrared spectral ranges, with high single-photon purity \cite{Martinez2016, Exarhos2017, Grosso2017} (Figs. \ref{fig:2D SPE}a,b). Many of these emitters exhibit optically detected magnetic resonance (ODMR) at room temperature, making them attractive for magnetic sensing  (Fig. \ref{fig:2D SPE}c). While there is significant progress, the microscopic origins, electronic structure, and vibronic coupling of many of these emitters remain a challenge. Advanced theory calculations and correlated structural--optical measurements are now being developed to narrow the range of candidate defect configurations \cite{hayee2020, Linderalv2021, Mendelson2021}. As a result, the field has not yet reached the level of deterministic defect identification and control achieved for the color centers in diamond \cite{Aharonovich2014, Gottscholl2020, Chejanovsky2021}.

Transition-metal dichalcogenides, such as WSe$_2$, form a second family of 2D hosts for single-photon emitters. In monolayer TMDs, strong excitonic effects, spin--valley locking, and large spin--orbit coupling give rise to bright exciton absorption and emission with rich spin-valley physics \cite{Mak2012, Xiao2012, Zeng2012, Jones2013, Wang2018}. To date, single photon emitters in TMD operate at cryogenic temperatures and often are dimmer as compared to hBN emitters \cite{Chakraborty2015, He2015, Koperski2015, Srivastava2015, Tonndorf2015}. Their origins are often thought to be localized excitons \cite{Koperski2015, Srivastava2015, Tonndorf2015}. A recent critical review evaluates the competing proposed origins of these emitters and benchmarks reported figures of merit across the field \cite{Chhaperwal2026}. These excitons may be confined via different mechanisms, such as by nearby defects, strain gradients, moir\'e potentials, or electrostatic potentials. For example, local strain can create exciton-localizing potentials, with work suggesting exciton funneling and activation of local defects as a possible mechanism \cite{Branny2017, PalaciosBerraquero2017}. Moir\'e potentials in twisted or lattice-mismatched heterostructures can also confine excitons and have enabled single-photon emission, possibly even forming quantum-emitter arrays \cite{Alexeev2019, Jin2019, Seyler2019, Tran2019}. Ferroelectric domain walls provide another promising route, where strong in-plane electric fields can confine excitons and modify their optical response, although single-photon emission from such electrostatically confined excitons has not yet been demonstrated \cite{Kim2025, Gu2025confining, Soubelet2025}. Some of them can produce single-photon emission while retaining some attractive features of free excitons, such as valley-selective optical transitions, enabling possibly efficient spin-photon interface \cite{Chakraborty2015, PalaciosBerraquero2017}.

The large tunability of TMD emission energy by strain, electric field, magnetic field, dielectric environment, and spin-valley selection rules makes these emitters attractive for integrated quantum photonics and valley-selective quantum optics \cite{Chakraborty2015, PalaciosBerraquero2017}. Yet the same environmental sensitivity also makes them vulnerable to charge noise, spectral diffusion, inhomogeneous broadening, and sample-to-sample variability. Real-time feedback control needs to be developed to stabilize these emitters for quantum communication applications \cite{Peyskens2019, Parto2022}.

Broadly speaking, a key potential advantage of 2D materials is the absence of surfaces with dangling bonds. In principle, this allows quantum emitters to be placed near the surface without the severe coherence degradation often associated with surface disorder in conventional hosts. This near-surface geometry can improve photon extraction, enhance near-field coupling to photonic structures, and increase sensitivity for quantum sensing applications \cite{tran2016, Exarhos2017, Grosso2017, Das2024}. However, in practice, two-dimensional surfaces are often contaminated by hydrocarbons, polymer residues, adsorbates, and processing-related disorder. Surface treatment and passivation strategies therefore need to be developed carefully, especially because the same surface accessibility that benefits sensing also makes the emitters sensitive to unwanted environmental perturbations.

Another open fundamental question is how the coherence properties of emitters in 2D depend on their depth within the host. Nuclear spins in the host material, charge fluctuations, and surface adsorbates can all contribute to spin and optical decoherence. Techniques such as isotopic purification, cleaner transfer methods, encapsulation, surface passivation, and dynamical decoupling may help improve coherence, but the microscopic mechanisms remain incompletely understood \cite{Gottscholl2020, Akbari2021, Chejanovsky2021, Mendelson2021, Horder2024, stern2024, Paralikis2025}. Room-temperature coherent control of hBN spin defects, evidenced by Rabi oscillations and Hahn-echo spin coherence, has since been demonstrated \cite{Guo2023} (Fig. \ref{fig:2D SPE}d). Furthermore, resonantly driven hBN emitters have shown optical coherence signatures, including Mollow triplets and photon indistinguishability via Hong-Ou-Mandel interference \cite{Gerard2026} (Fig. \ref{fig:2D SPE}e).

Single-photon emitters in two-dimensional materials also face many challenges shared with other solid-state systems. Spectral reproducibility and drift, linewidth broadening, photon indistinguishability, deterministic positioning, charge-state control, and scalable integration remain major obstacles to practical devices. A key question is whether emitters can be created with deterministic spatial placement and sufficiently reproducible spectral properties, such that any remaining variability can be compensated by external tuning. Ion implantation, electron irradiation, plasma treatment, laser writing, strain engineering, chemical functionalization, and thermal processing have all been explored as routes to emitter generation \citep{Shotan2016, Branny2017, hayee2020}. However, no method has yet been shown to achieve this desired level of control.

For QIS applications, the field must move from discovery-driven emitter generation to defect-by-design synthesis. This requires correlating synthesis conditions, atomic structure, local strain and electrostatic environment, and quantum optical performance. Progress will depend on developing reproducible synthesis and processing methods, along with characterization tools that link atomic-scale structure to photon statistics, linewidth, spectral stability, and coherence \citep{Mendelson2021, Ping2021, Tsai2024}.

Placing 2D-material emitters in this broader context, it is useful to compare them with the leading solid-state single-photon emitter platforms developed over roughly the same period. Nitrogen-vacancy (NV) centers in diamond remain the benchmark spin--photon interface: they combine millisecond-scale room-temperature spin coherence with a mature microwave and optical control toolbox and have enabled multi-node entanglement distribution over metropolitan-scale fiber networks, making them the most mature platform for quantum networking \citep{Pompili2021, OrphalKobin2025}. Their main drawbacks are a weak, phonon-broadened zero-phonon line and the need for cryogenic operation to obtain spectrally indistinguishable photons, which limits collection efficiency and multiplexing. Group-IV color centers in diamond (SiV, GeV, SnV) have inversion-symmetric electronic structure that suppresses spectral diffusion and gives near-transform-limited optical linewidths, but their electron spin coherence is comparatively short and strongly phonon-limited, again requiring operation at millikelvin-to-few-kelvin temperatures for high-fidelity spin--photon entanglement \citep{Glazov2014, Harris2024}. Silicon carbide divacancies and related defects combine NV-like spin coherence with more mature, wafer-scale, CMOS-compatible growth infrastructure, and recent work has improved their photostability under resonant excitation, though native emitters still largely emit in the visible/near-infrared rather than the telecom band \citep{he2024}. III--V epitaxial quantum dots (e.g., InAs/GaAs) are engineered heterostructures rather than atomic defects; embedded in optical microcavities, they provide the brightest and most indistinguishable deterministic single-photon sources demonstrated to date, but they lack an easily addressable, long-lived spin degree of freedom and require non-scalable epitaxial growth and deterministic device fabrication \citep{Tomm2021}.

Against this backdrop, single-photon emitters in 2D materials occupy a distinct point in this trade-off space. Because they live at the surface of an atomically thin, van der Waals host, they are easier to integrate heterogeneously with photonic circuits and other quantum systems, and their emission is more readily tuned by strain, electric field, and near-field coupling than that of defects buried in diamond or SiC. Many h-BN emitters also operate at or near room temperature, unlike epitaxial quantum dots and most cryogenic color centers. Their principal disadvantages, discussed above, are the still-uncertain microscopic origin of many emitters, greater sensitivity to surface and environmental noise, and spin and optical coherence properties that, while improving rapidly \citep{Guo2023, Gerard2026}, have not yet matched the benchmarks set by NV centers or group-IV defects.

\begin{figure}[htbp]
    \centering
    \includegraphics[width=0.75\textwidth]{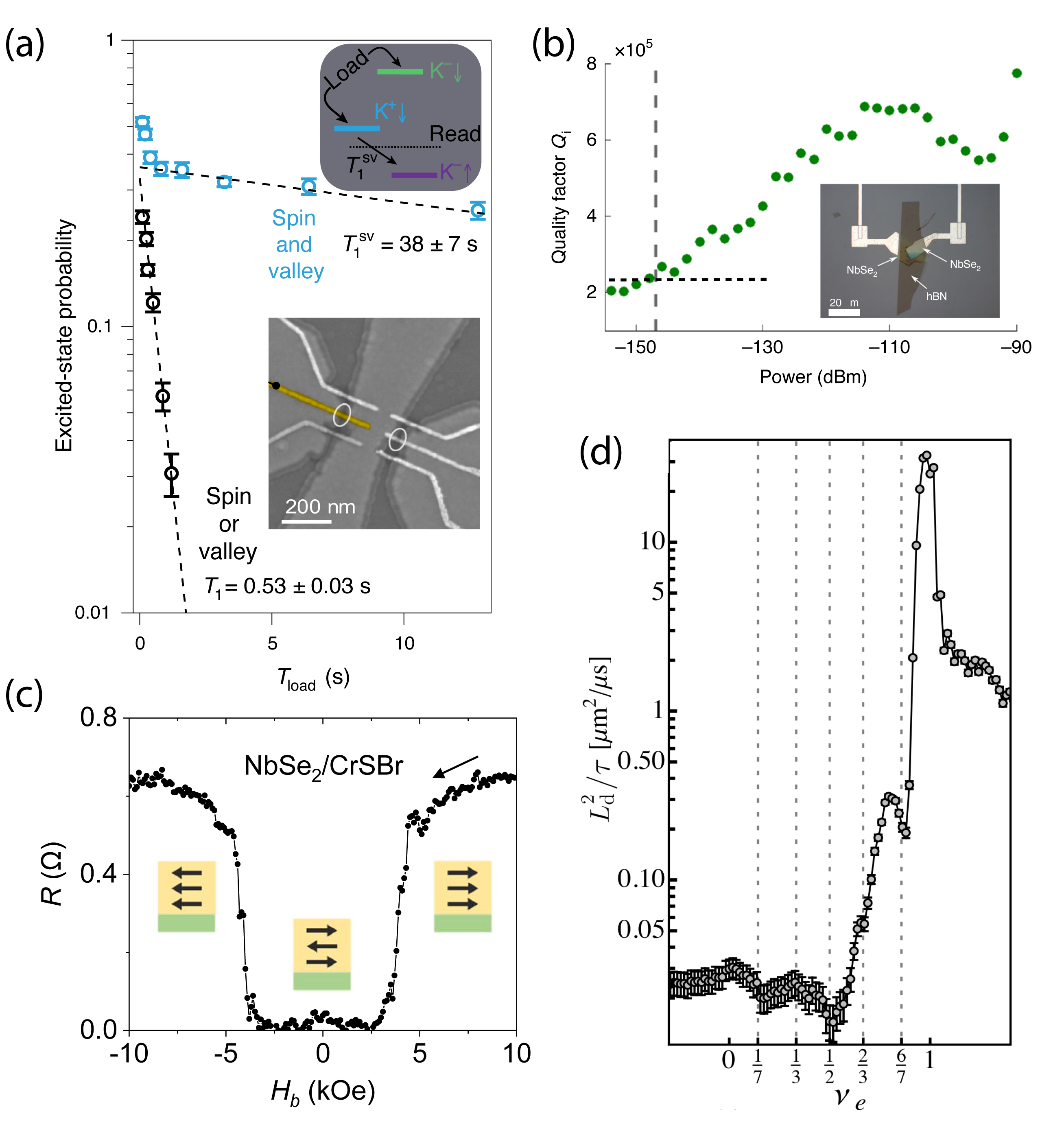}
    \caption{Van der Waals heterostructures for quantum technologies: gate-defined qubits, superconducting circuits, proximitized systems, and moiré quantum simulators. (a) Spin–valley-protected Kramers qubit in a gate-defined bilayer graphene quantum dot. Single-shot relaxation measurements at 30 mK show a spin–valley relaxation time T$_1$ = 38 ± 7 s for the Kramers pair, two orders of magnitude longer than for spin-blocked states (T$_1$ = 0.40 ± 0.03 s), because relaxation requires a simultaneous flip of spin and valley. Inset: an SEM image of the quantum dot device. (b) Crystalline hBN as a low-loss dielectric for superconducting quantum circuits: internal quality factors Q$_i$ of microwave resonators incorporating NbSe$_2$–h-BN–NbSe$_2$ parallel-plate capacitors exceed 105 in the single-photon regime, indicating a low density of two-level-system losses in the van der Waals dielectric. Inset: an optical image of the capacitor device. (c) Superconductor–magnet proximity effect in a NbSe$_2$/CrSBr van der Waals heterostructure. The channel switches between the superconducting and normal states as magnetic layer transition from the antiparallel to the parallel state (insets: magnetic configurations of the CrSBr layers). (d) Quantum simulation of mixed Fermi-Bose system. In an angle-aligned WS$_2$/WSe$_2$ heterobilayer, the diffusivity D of interlayer excitons is measured to be enhanced by more than three orders of magnitude as the electron filling approaches the Mott-insulating state, reflecting non-monogamous exciton diffusion. Images taken adapted with permission from: \cite{Denisov2025}(a), \cite{Wang2022_hBN}(b), \cite{Jo2023}(c), and \cite{Upadhyay2026}(d). }
    \label{fig:2D devices}
\end{figure}

\subsection{Quantum Dots and Spin Qubits}

2D materials also offer a promising route toward electrically defined quantum dots and spin-based qubits. Graphene is attractive because of its low nuclear spin density, weak intrinsic spin--orbit coupling, and high carrier mobility \citep{Trauzettel2007, Recher2010}. These features can, in principle, support long spin coherence. However, the absence of a band gap in pristine graphene complicates electrostatic confinement. This has motivated the use of bilayer graphene, which can be gated to open a tunable gap, as well as etched nanostructures, moir\'e superlattices, and hybrid graphene heterostructures. In particular, gate-defined bilayer graphene quantum dots have now demonstrated single-carrier control, spin and valley Pauli blockade, and spin--valley relaxation times extending to tens of seconds in protected Kramers-pair states, highlighting graphene's potential as a spin- and valley-qubit platform \citep{Volk2011, Guettinger2012, Eich2018coupled, Eich2018spin, Banszerus2020, Banszerus2022, Denisov2025} (Fig. \ref{fig:2D devices}a).

TMDs provide a different route because they possess an intrinsic band gap. In gated TMD quantum dots, the qubit basis can be formed from spin-valley-locked states, potentially enabling magnetic, electric, or even optical qubit control \citep{Kormanyos2014, Pisoni2017, Thureja2024, Pawlowski2025}. One major challenge in realizing TMD quantum dots is the large contact resistance often encountered in TMD devices, which reduces contact transparency and makes transport measurements exceedingly difficult at low temperatures. Recent contact-engineering approaches, such as epitaxially grown semimetal and crystalline antimony contacts, have substantially reduced ohmic contact resistance in 2D semiconductor devices and may help extend these strategies to gated TMD quantum dots \citep{Du2025}.

Graphene/TMD heterostructures provide another important example of proximity engineering. When graphene, with its weak intrinsic spin--orbit coupling, is placed on a TMD with strong spin--orbit coupling, interfacial hybridization can induce spin--orbit effects in graphene while preserving its high mobility. This effect has been studied extensively in spin-transport devices and may be relevant for future spin-qubit and spintronic quantum devices \citep{Wang2015, Avsar2014, Gmitra2015, Yang2017, Ghiasi2017, Island2019, Zollner2022, Zollner2023}.

Scalable spin qubits place stringent requirements on materials quality: clean interfaces, low magnetic noise, charge-trap density, controlled dielectric environments, and reproducible contacts. The qubit performance is therefore not only determined by the 2D semiconductor itself, but also a property of the full materials stack, including encapsulation layers, gates, contacts, substrate, and their interfaces, all of which must be carefully engineered.

\subsection{Proximitized Systems and Topological Quantum Materials}

A particularly intriguing role for 2D heterostructures in QIS is the ability to engineer proximity effects. In a proximitized system, one material inherits properties from another through interfacial coupling. For example, graphene can acquire spin--orbit coupling from a nearby TMD, while coupling a topological material to a superconductor can enable searches for Majorana modes. This approach is attractive for topological quantum computing because the desired ingredients---superconductivity, magnetism, and strong spin--orbit coupling---rarely coexist naturally in a single material. 2D heterostructures provide a natural pathway to realize this by enabling close contact between different materials at atomically sharp interfaces \citep{Gmitra2015, Island2019, Xi2016}.

Atomically thin superconductors such as NbSe$_2$ and NbS$_2$ have been integrated with other 2D semiconductors and insulators such as TMDs and hBN to form vdW Josephson and tunneling devices for studying unconventional and proximity-induced superconducting states \citep{Xi2016, Saito2016, Lu2015, Costanzo2016, Tsen2016, Ugeda2016, Wang2017}. Experiments using crystalline hBN as a low-loss parallel-plate capacitor dielectric in superconducting circuits have realized resonator quality factors exceeding $2 \times 10^{5}$ in the single-photon regime, suggesting a reduced density of two-level-system losses due to the atomically clean van der Waals interfaces~\cite{Wang2022_hBN} (Fig.~\ref{fig:2D devices}b). Superconductor--magnet heterostructures such as NbSe$_2$/CrBr$_3$ and NbSe$_2$/CrSBr can couple magnetic order to superconductivity and exhibit spin-valve behavior, nonreciprocal transport, or possible signatures of topological superconductivity \citep{Gong2017, Huang2017, Burch2018, Gibertini2019, Mak2019, Jo2023, Li2024} (Fig. \ref{fig:2D devices}c).

The materials challenge is again interfacial. Topological superconductivity is exquisitely sensitive to disorder, transparency, induced gap uniformity, magnetic texture, and electrostatic inhomogeneity. Microscopic features, such as adsorbates, bubbles, twist-angle disorder, and strain gradients, can create trivial low-energy states that mimic topological signatures. Therefore, future progress requires not only transport and tunneling signatures, but also direct correlation among interface structure, local electronic properties, and device-level quantum behavior.

\subsection{Twistronics and Moir\'e quantum systems}

Twistronics extends van der Waals engineering by using relative rotational alignment to create new electronic states. When two 2D layers are stacked with a small twist angle or lattice mismatch, a moir\'e superlattice forms. This superlattice produces a periodic moir\'e potential that can generate flat bands and enhance correlations \citep{Bistritzer2011, Carr2017, Weston2020}.

For QIS, moir\'e systems provide a powerful route to programmable quantum matter and reconfigurable quantum simulators \citep{Zhang2023}. Substantial experimental progress has already been made in this direction. Twisted bilayer graphene and related graphene moir\'e structures have revealed correlated insulating states \citep{Cao2018a}, superconductivity \citep{Cao2018b, Lu2019, Yankowitz2019}, orbital magnetism \citep{Lu2019, Sharpe2019}, and topological electronic phases \citep{Serlin2020, Nuckolls2020} that can be tuned by twist angle, carrier density, and displacement field. TMD moir\'e heterostructures have further enabled studies of generalized Wigner crystals \citep{Regan2020, Li2021}, superconductivity \citep{Wang2020}, quantum anomalous Hall effects \citep{Xu2023}, and fractional Chern insulators \citep{Cai2023, Zeng2023, Park2023, Xu2023}. Optical spectroscopy has also shown that moir\'e potentials can localize excitons and create arrays of quantum emitters \citep{Yu2017, Jin2019, Seyler2019, Tran2019, Alexeev2019}.

These advances make moir\'e materials attractive for simulating lattice Hamiltonians that are difficult to realize in conventional solids. For example, electrons in moir\'e superlattices can be tuned through global parameters such as tunneling strength, interaction strength, and filling factor, allowing for the study of Fermi-Hubbard physics \citep{Wu2018, Tang2020, Regan2020}. On the other hand, optical pumping can generate strongly interacting excitons, simulating Bose-Hubbard systems in which collective states such as Bose Mott insulators and superfluids may emerge from exciton interactions \citep{Jin2019, Seyler2019, Shimazaki2020, Gao2024}. More broadly, combining electrons and excitons could enable simulations of mixed Fermi-Bose Hubbard models \citep{Upadhyay2026, Yan2025}(Fig. \ref{fig:2D devices}d). Together, these capabilities establish moir\'e materials as a versatile platform for studying interaction-driven quantum phases and for exploring a new regime of quantum simulation \citep{Wu2018, Regan2020, Li2021, Smolenski2021, Zhou2021}.

A major challenge is reproducibility. Many moir\'e parameters and phenomena are highly sensitive to the twist angle, and the local twist angle often varies within a device, as evidenced by SEM, AFM, TEM and various other techniques \citep{Uri2020, Yoo2019, Weston2020, Andersen2021, McGilly2020, Rosenberger2020, Benschop2021, Sung2022}. For QIS applications, where device-to-device reproducibility and long-term stability are essential, twist angle must become a controlled manufacturing parameter rather than an artisanal fabrication variable.

\subsection{Challenges and Outlook in Emerging 2-D Materials for QIS}

Two-dimensional materials have already demonstrated a remarkable range of Quantum 2.0 applications by providing a uniquely flexible materials platform for engineering quantum states at interfaces. Their atomically thin geometry, clean van der Waals surfaces, and wide tunability make them especially attractive for quantum emitters, spin qubits, proximitized superconducting devices, and Moir\'e quantum systems \citep{Novoselov2016, Wang2012, Chhowalla2013, Aharonovich2016}. At the same time, these advantages also expose a central challenge: quantum performance in 2D systems is often extremely sensitive to local disorder, interface contamination, strain inhomogeneity, and device-to-device variability. Many current demonstrations still rely on individually optimized devices, often fabricated in a non-scalable fashion. This discovery-driven mode has been essential, but it is not sufficient for scalable Quantum Evolution 2.0.

The next phase of the field will require moving from isolated demonstrations toward reproducible materials growth, assembly, and fabrication protocols. Scalable QIS based on 2D materials requires wafer-scale control over layer thickness, twist angle, defect density, strain, dielectric environment, and interface contamination. Interface engineering is particularly important, since poorly controlled interfaces can limit optical stability, charge noise, coherence, and device yield, all of which are critical for any Quantum 2.0 applications. For example, although hBN encapsulation has become a standard strategy for improving 2D device quality, it is still often performed via exfoliation and transfer methods in a hardly scalable way. Scalable hBN growth, clean wafer-scale transfer, direct heterogeneous growth, or alternative encapsulation strategies will be essential for foundry-style fabrication \citep{Banszerus2015, Lee2014, Lee2018, Kim2012, Kang2015, Yu2015, Li2015}.

For single-photon emitters in hBN and TMDs, synthesis must be linked more directly to defect identity, charge stability, emission properties, and optical coherence. Emitters generated by irradiation, plasma treatment, strain engineering, or native defects often have broad distributions in wavelength \citep{tran2016, Mendelson2021, Koperski2015, He2015}. Better control over intrinsic defects and individual dopants is therefore critical. Growth parameters such as precursor purity, growth temperature, isotope composition, substrate choice, and post-growth annealing should be optimized together with optical metrics. Deterministic placement may require combining bottom-up growth with top-down patterning, local strain engineering, or templated nucleation to enable reproducible integration of quantum emitters with phononic structures \citep{PalaciosBerraquero2017, Branny2017, Tsai2024}.

Scalable growth and patterning are equally important for electrically defined quantum dots, proximitized heterostructures, and moir\'e quantum systems. In graphene and TMD spin or valley qubits, charge noise and contact disorder are often governed by dielectric disorder, trapped contaminants, edge roughness, and processing-induced defects. In proximitized systems, the induced superconducting or magnetic interaction is highly sensitive to interface transparency, lattice registry, air exposure, and damage during metallization. In moir\'e materials, even sub-degree twist-angle inhomogeneity or nanoscale strain relaxation can change the local band structure and interaction landscape, limiting reproducibility across devices. These examples make it clear that the entire growth and device fabrication processes must be optimized as a single integrated materials workflow \citep{Banszerus2020, Wang2015, Gmitra2015, Xi2016, Uri2020, Weston2020}.

A practical path forward is closed-loop materials growth and characterization. In situ diagnostics such as RHEED and optical spectroscopy should be paired with ex situ probes of atomic structure, strain, and contamination, as well as measurements of quantum properties. Growth and fabrication metadata should be reported alongside relevant device metrics, such as optical linewidth, coherence time, oscillator quality factor, and device yield. Benchmark samples and results shared across different labs would help separate intrinsic materials limits from processing artifacts. This type of integrated dataset would enable data-driven optimization, where machine-learning models help identify hidden correlations between synthesis/fabrication conditions and quantum performance \citep{Jain2013, Butler2018, Schmidt2019, Schleder2019}. For instance, AI-assisted approaches could also accelerate the search for growth windows, defect-generation recipes, and processing conditions that are difficult to optimize by intuition alone \citep{Butler2018, Tsai2024}. To enable this feedback loop, it is also critical to develop rapid, high-throughput test methods for quantum properties, such as coherence, which often require low-temperature measurements. With this level of standardization, 2D materials promise to move from flexible research platforms toward reproducible materials systems for scalable QIS.

%% file: 6.materials_characterization.tex


\section {Materials Characterization for Quantum Systems}

\subsection{The Characterization Challenge}
Understanding how atomic-scale structure governs qubit decoherence requires bridging two worlds that operate on vastly different scales. On one side are the structural and chemical tools of materials science, such as electron microscopy, surface spectroscopy, and scanning probes, which yield information at the nanometer to sub-Angstrom scale with chemical and structural specificity. On the other side are the quantum properties that matter for device performance, among them T$_1$, T$_2$, quality factors, and TLS coupling strengths. These are governed by energy scales of a few $\mu$eV and involve defect densities as low as a single active TLS per GHz per micrometer of device perimeter. However, the energy resolution and system-size sensitivity of established surface analysis techniques do not yet reach the regime of coherent TLS. The two worlds are separated by several orders of magnitude in both energy and length scale simultaneously (Fig. 10a)\cite{de_graaf_chemical_2022}.
\begin{figure}[htbp]
    \centering
    \includegraphics[width=1.0\textwidth]{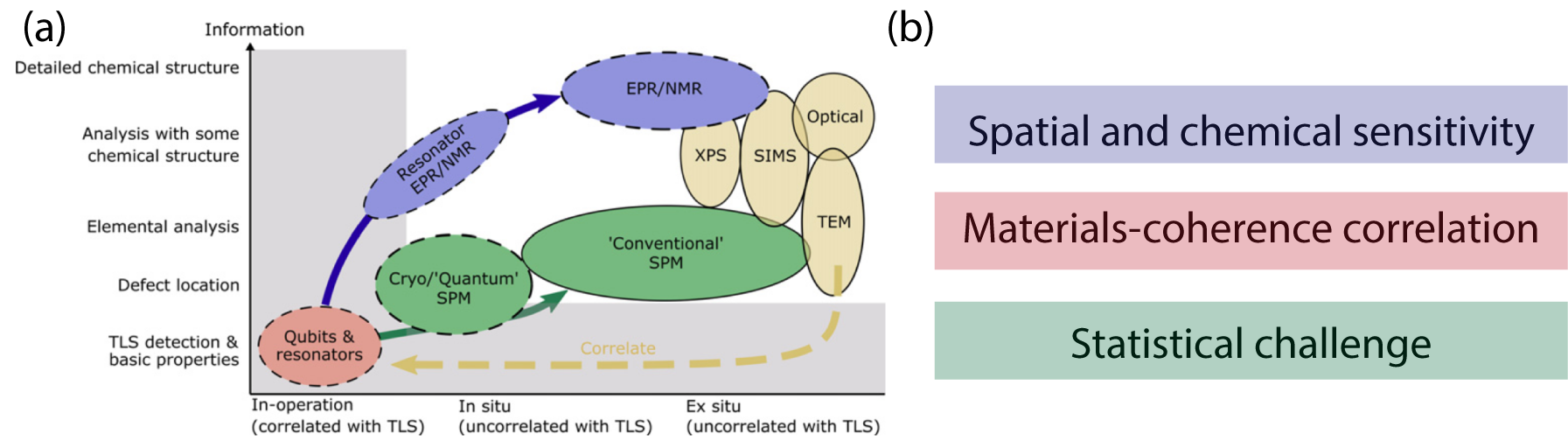}
    \caption{(a) Diagram of material analysis techniques towards understanding TLS defects as a function of information content and operating conditions. Established groups of surface analysis techniques (yellow) at present used to correlate properties with qubit performance. (b) Three gaps identified in this review for the structure-coherence correlation. Images adapted with permission from \cite{de_graaf_chemical_2022} (a).}
\end{figure}

This gap has three distinct dimensions (Fig. 10b), each of which limits the field in a different way. The first, hereafter Gap 1, is an access problem, and it is chemical as well as spatial. The most decoherence-relevant structures, including native oxides a few nanometers thick, buried metal-substrate interfaces, and surface adsorbates, are either too thin, too buried, or too chemically subtle for most non-destructive probes to resolve with sufficient specificity. The most likely culprit is also the element hardest to see. Hydrogen, the prime suspect for flux noise and TLS activity in AlO$_x$, is invisible to XPS, and its low electron scattering cross-section makes it difficult to detect by EELS. The second, Gap 2, is a correlation problem. Even when structural or chemical information is obtained with high fidelity, establishing a direct quantitative connection to a quantum property such as T$_1$ or the internal quality factor Q requires that the same or equivalent sample be measured by both structural and quantum techniques. In practice this is rarely achieved, and the field largely operates with indirect correlations, in which better crystallinity is assumed to imply fewer TLS, which in turn is inferred to improve coherence, without ever closing the causal chain. The correlation problem also has a temporal dimension. The surface probed ex situ is not necessarily the surface present during operation. Even at pressures as low as 10-15 mbar inside a dilution refrigerator, a residual gas molecule strikes the $\sim 100 \times 100$ nm$^2$ interaction area of a TLS once every few minutes, so the defect landscape continues to evolve while the device operates. The third, Gap 3, is a statistics and workflow problem. Destructive high-resolution techniques such as TEM have inherently low throughput. Focus ion beam (FIB) lamella preparation requires several hours per sample, is irreversible, and introduces its own artifacts. Studies characterizing ten devices and drawing conclusions about a population are common, but given that coherence times in nominally identical devices on the same wafer can vary by a factor of three to five over hours to days, a fluctuation driven by TLS switching and environmental coupling, statistical conclusions from small sampling are unreliable.

Solving the characterization challenge therefore requires simultaneous advances on all three dimensions, namely better access to buried interfaces, sample-matched structural and quantum measurements, and higher-throughput workflows that can generate statistically meaningful datasets. The following subsections survey the characterization techniques available to address these challenges, organized by the nature of the information they provide and their proximity to closing the structure-coherence gap. The emphasis is on scanning and transmission electron microscopy, which currently provides the most direct structural and chemical access to the relevant interfaces, followed by surface spectroscopy, and finally scanning probe methods. We note throughout where sample-matched measurements have been achieved and where they remain aspirational.

\subsection{Scanning Transmission Electron Microscopy: Atomic-Scale Access to Qubit-Relevant Interfaces}
Scanning transmission electron microscopy, particularly in its high-angle annular dark-field (HAADF-STEM) and electron energy loss spectroscopy (EELS) implementations, is currently the most powerful technique for accessing the buried interfaces and thin oxide layers that dominate decoherence in solid-state qubit platforms. The combination of sub- Å spatial resolution, chemical specificity through core-loss EELS, and bonding-environment sensitivity through energy-loss near-edge structure (ELNES) makes STEM uniquely suited to the structural questions posed by QIS materials. Its principal limitation is that it is destructive and low-throughput, an inherent constraint that governs how the technique can be deployed in a characterization pipeline.

\subsubsection{Josephson Junction Tunnel Barriers and The AlO$_x$ Interface Problem \label{josephson_alox}}

The Al/AlO$_x$/Al Josephson junction (Figs. 11a-c) is the most studied interface in superconducting qubit materials, and for good reason. The amorphous AlO$_x$ tunnel barrier is simultaneously the nonlinear element essential for qubit operation and a suspected host of TLS defects. A decade of cross-sectional TEM and STEM work has established that this barrier is structurally heterogeneous at every length scale examined. Its thickness varies from grain to grain in the underlying polycrystalline Al film (Figs. 11a-b), with thicker oxide forming above grain boundaries due to enhanced oxygen diffusion along boundary triple junctions, a variation that translates directly into critical current spread and qubit frequency scatter \cite{zeng_direct_2015}. Its stoichiometry is typically sub-stoichiometric (AlO$_x$ with $x < 1.5$) when formed by room-temperature oxidation, with oxygen-deficient regions concentrated near the metal interfaces \cite{Murray2021}. Its defect and pinhole populations depend measurably on deposition and oxidation conditions, as established by TEM surveys combined with low-temperature dielectric capacitance measurements that link AlO$_x$ structural metrics to ensemble TLS properties \cite{fritz_correlating_2018, fritz_optimization_2019}. Its local bonding environment is also graded. ELNES of the Al L2,3 edge reveals a coordination gradient through the barrier, in which lower Al coordination, indicating a higher density of defective bonding states that can act as dissipation traps, is concentrated near the lower Al/AlO$_x$ interface (Figs. 11d-f)\cite{oh_correlating_2025,  lim_unique_2025}.

Two independent lines of evidence have nonetheless shifted the field's attention away from the tunnel barrier and toward the surrounding device interfaces. The first is device-side. Geometric participation-ratio engineering, through deep-trenched substrates and enlarged capacitor gaps that dilute surface electric fields, improves quality factors substantially without modifying the junction at all \cite{wang_surface_2015, Calusine2018, Woods2019}. Epitaxial tunnel barriers, long assumed to be the route to lower TLS density, meanwhile yielded little or no coherence benefit when realized\cite{weides_coherence_2011}. The second line of evidence is materials-side and more recent a systematic STEM study of Al deposition rate spanning 0.5 to 5 \AA{}/s found that higher rates produce larger Al grains, flatter surfaces, and more uniform tunnel barriers, yet median T$_1$ remains essentially unchanged ($\sim 43-46$ $\mu$s) across all conditions \cite{oh_correlating_2025}. This decoupling of AlO$_x$ morphology from coherence carries a significant implication. At current performance levels, loss at the metal-substrate (Fig. 11c) and metal-air interfaces dominates over tunnel barrier thickness variation, redirecting engineering attention from the junction to the broader device surface. The same study identified a stress-induced grain boundary sliding failure mechanism, in which thermally driven plastic deformation of the lower Al electrode produces a grain tip that pierces the AlO$_x$ layer. This short-circuit mode is visible only by cross-sectional TEM and absent from surface inspection. In parallel, structurally well-defined van der Waals tunnel barriers have emerged as an alternative junction technology whose atomically sharp interfaces may sidestep the amorphous oxide problem entirely \cite{wang_coherent_2019, antony_making_2021}.

The current frontier is multimodal characterization of actual devices rather than model films. Lim et al. \cite{lim_unique_2025} applied HAADF imaging, EDS, and EELS/ELNES across both the Nb coplanar waveguide resonator and the Al/AlO$_x$/Al junction of the same qubit architecture (Figs. 11a-c), revealing that multiple structurally distinct loss channels coexist within a single device. At the CPW resonator, the NbO$_x$ surface oxide thickness varies with sidewall geometry, approximately 16 nm on one facet and 11 nm on another within the same chip, demonstrating that a single lamella cross-section cannot capture the spatial heterogeneity of dielectric loss in a real circuit. At the metal-substrate interface, distinct amorphous interlayers are identified, an NbSi$_x$ phase of approximately 2 nm beneath the Nb film and an AlSi$_x$ phase of approximately 3 nm beneath the Al electrodes. Both constitute independent loss channels at the substrate-to-metal interface, which the participation ratio framework predicts to carry significant electric field energy. Pham et al. \cite{pham_structure_2024} extended the multimodal approach to cross-sectional lamellae of transmon qubits measured at both room temperature and sub-6 K, combining ADF-STEM, core-loss EELS at the Al, Si, and O K-edges, and 4D-STEM electron pair distribution function (ePDF) analysis. The EELS maps confirm oxide formation at the Al/Si substrate interface as an independent TLS-hosting site. The ePDF analysis supplies information that conventional imaging cannot, resolving pair distances and coordination environments within the amorphous oxide regions where TLS reside. The sub-6 K capability brings structural characterization physically closer to device operating conditions, reducing artifacts from thermal expansion and surface contamination that affect room-temperature lamellae.

\begin{figure}[htbp]
    \centering
    \includegraphics[width=1.0\textwidth]{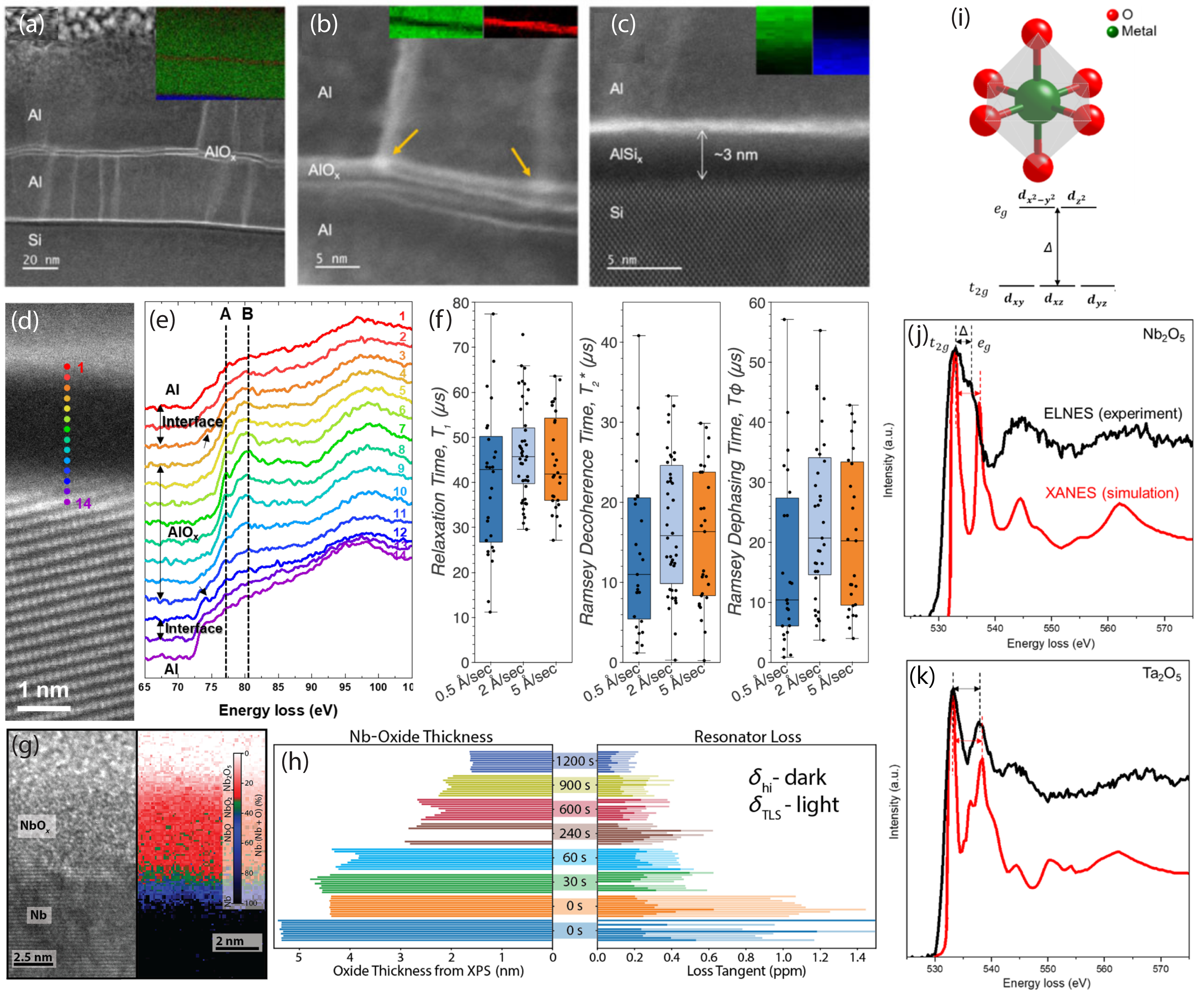}
    \caption{STEM characterization of quantum materials and quantum devices. (a-c) HAADF-STEM images of several key components of a JJ: (a) an overall image of a JJ showing top and bottom Al electrodes and AlO$_x$ tunnel barrier, zoom-in image of (b) a tunnel barrier, and (c) bottom Al electrode-Si substrate interface. Inset: EELS elemental mappings showing distribution of Al (green), O (red) and Si (blue) across the JJ. (d-e) EELS line profile showing changes of the local chemical environment of Al-O bonds across a tunnel barrier Al/ AlO$_x$ / Al, indicative by Al L$_{23}$ edge. (f) Comparison in relaxation time (T$_1$), Ramsey decoherence time (T$_2^*$), and Ramsey dephasing time (T$_\phi$) of three sets of qubits fabricated with Al deposition rates of 0.5, 2, and 5 \AA{}/s. Each data point represents a median characteristic for a single qubit where $n = 28$ for 0.5 and 5 \AA{}/s, and $n = 43$ for 2 \AA{}/s. (g) HRTEM image of Nb-air interface and corresponding EELS quantitative mapping showing O concentration across the inferface. (h) Grouped histograms correlating the resonator loss (right) and the NbO$_x$ thickness from XPS (left) for all resonators. (i) Schematic of the octahedrally coordinated transition metal ion, causing a splitting $\Delta$ in the energy levels of the $d$-orbitals. (j) O K edges of integrated experimental ELNES of amorphous Nb$_2$O$_5$ and simulated XANES of crystalline Nb$_2$O$_5$. (k) O K edges of integrated experimental ELNES of amorphous Ta$_2$O$_5$ and simulated XANES of crystalline Ta$_2$O$_5$.  Images adapted with permission from refs \cite{lim_unique_2025} (a-c), \cite{oh_correlating_2025} (d-f), \cite{Altoe2022} (g-h) and \cite{oh_structure_2024}.}
\end{figure}

\subsubsection{Niobium and Tantalum Oxide: Connecting Surface Chemistry to Resonator Loss \label{niobium_tantalum_oxide}}

Nb-based circuits dominate large-scale processors, and the Nb surface oxide is correspondingly the most intensively characterized loss source in the field. It is chemically more complex than AlO$_x$, since Nb forms multiple co-existing oxide phases (Nb$_2$O$_5$, NbO$_2$, and NbO) with different dielectric properties and different TLS activities. Its behavior under processing is also dynamic. The oxide regrows within hours to days after chemical removal \cite{verjauw_investigation_2021}, and both buffered etching and nitrogen plasma passivation have been shown to improve resonator quality factors \cite{Kowsari2021, Zheng2022}. Additional Nb-specific defect populations complicate the picture. Hydride precipitates form in Nb films and cavities and contribute to both TLS and quasiparticle loss \cite{TorresCastanedo2024}, and theoretical work attributes part of the loss to magnetic disorder from oxygen vacancies in the suboxide layers \cite{sheridan_microscopic_2021}. Against this backdrop, Altoe et al. \cite{Altoe2022} demonstrated the power of correlating STEM/EELS characterization (Figs. 11g-h) with microwave resonator measurements on the same Nb film samples, localizing the dominant loss contribution to the outermost Nb$_2$O$_5$ layer and showing that selective oxide removal improved Q. This remains one of the closest existing examples of a closed structure-loss correlation in the literature.
Murthy et al. \cite{murthy_developing_2022} sharpened the structural picture by combining TOF-SIMS depth profiling with 4D-STEM fluctuation electron microscopy (FEM) on Nb transmon test devices. TOF-SIMS reveals a stoichiometric gradient through the NbO$_x$ depth, with Nb$_2$O$_5$ at the surface and NbO$_2$ and NbO deeper, and establishes that lithography and etching steps drive additional oxygen diffusion into the Nb film beyond what ambient exposure alone produces. This fabrication-process contribution to oxide disorder is one that post-fabrication imaging cannot isolate. FEM maps the relative crystallinity of the Nb$_2$O$_5$ region at nanometer resolution, finding a semicrystalline structure of 1-3 nm Nb$_2$O$_5$ crystallites in an amorphous matrix. Highly disordered regions correlate with reduced Nb-O bond amplitude and increased bond distance in the radial distribution function (RDF), consistent with elevated oxygen vacancy concentration.

The comparative study of Nb and Ta oxides by Oh et al. \cite{oh_structure_2024} supplies the mechanistic centerpiece of this picture and the most thorough atomic-scale explanation for why Ta-based circuits outperform bare Nb (Figs. 11i-k). Aberration-corrected STEM-EELS reveals that Nb oxide undergoes a gradual valence state transition from Nb$_2$O$_5$ through NbO$_2$ and NbO to metallic Nb, spanning approximately 1.6 nm at the oxide–metal boundary, while Ta oxide transitions abruptly from Ta$_2$O$_5$ to metallic Ta within $\sim$ 0.8 nm with no stable suboxides. The difference is thermodynamically grounded. CALPHAD modeling \cite{oh_structure_2024} shows a single discontinuity in oxygen chemical potential for Ta-O versus three for Nb-O, corresponding to the single stable Ta$_2$O$_5$ phase versus the three Nb oxide phases. Beyond the suboxide argument, ELNES of the O K-edge shows that amorphous Nb$_2$O$_5$ exhibits a crystal-field splitting energy reduced by $\sim$ 40\% relative to its crystalline reference (Figs. 11i-j), indicating highly distorted octahedral coordination that lowers the tunneling barrier between bonding configurations and promotes TLS activity. Amorphous Ta$_2$O$_5$, in contrast, retains $\sim$ 95 \% of its crystalline crystal-field splitting, suppressing the same pathway. TOF-SIMS further shows hydrogen accumulating an order of magnitude (Fig. 11k) more at the oxide-metal interface in uncapped Nb than in Ta-capped films, with the more ordered Ta$_2$O$_5$ network proposed to impede hydrogen diffusion toward that interface. Together with the previous studies \cite{Altoe2022, murthy_developing_2022}, this establishes a multi-mechanism picture of Ta$_2$O$_5$ lossiness, comprising suboxide magnetism, structural disorder, and hydrogen accumulation, each independently addressable and each grounded in observables measured on actual device lamellae. The diagnosis is increasingly validated from the mitigation side. Encapsulating the Nb surface with capping layers that prevent Nb$_2$O$_5$ formation produces systematic improvements in transmon coherence \cite{Bal2024}, precisely as the structural picture predicts.

Tantalum itself illustrates how rapidly a multi-technique characterization response can assemble once device results motivate it. Following the demonstration of transmons with T$_1$ exceeding 0.3 ms on Ta capacitor pads \cite{Place2021}, the Ta surface oxide has been examined from four directions in as many years. Variable-energy XPS has depth-profiled the film, resolving the distribution of oxidation states and their response to chemical treatments \cite{McLellan2023}. STEM imaging combined with computational modeling has established the oxide formation mechanism \cite{oh_structure_2024}. Resonator measurements spanning temperature, photon number, and geometry have disentangled surface from bulk TLS contributions, showing that the surface TLS bath can be reduced roughly twofold by buffered oxide etching \cite{Crowley2023}. Most recently, self-assembled organic monolayer passivation \cite{Gupta2026} has been shown to suppress oxide regrowth and improve single-photon quality factors by $\sim$140\%. What no study in this rapidly assembled portfolio yet provides is the closed loop emphasized throughout this section, in which structural, chemical, and quantum measurements are performed on the same high-coherence device.

\subsubsection{Color Centers and 2D Materials: STEM At Different Length Scales \label{color_centers_2d}}

\begin{figure}[htbp]
    \centering
    \includegraphics[width=0.8\textwidth]{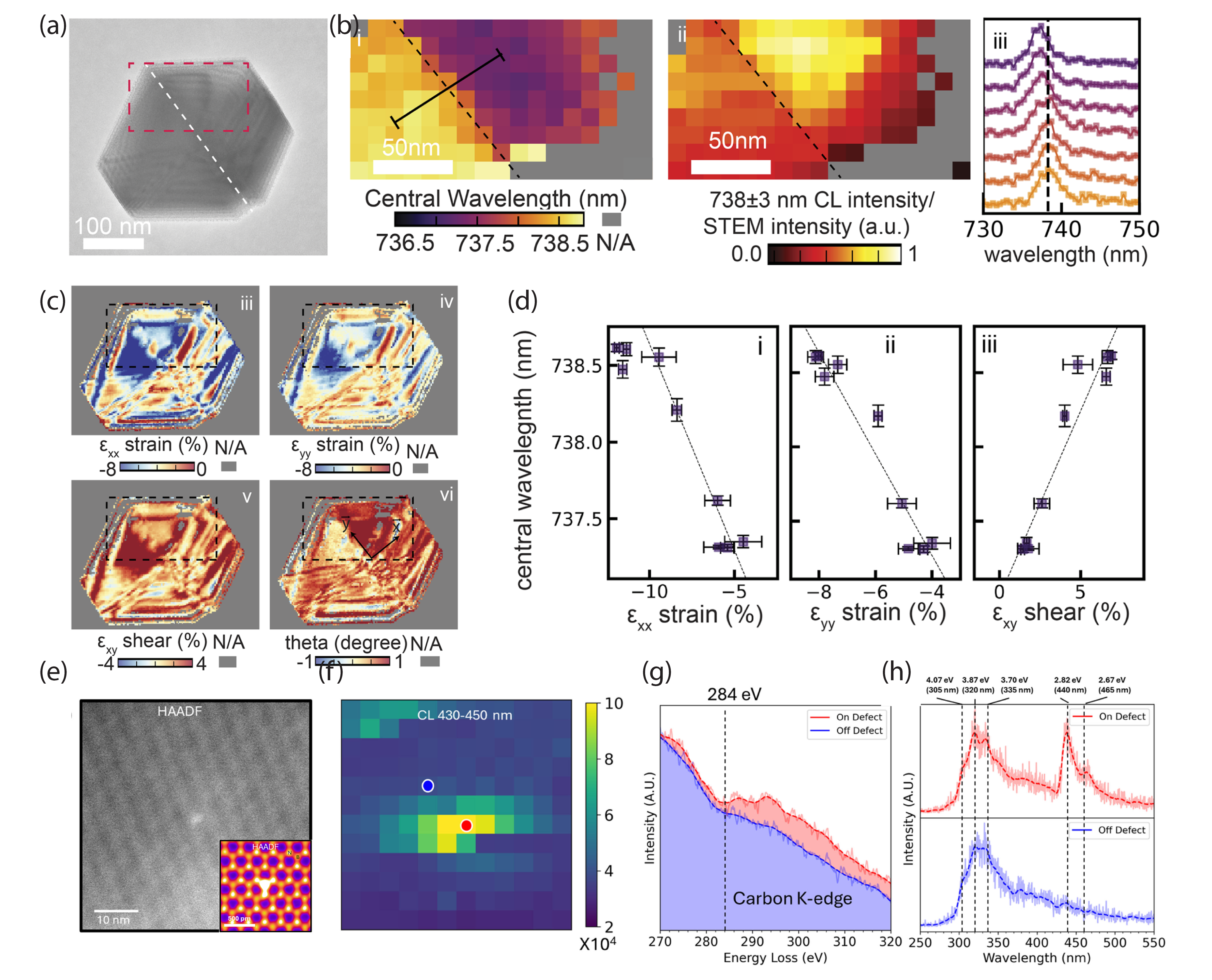}
    \caption{Application of STEM-CL to correlate structure-SPE photophysics.(a-d) SiV- optical properties correlate with strain at the nanoscale.(a) TEM image of a nanodiamond particle embedded with SiV- color centers. (b) 2D hyperspectral maps taken at the red dashed box in (a), (i) Central wavelength of lorentz fit, (ii): 738 ± 3 nm summed intensity, (iii): CL point spectra taken along black line in (b, i). (c) exx strain, eyy strain, exy shear, rotation by 4D-STEM. (d) ZPL wavelength vs exx strain, eyy strain, and exy shear. (e-h) Correlative nanoscale CL microscopy and corresponding STEM-EELS of the blue emitter in twisted h-BN. (e) HAADF image of a folded area in thin h-BN with multiple twist interfaces. The bright contrast in the HAADF image indicates the larger thickness-mass contrast at the emitter. Inset: Atomic-resolution STEM image showing a C substitutional atom within hexagonal BN lattice. (f) corresponding CL mapping of the 440 nm blue emitter.(g) EELS spectra taken on and off the emitter reveal trace amounts of carbon at the same emitter. (h) CL Spectra acquired on and off the emitter, showing the 440 nm emission. The 465 nm peak is the PSB of the 440 nm emitter. Images adapted with permission from refs. \cite{angell_unraveling_2024} (a-d) and \cite{hou_nanometer_2025} (e-h).}
\end{figure}

For color-center qubit platforms, the structural characterization challenge is different in character. The relevant defects are atomic, with an NV center occupying a volume of only a few lattice sites, and the substrate is single-crystal diamond rather than a polycrystalline thin film. STEM is not the primary tool for locating individual NV centers, but it plays an important role in characterizing the extended defect landscape that governs ensemble coherence (Fig. 12). This landscape includes strain fields from dislocations and stacking faults, the lattice damage introduced by ion implantation, which generates the electric and magnetic field noise responsible for short coherence times and spectral diffusion in implanted qubits \cite{Leon2021}, and the structural quality of the near-surface region on which shallow NV performance depends.
A qualitatively more direct form of structure–emission correlation has recently been achieved for the silicon vacancy (SiV$^-$) center in diamond (Figs. 12a-d). Building on the earlier demonstration that STEM-cathodoluminescence can detect single-photon emission from point defects in solids \cite{bourrellier_bright_2016}, Angell et al. \cite{angell_unraveling_2024} combined cryo-STEM with CL spectroscopy and 4D-STEM strain mapping on CVD nanodiamonds, simultaneously acquiring structural and hyperspectral optical data on the same sub-particle volume at $\sim$ 5 nm spatial resolution (Figs. 12a-d). Individual sub-crystallites within a single nanodiamond exhibit ZPL energy shifts of up to 2 nm and brightness variations exceeding 70\%, with grain boundaries acting as sharp barriers to both carrier diffusion and emission uniformity (Fig. 12b-). 4D-STEM CBED analysis (Fig. 12c) identifies the mechanism. Facet-dependent SiV$^-$ incorporation rates produce different defect densities across crystallite boundaries, generating a static strain field in which tensile lattice expansion correlates with ZPL blueshift and increased brightness while compressive strain produces the reverse. Sub-50 nm nanodiamonds, too small to develop distinct crystallite domains, show no spatial ZPL inhomogeneity, a finding that points toward crystallite size control as a practical route to more uniform photon sources. This work achieves the most complete Gap 2 closure demonstrated for any color center platform, with structural and optical quantum properties measured simultaneously on the same nanoscale volume.
Hou et al. \cite{hou_nanometer_2025} extended STEM-CL to hexagonal boron nitride (h-BN) (Figs. 12e-h) by exploiting a CL signal enhancement of up to $120\times$ at twisted h-BN interfaces, which arises from Moir\'{e}-induced modification of the electronic density of states, to achieve sub-nanometer emitter localization in samples thin enough for atomic-resolution STEM and EELS. Through quantitative HAADF intensity analysis (Fig. 12e) combined with EELS carbon K-edge detection (Figs. 12g-h) and DFT defect level calculations, the 440 nm blue emitter is identified as a vertically aligned carbon dimer (VACD), a split interstitial defect substituting at a boron site. The study further demonstrates deterministic creation of isolated blue emitters at targeted sites by focused electron beam irradiation of carbon-coated h-BN with real-time CL feedback, establishing that structural understanding of an emitter can immediately inform its controlled engineering. Together, these studies \cite{angell_unraveling_2024, hou_nanometer_2025} establish cryo-STEM-CL as the emerging tool of choice for closing the structure–emission gap in color center and 2D emitter platforms. It occupies a role analogous to what multimodal STEM now occupies for superconducting qubit interfaces and points toward a shared methodological framework that transcends platform boundaries.

Additionally, the results from Angell et al. \cite{angell_unraveling_2024} also resolves a dimension of the color-center decoherence problem that surface spectroscopy cannot access, namely the contribution of internal crystal structure and grain boundary strain to emission heterogeneity. It thereby complements the sample-matched surface spectroscopy strategy \cite{sangtawesin_origins_2019} discussed in Section \ref{surf_chem_spec}, and together these studies define the spatial hierarchy, from surface chemistry to internal microstructure, across which structure-coherence correlation must eventually be established for color center platforms.

\subsubsection{High-Throughput STEM and The Statistical Bottleneck \label{high-throughput}}

A recurring constraint across all of the STEM studies discussed above is sampling size. The FIB lamella preparation that enables cross-sectional STEM of buried interfaces requires several hours per sample, is irreversible, and is currently the rate-limiting step in any characterization pipeline \cite{mayer_tem_2007}. Studies correlating structural observations with quantum properties typically involve fewer than twenty devices, and given the documented device-to-device variability in coherence times, which can span a factor of three to five on a single wafer, statistical conclusions from such datasets are inherently limited \cite{oh_correlating_2025, Kopas2024, burnett_decoherence_2019}. Detecting a weak but reproducible TLS-count dependence required measuring median coherence across 28-43 qubits per condition, and even that dataset leaves the dominant loss mechanism at other interfaces unresolved.
Addressing this requires both workflow innovation and community coordination. On the workflow side, plasma FIB systems \cite{zhong_comparing_2021, burnett_large_2016} using Xe$^+$ ions rather than Ga$^+$ offer reduced implantation damage and faster bulk material removal, making multi-lamella preparation from a single wafer more practical. Automated multi-site FIB preparation protocols \cite{klumpe_modular_2021}, adapted from semiconductor process control workflows, could increase throughput substantially. On the coordination side, the field needs pre-characterized sample libraries, meaning devices with documented T$_1$, T$_2$, and Q values measured under standardized conditions and made available to the characterization community as shared resources \cite{McRae2020_resonators}. Some national laboratory programs, the Superconducting Quantum Materials and Systems (SQMS) Center and Q-NEXT among them, are developing infrastructure in this direction, but the coordination remains largely ad hoc. The goal, analogous to the wafer-scale process control infrastructure of the semiconductor industry, is a systematic feedback loop in which quantum measurement, structural characterization, process modification, and re-measurement iterate rapidly enough to be scientifically useful.

\subsection{Surface and Chemical Spectroscopy \label{surf_chem_spec}}

Surface analytical techniques provide chemical information complementary to what STEM offers. Where STEM excels at spatial resolution and local structural characterization, surface spectroscopy techniques, including XPS, SIMS, XAS, and related methods, provide ensemble chemical composition, elemental depth profiles, oxidation state specificity, and in some cases trace-element sensitivity far below what EELS can achieve.

\begin{figure}[htbp]
    \centering
    \includegraphics[width=0.8\textwidth]{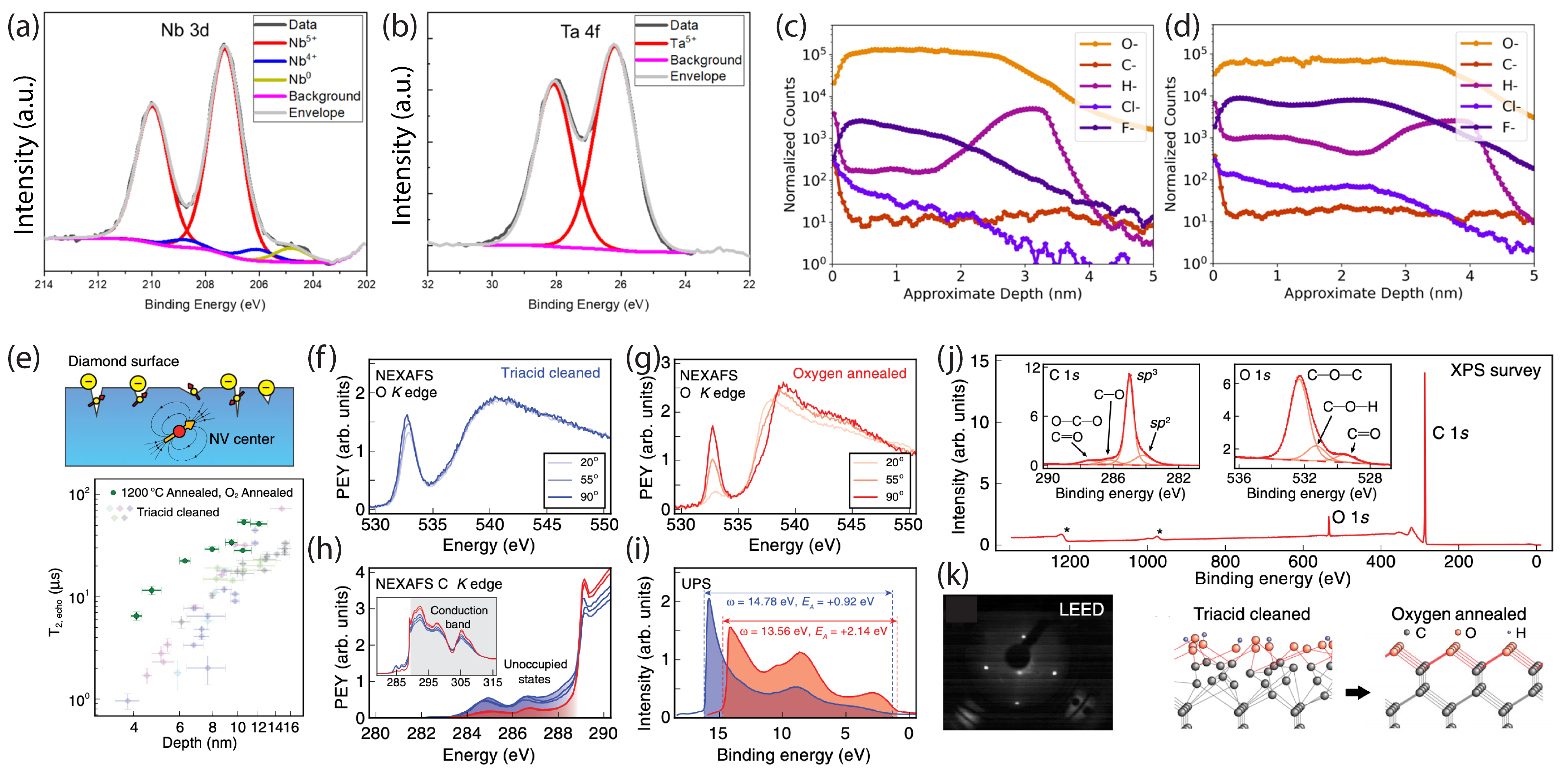}
    \caption{Surface and chemical spectroscopy techniques for quantum devices and quantum materials. (a-b) XPS and (c-d) ToF-SIMS characterization of Nb$_2$O$_5$ / Nb and Ta$_2$O$_5$ / Ta. (a) Nb 3d intensity consists of Nb$_2$O$_5$, NbO$_2$, and Nb (only Nb 3d 3/2 is shown). b, Ta 4f intensity only consists of Ta$_2$O$_5$. Black, violet, and gray lines represent the raw spectrum, background, and envelope, respectively. Normalized ToF-SIMS depth profiles showing O-, C-, H-, Cl-, and F- of bare Nb (c), and Ta-capped Nb (d) thin films. (e-k) Origins of diamond surface noise probed by correlating single-spin measurements  with surface spectroscopy. (e) (Top) Schematic showing an NV center near the diamond surface. The surface can host defects that produce electric and magnetic field noise. (Bottom) Hahn echo coherence time T2;echo as a function of NV depth, measured across six samples with different surface conditions. (f-g) Polarization dependence of NEXAFS spectra at the oxygen K edge compared between triacid-cleaned (blue) and oxygen-annealed (red) surfaces. (h) NEXAFS spectra at the carbon K edge from three triacid-cleaned samples (blue curves) and two oxygen-annealed samples (red curves). Triacid-cleaned surfaces show a higher density of unoccupied states (shaded regions). (i) UPS spectra with excitation energy 21.2 eV of the triacid-cleaned (blue) and oxygen-annealed (red) surfaces. The oxygen-annealed surface exhibits a higher positive electron affinity E$_A$ (+2.14 eV) than the triacid-cleaned surface (+0.92 eV).(j-k) XPS (j) and LEED (k) suggests the surface atomic structure of the two samples, triacid-cleaned and oxygen-annealed diamond. Images adapted with permission from refs. \cite{oh_structure_2024} (a-d), \cite{sangtawesin_origins_2019}(e-k). }
    \label{scanning_probes}
\end{figure}

X-ray photoelectron spectroscopy (XPS) has become the most widely applied surface analysis tool in the SC qubit community, used to identify the chemical composition and oxidation states of superconductor surface oxides across essentially every metallization in use, including Nb, NbTi, NbTiN, TiN, Al, and Ta, and to track oxide regrowth after surface treatments and the effect of substrate cleaning on oxide composition \cite{verjauw_investigation_2021, TorresCastanedo2024, Altoe2022, davies_atmospheric_2020, melville_comparison_2020} (Figs. 13a-b). Two recent applications illustrate the technique's evolving sophistication. The first is depth resolution. Variable-energy XPS using synchrotron radiation, spanning incident photon energies from 630 to 6000 eV, resolves not just which Ta oxidation states are present but how they are distributed through the film depth and how that distribution responds to chemical treatments, converting XPS from a surface-averaged fingerprint into a non-destructive depth profile \cite{McLellan2023}. The second is discriminating power. The comparative XPS study of Nb and Ta surface oxides \cite{oh_structure_2024} shows that the Nb $3d$ spectrum requires Nb$^{5+}$, Nb$^{4+}$, and Nb$^0$ components while the Ta 4f spectrum is entirely explained by a single Ta$^{5+}$ component, corroborating the EELS-derived suboxide picture of Section \ref{niobium_tantalum_oxide} and illustrating how XPS and STEM-EELS provide mutually reinforcing evidence when applied to the same system. XPS remains limited for the lightest elements. Hydrogen is completely invisible to it \cite{TorresCastanedo2024}, and detection limits for other light elements are typically at the parts-per-thousand level. This blindness matters because surface hydroxylation of Al$_2$O$_3$ upon ambient exposure produces a complex cocktail of OH-terminated and hydrogen-bonded surface defects whose TLS activity is plausible but unconfirmed at the single-defect level. This is a defect population that the field's most widely used surface analysis tool cannot see at all.

Secondary ion mass spectrometry (SIMS) addresses the hydrogen detection problem directly. SIMS has parts-per-million sensitivity for most elements including hydrogen, and its depth profiling capability allows compositional mapping through multilayer structures with nanometer depth resolution (Fig.s 13c-d). The previous TOF-SIMS results \cite{murthy_developing_2022, oh_structure_2024} discussed in Section \ref{niobium_tantalum_oxide}, namely the fabrication-process contribution to oxygen diffusion in NbO$_x$ and the order-of-magnitude hydrogen accumulation at the uncapped Nb oxide-metal interface, were accessible only through this depth-profiling and light-element sensitivity. They illustrate the role SIMS plays in a multimodal pipeline, supplying, on the same device materials, the chemical species and process-history information that electron microscopy cannot.

Synchrotron-based techniques, including x-ray absorption spectroscopy (XAS), x-ray magnetic circular dichroism (XMCD), and resonant inelastic x-ray scattering (RIXS), provide capabilities beyond what laboratory instruments can achieve (Figs. 13e-k). XAS at higher pressures than UHV methods allows characterization of surface layers forming under near-ambient conditions, relevant to understanding oxide growth during device processing. XMCD has been used to study cryogenic surface adsorption on Al and Nb films and to correlate the presence of specific adsorbates with SQUID $1/f$ flux noise \cite{kumar_origin_2016}. Premkumar et al. \cite{Premkumar2021} combined XPS with RIXS using synchrotron radiation to investigate Nb surface oxide composition in transmon circuits, revealing a correlation between qubit relaxation times, grain boundary oxygen diffusion, and near-surface oxide content. This remains one of the more comprehensive structure–coherence correlation studies in the SC qubit literature to date.

The most instructive example of Gap 2 closure through surface spectroscopy comes from the color center community. Sangtawesin et al. \cite{sangtawesin_origins_2019} correlated XPS measurements (Figs. 13e, j) of diamond surface chemistry with NV spin coherence measurements on the same sample, establishing that rough surface morphology, dangling bonds, and disorder in surface termination produce electronic traps that generate broadband magnetic noise. By systematically varying surface termination and measuring both XPS signatures and NV coherence, they established a causal chain from specific surface chemical states to quantum decoherence. This methodology, in which one variable is changed and both structure and quantum property are measured at each step, is the template that the broader QIS characterization community needs to adopt, regardless of platform.

\subsection{Scanning Probe Characterization: Spatially Resolved Physical Measurements \label{scanning_probe_characterization}}

Scanning probe microscopy techniques occupy a conceptually distinct position in the QIS characterization landscape. Unlike STEM and surface spectroscopy, which measure structural or chemical properties and infer quantum implications, several scanning probe methods measure physical quantities such as magnetic field, microwave impedance, and electric potential with spatial resolution, and can therefore be positioned closer to quantum properties on the information axis defined by de Graaf et al. \cite{de_graaf_chemical_2022} (Figure \ref{scanning_probes}).

Scanning SQUID-on-tip and scanning Hall bar microscopy provide spatially resolved measurements of local magnetic flux with single-spin sensitivity, enabling detection of magnetic disorder on surfaces and near interfaces. Notably, Vasyukov et al. \cite{vasyukov_scanning_2013} demonstrated scanning SQUID with sub-electron sensitivity sufficient to detect individual magnetic moments at surfaces. For qubit applications, this translates to the ability to map surface spin densities and spatial distributions, information directly relevant to flux noise in tunable qubits. The Pelliccione et al. \cite{pelliccione_scanned_2016} scanning NV magnetometry study demonstrated imaging of magnetic vortices in a superconductor at cryogenic temperatures, establishing that NV-based scanning probes can resolve field features at the length scales relevant to superconducting devices.

Microwave impedance microscopy (MIM) provides spatially resolved dielectric permittivity and conductivity at microwave frequencies, making it sensitive to the same TLS-hosting dielectric layers that limit qubit coherence. Applied to qubit-relevant substrates and thin films, MIM can in principle map spatial variations in dielectric loss, identifying regions of elevated TLS density, at length scales comparable to the qubit's electric field distribution. This would directly address the spatial dimension of the TLS problem and is a more natural complement to STEM than broad-area surface spectroscopy, since both yield spatially resolved information.

Scanning gate microscopy applied to semiconductor quantum dot devices has demonstrated the ability to map the local electric field profile of a device while monitoring its charge state in situ. Oh et al. \cite{oh_cryogen-free_2021} implemented this at millikelvin temperatures on Si/SiGe quantum dot devices, using the device itself as a local charge sensor to reconstruct the electrostatic environment with nanometer spatial resolution. This approach represents a qualitatively different strategy for Gap 2 closure, using the qubit as its own probe, and is conceptually related to the in-operation TLS spectroscopy techniques discussed in Section \ref{mats_q_photonics}, but with the added dimension of spatial mapping.

The frontier of scanning probe characterization for QIS lies in techniques that combine quantum-coherent probes with scanning capability. The scanning transmon proposed and prototyped by Shanks et al. \cite{shanks_scanning_2013} uses a high-coherence superconducting circuit as the scanning element, potentially enabling direct detection of individual TLS at their location in the device. Near-field scanning microwave microscopy at single-photon power levels \cite{geaney_near-field_2019} extends conventional microwave microscopy into the quantum regime where TLS coupling can be directly probed. These techniques remain at an early stage of development, with mechanical stability at millikelvin temperatures the primary engineering challenge, but they represent the most direct route that currently exists to closing Gap 1, the spatial access to individual TLS.

\subsection{Toward Direct Structure-Coherence Correlation: State of The Art and The Path Forward \label{toward_structure-coherence correlation}}

Surveying the techniques above, a consistent pattern emerges. The characterization community has made substantial progress in identifying the candidates for decoherence, including surface oxides, interface disorder, paramagnetic species, and physisorbed molecules, and in developing techniques that reveal their structural and chemical character. What remains largely unachieved is the direct, quantitative, sample-matched connection from a specific structural or chemical observation to a specific change in a quantum property.

The studies surveyed in this section define a hierarchy of proximity to that goal. At the ensemble level, the correlation of Nb oxide chemistry with resonator quality factor \cite{Altoe2022} and of diamond surface chemistry with NV spin coherence \cite{sangtawesin_origins_2019} remain the field's clearest demonstrations of the causal chain. These are now bracketed from the device side by resonator measurements that quantitatively apportion loss between surface and bulk TLS baths \cite{Crowley2023} and by mitigation strategies, such as surface encapsulation and oxide passivation, whose success confirms the structural diagnosis \cite{Bal2024}. One step deeper, mechanistic explanation has begun to replace correlation. The crystal-field splitting argument for why amorphous Ta$_2$O$_5$ largely resists TLS formation while amorphous Nb$_2$O$_5$ promotes it, together with its CALPHAD thermodynamic grounding \cite{oh_structure_2024} [n], moves the field from observing a coherence advantage to explaining one, and fluctuation electron microscopy of disordered Nb$_2$O$_5$21 supplies the most specific structural hypothesis yet proposed for the atomic identity of the TLS in that system. Spatial specificity has advanced in parallel. Multimodal STEM of transmon devices at sub-6 K \cite{pham_structure_2024} and the resolution of multiple coexisting loss channels within a single device architecture, namely surface oxide, metal-substrate silicide, and tunnel barrier electronic structure variation \cite{lim_unique_2025}, tie the structural picture to specific interfaces of specific devices rather than to model films. The most complete closures to date have come at the single-emitter level, where cryo-STEM-CL has mapped ZPL energy and brightness simultaneously with grain boundary structure and nanoscale strain on the same nanodiamond \cite{angell_unraveling_2024}, and has identified the carbon dimer responsible for the 440 nm blue emission in h-BN while demonstrating its deterministic creation \cite{hou_nanometer_2025}.

The cross-platform view this survey enables is itself informative. In superconducting qubits, surfaces and interfaces host TLS that limit microwave quality factors. In color centers, surface chemistry and near-surface structural disorder degrade spin coherence and optical linewidths. In 2D material heterostructures, interface contamination, local strain, and edge termination affect both spin and photonic properties. The structural character of these decoherence sources differs across platforms, but the methodological challenge is identical. The cryo-STEM-CL approach \cite{angell_unraveling_2024, hou_nanometer_2025}, the multimodal STEM pipeline \cite{lim_unique_2025, pham_structure_2024}, and the sample-matched spectroscopy strategy \cite{sangtawesin_origins_2019} are all transferable across platform boundaries, and cross-platform collaboration between characterization groups represents an underutilized opportunity to accelerate progress in ways that platform-siloed efforts cannot.

The path toward genuine structure-coherence correlation requires three converging developments. First, sample-matched measurement workflows must become standard rather than exceptional, with devices carrying documented quantum properties distributed to structural characterization groups and results compared on the same region of the same device wherever possible. The condensed matter cryo-EM community demonstrated this workflow first, for instance, in the combined STEM/EELS and transport study of interface-enhanced superconductivity in FeSe/SrTiO$_3$ \cite{zhu_cryogenic_2021}. Second, characterization throughput must increase to enable statistically meaningful datasets. For STEM, this requires faster and more automated FIB preparation workflows, and adoption of plasma FIB systems that reduce preparation artifacts. For the broader pipeline, pre-characterized sample libraries with standardized quantum metrics, developed as community resources through national laboratory programs, are necessary to connect individual characterization results into a coherent statistical picture. Third, techniques operating at the correct energy and length scale must be developed further. High-field EPR combined with DFT, as demonstrated by Un et al. \cite{un_nature_2022} for surface radicals on $\alpha$-Al$_2$O$_3$, provides chemical identity for paramagnetic TLS candidates at energy scales approaching the quantum circuit regime. Cryo-STEM at sub-10 K \cite{pham_structure_2024} brings structural characterization physically closer to device operating conditions. Cryo-STEM-CL \cite{angell_unraveling_2024, hou_nanometer_2025} offers the possibility of directly correlating optical quantum properties with the structural features responsible for them at nanometer resolution. Its extension to near-surface NV centers, shallow SiV$^-$ implants, and 2D heterostructure emitters represents an immediate frontier. Quantum scanning probe techniques such as scanning transmons and near-field microwave microscopy at single-photon power offer the possibility of directly localizing individual TLS and correlating their positions with structural features visible to conventional STEM or AFM on the same sample.

The characterization challenge in QIS is ultimately a coordination challenge as much as a technical one. The tools to make progress exist. What is needed is the infrastructure, the community practice, and the explicit commitment to close the causal chain from atomic structure to quantum coherence, rather than merely accumulating correlations and inferring the connection.

%% file: 7.outlook.tex

\section{Summary and Outlook}

The first phase of quantum information science established that coherent control of individual quantum states is physically realizable. The surveys assembled in this review document, each from its own vantage point, that the second phase will be decided by advances in materials. Superconducting qubit performance is now set by two-level system loss in disordered dielectrics at surfaces and interfaces rather than by circuit design (Section 2). Quantum defect platforms are limited less by the intrinsic properties of known centers than by deterministic fabrication, charge-state stability, and surface noise (Section 3). Quantum photonic devices meet their hardest constraints in the hydrogen chemistry, electro-optic drift, and domain uniformity of their constituent thin films (Section 4). Two-dimensional heterostructures offer atomically clean interfaces in principle but remain limited by contamination, strain, and twist-angle inhomogeneity in practice (Section 5). And across all of these platforms, the causal chain from a specific structural feature to a specific quantum property remains largely inferred rather than measured (Section 6). Materials science is no longer a supporting discipline for QIS. It has become the rate-limiting one, and the Quantum Evolution 2.0 will advance at the pace at which its materials problems are solved.

The platform surveys also reveal how much each community has already built that the others need. The superconducting community possesses the most mature loss metrology, including resonator proxy devices, participation-ratio analysis, and methods that apportion loss among specific interfaces, all of which could quantify dielectric and interface loss in photonic and 2D devices. The color-center community holds the deepest experience in surface termination chemistry and isotopic engineering, expertise directly applicable to shallow spin defects in 2D hosts and to surface-limited superconducting resonators. Photonic materials development proceeds under CMOS-compatible thermal budgets and wafer-scale process discipline that qubit fabrication is only beginning to adopt. From the 2D community comes the demonstration that van der Waals assembly can produce interfaces free of the bonding defects that plague deposited films, and the adoption of crystalline h-BN as a low-loss dielectric in superconducting circuits shows that such transfers already yield measurable coherence gains (Section 5). The characterization community, finally, has established the correlative and sample-matched workflows that every platform requires (Section 6). These exchanges have so far been episodic. Making them systematic, through shared benchmark samples, common reporting standards, and deliberately cross-platform collaboration, is among the least expensive accelerations available to the field.

The deeper unity lies in the problem itself. The loss inventories compiled independently by each community converge on the same materials physics. Surfaces and interfaces dominate decoherence in every solid-state platform. The implicated chemistry recurs with striking regularity, whether as hydrogen in niobium and tantalum oxides and in silicon nitride, oxygen vacancies in tantala and lithium niobate, or carbon-related species in silicon, diamond, and h-BN. These are light elements at low concentrations in disordered or buried environments, precisely the regime in which conventional structural probes perform worst. We refer to this convergence as the structure-coherence problem. In no platform can the field yet point to a specific atomic-scale structure observed in a specific device and connect it quantitatively to a measured change in coherence. Until that connection is routine, materials optimization will remain empirical, gains will remain difficult to transfer between laboratories, and nominally identical devices will continue to perform differently for reasons no measurement can articulate.

Resolving the structure-coherence problem defines what materials science must deliver for the Quantum Evolution 2.0, and three needs stand out. The first is mechanistic understanding, at the atomistic level, of which materials features drive decoherence. Candidate structures now exist, among them hydrogen tunneling centers in Nb$_2$O$_5$, paramagnetic surface species, and specific defect complexes in h-BN, but candidates must become identified mechanisms through the combination of first-principles theory, targeted synthesis, and atomic-resolution characterization described in Sections 3 and 6. The second is proxy metrology. Device performance today is known only after complete fabrication and cryogenic measurement, a cycle that consumes weeks per iteration and suppresses the statistics that materials optimization requires. The field needs materials properties measurable at high throughput, and at room temperature wherever possible, that predict quantum performance before a device is built. Residual resistivity ratio, terahertz conductivity, and resonator-based screening are early examples (Section 2), but no descriptor has yet been validated across fabrication facilities, let alone across platforms. The third is characterization developed specifically for QIS materials. The required tools must detect light elements at part-per-million densities in buried interfaces, operate at cryogenic and ideally operando conditions, and measure structure and quantum properties on the same sample. Cryo-STEM, correlated cathodoluminescence, and quantum scanning probes (Section 6) mark the frontier, and their extension from demonstration experiments to routine practice is itself a materials-instrumentation challenge.

Meeting these needs would pay out against the three challenges that define the coming decade of quantum hardware. The first is performance. Coherence times in every platform now terminate in materials-based loss, so each mechanism identified and eliminated converts directly into longer-lived qubits, narrower emitter linewidths, and lower-loss photonic circuits. The second is scale. Wafer-level uniformity, device-to-device reproducibility, and industrially compatible processes are materials problems before they are engineering ones, as the frequency-yield and twist-angle examples of Sections 2 and 5 make explicit. The third, and least developed, is integration. A quantum internet or a modular processor will couple dissimilar platforms on one chip, for instance a superconducting qubit exchanging quantum states with a color center through a microwave-to-optical transducer. Every such interface stacks the materials constraints of both platforms, from incompatible thermal budgets to chemically abrupt heterointerfaces between oxides, nitrides, semiconductors, and superconductors, and the transduction example of Section 4 shows how severely materials nonidealities already tax these hybrid devices.

This convergence is what makes the present survey timely. Over the past five years \cite{Leon2021} the platform communities have arrived independently at the same recognition, that progress is limited by surfaces, interfaces, and defects they cannot yet fully identify, yet that understanding remains dispersed across a literature still organized by platform. Collecting the evidence in one place and stating the shared problems explicitly serves two purposes. It lowers the entry barrier for the materials scientists, microscopists, and theorists whose expertise the field now needs most, and it gives the platform communities a common frame for problems that none of them will solve alone. The timing also matters for a practical reason. The coming years will settle long-lived choices, among them which materials systems are carried into foundry-scale fabrication, which proxy metrics become accepted standards, and which characterization infrastructure is built as shared facilities. Those decisions will be sounder if they are made with the full cross-platform materials picture in view, and providing that picture is the purpose of this review.